\documentclass[11pt]{article} 

\usepackage{textcomp, amssymb}  
\usepackage{anysize}            
\usepackage{amsfonts}
\usepackage{amsmath}
\usepackage{dsfont}
\usepackage{MnSymbol}
\usepackage{mathtools}
\usepackage{graphicx}
\usepackage{epsfig}
\usepackage{epstopdf}
\usepackage{tikz}
\usepackage{amsthm}
\usepackage{mathrsfs}
\usepackage{lmerge}
\usepackage{ifpdf}
\usepackage{thm-restate}
\usepackage{stmaryrd}
\usepackage{cleveref}
\usepackage{subcaption}
\usepackage{forest}
\usepackage{longtable}

\usetikzlibrary{calc,decorations.pathmorphing,shapes,graphs,positioning,automata,angles,cd,trees,shapes.geometric}

\newcounter{sarrow}

\newcounter{sarrow1}
\newcommand\xnrsquigarrow[1]{%
\stepcounter{sarrow1}%
\mathrel{\begin{tikzpicture}[baseline= {( $ (current bounding box.south) + (0,-0.5ex) $ )}]
\node[inner sep=.5ex] (\thesarrow) {$\scriptstyle #1$};
\path[draw,<-,decorate,
  decoration={zigzag,amplitude=0.7pt,segment length=1.2mm,pre=lineto,pre length=4pt}]
    (\thesarrow1.south east) -- (\thesarrow1.south west);
    $\slashedarrowfill@\relbar\relbar/$
    \end{tikzpicture}}%
}

\makeatletter
\def\slashedarrowfill@#1#2#3#4#5{%
  $\m@th\thickmuskip0mu\medmuskip\thickmuskip\thinmuskip\thickmuskip
   \relax#5#1\mkern-7mu%
   \cleaders\hbox{$#5\mkern-2mu#2\mkern-2mu$}\hfill
   \mathclap{#3}\mathclap{#2}%
   \cleaders\hbox{$#5\mkern-2mu#2\mkern-2mu$}\hfill
   \mkern-7mu#4$%
}
\def\rightslashedarrowfillb@{%
  \slashedarrowfill@\relbar\relbar/\rightarrow}
\newcommand\xnrightarrow[2][]{%
  \ext@arrow 0055{\rightslashedarrowfillb@}{#1}{#2}}

\def\rightslashedarrowfille@{%
  \slashedarrowfill@\relbar\relbar/\twoheadrightarrow}
\newcommand\xntworightarrow[2][]{%
  \ext@arrow 0055{\rightslashedarrowfille@}{#1}{#2}}

\def\rightslashedarrowfillg@{%
  \slashedarrowfill@\relbar\relbar{\raisebox{.12em}{}}\twoheadrightarrow}
\newcommand\xtworightarrow[2][]{%
  \ext@arrow 0055{\rightslashedarrowfillg@}{#1}{#2}}

\def\rightslashedarrowfillx@{%
  \slashedarrowfill@\Relbar\Relbar/\rightrightarrows}
\newcommand\xnTworightarrow[2][]{%
  \ext@arrow 0055{\rightslashedarrowfillx@}{#1}{#2}}

\def\rightslashedarrowfilly@{%
  \slashedarrowfill@\Relbar\Relbar{\raisebox{.12em}{}}\rightrightarrows}
\newcommand\xTworightarrow[2][]{%
  \ext@arrow 0055{\rightslashedarrowfilly@}{#1}{#2}}

\pgfdeclareshape{slash underlined}
{
  \inheritsavedanchors[from=rectangle] 
  \inheritanchorborder[from=rectangle]
  \inheritanchor[from=rectangle]{north}
  \inheritanchor[from=rectangle]{north west}
  \inheritanchor[from=rectangle]{north east}
  \inheritanchor[from=rectangle]{center}
  \inheritanchor[from=rectangle]{west}
  \inheritanchor[from=rectangle]{east}
  \inheritanchor[from=rectangle]{mid}
  \inheritanchor[from=rectangle]{mid west}
  \inheritanchor[from=rectangle]{mid east}
  \inheritanchor[from=rectangle]{base}
  \inheritanchor[from=rectangle]{base west}
  \inheritanchor[from=rectangle]{base east}
  \inheritanchor[from=rectangle]{south}
  \inheritanchor[from=rectangle]{south west}
  \inheritanchor[from=rectangle]{south east}
  \inheritanchorborder[from=rectangle]
  \foregroundpath{
    \southwest \pgf@xa=\pgf@x \pgf@ya=\pgf@y
    \northeast \pgf@xb=\pgf@x \pgf@yb=\pgf@y
    \pgf@xc=\pgf@xa
    \advance\pgf@xc by .5\pgf@xb
    \pgf@yc=\pgf@ya 
    \advance\pgf@xc by -1.3pt
    \advance\pgf@yc by -1.8pt
    \pgfpathmoveto{\pgfqpoint{\pgf@xc}{\pgf@yc}}
    \advance\pgf@xc by  2.6pt
    \advance\pgf@yc by  3.6pt
    \pgfpathlineto{\pgfqpoint{\pgf@xc}{\pgf@yc}}
    \pgfpathmoveto{\pgfqpoint{\pgf@xa}{\pgf@ya}}
    \pgfpathlineto{\pgfqpoint{\pgf@xb}{\pgf@ya}}
 }
}
\tikzset{nomorepostaction/.code=\let\tikz@postactions\pgfutil@empty}

\newcommand{\makecircled}[2][\mathord]{#1{\mathpalette\make@circled{#2}}}
\newcommand{\make@circled}[2]{%
  \begingroup\m@th
  \vphantom{\bigcirc}%
  \ooalign{$#1\bigcirc$\cr\hidewidth$#1\make@smaller{#1}{#2}$\hidewidth\cr}%
  \endgroup
}
\newcommand{\make@smaller}[2]{%
  \vcenter{\hbox{\scalebox{0.7}{$\m@th#1#2$}}}%
}
\newcommand{\osharp}{\makecircled[\mathbin]{\#}}
\newcommand{\obullet}{\makecircled[\mathbin]{\bullet}}

\newcommand{\osum}{ 
  \mathop{
    \mathchoice
      {\buildosum{\displaystyle}{0.1}}
      {\buildosum{\textstyle}{0.075}}
      {\buildosum{\scriptstyle}{0.075}}
      {\buildosum{\scriptscriptstyle}{0.075}}
  }\displaylimits 
}
\newcommand\buildosum[2]{%
  \begin{tikzpicture}[baseline=(char.base), inner sep=0, outer sep=0]
    \draw (-0.3ex,0) circle (#2);
    \node (char) at (0,0) {$#1\sum$};
  \end{tikzpicture}%
}
\DeclareMathOperator{\ucup}{\mathcal{U}}

\newcommand{\sembrack}[1]{\llbracket #1\rrbracket}
\newcommand*{\rmbrace}{|\mskip-4mu\}}
\newcommand*{\lmbrace}{\{\mskip-4mu|}
\newcommand*{\mset}[1]{\lmbrace#1\rmbrace}
\newcommand*{\rsbrace}{|\mskip-4mu\rangle}
\newcommand*{\lsbrace}{\langle\mskip-4mu|}
\newcommand*{\step}[1]{\lsbrace#1\rsbrace}

\newcommand{\pretesting}{\mathrel{\ooalign{\raise0.13ex\hbox{$\sqsubset$}\cr\hidewidth\raise-0.6ex\hbox{\scalebox{0.9}{$\sim$}}\hidewidth\cr}}}

\newcommand{\pretestingeq}{\bar{\sim}}

\newcommand{\pretestingmust}{\pretesting_{\mathrm{MUST}}}

\newcommand{\pretestingmusteq}{\pretestingeq_{\mathrm{MUST}}}

\newcommand{\pretestingmay}{\pretesting_{\mathrm{MAY}}}

\newcommand{\pretestingmayeq}{\pretestingeq_{\mathrm{MAY}}}

\newcommand{\prepomset}{\lesssim_{\mathrm{P}}}

\newcommand{\prestep}{\lesssim_{\mathrm{S}}}

\newcommand{\prehp}{\lesssim_{\mathrm{HP}}}

\newcommand{\prehhp}{\lesssim_{\mathrm{HHP}}}

\newcommand{\prepomsets}[1]{\lesssim_{\mathrm{(P,#1)}}}

\newcommand{\presteps}[1]{\lesssim_{\mathrm{(S,#1)}}}

\newcommand{\prehps}[1]{\lesssim_{\mathrm{(HP,#1)}}}

\newcommand{\prehhps}[1]{\lesssim_{\mathrm{(HHP,#1)}}}

\newcommand{\preeq}{{\mathrel{\ooalign{\raise0.13ex\hbox{$=$}\cr\hidewidth\raise-0.6ex\hbox{\scalebox{0.9}{$\sim$}}\hidewidth\cr}}}}

\newcommand{\prepomseteq}{\preeq_{\mathrm{P}}}

\newcommand{\prestepeq}{\preeq_{\mathrm{S}}}

\newcommand{\prehpeq}{\preeq_{\mathrm{HP}}}

\newcommand{\prehhpeq}{\preeq_{\mathrm{HHP}}}

\newcommand{\preweakeq}{{\mathrel{\ooalign{\raise0.13ex\hbox{$=$}\cr\hidewidth\raise-0.6ex\hbox{\scalebox{0.9}{$\approx$}}\hidewidth\cr}}}}

\newcommand{\lub}{\mathord{\scalebox{1.3}{$\bigvee$}}}
\newcommand{\must}{~\mathsf{must}~}
\newcommand{\may}{~\mathsf{may}~}
\newtheorem{theorem}{Theorem}[section]
\newtheorem{definition}[theorem]{Definition}
\newtheorem{proposition}[theorem]{Proposition}
\newtheorem{lemma}[theorem]{Lemma}

\newtheorem{corollary}[theorem]{Corollary}

\newcommand{\Fin}{\mathrm{Fin}}
\newcommand{\App}{\mathrm{App}}
\newcommand{\Abs}{\mathrm{Abs}}
\begin{document}

\begin{titlepage}
\thispagestyle{empty}

\hrule
\begin{center}
{\bf\LARGE Theories of Truly Concurrent Processes\\}

\vspace{0.5cm}
--- Yong Wang ---

\vspace{2cm}

\end{center}
\end{titlepage}

\newpage 

\setcounter{page}{1}\pagenumbering{roman}

\tableofcontents

\newpage
\setcounter{page}{1}\pagenumbering{arabic}

        \section{Introduction}\label{intro}

A process is able to execute a set of actions with a predefined manner, while a truly concurrent process executes this set of actions with a manner with the flavour of true concurrency. 

The so-called truly concurrent process algebras \cite{APTC} \cite{APTC2} bridge the true concurrency (such as Petri nets \cite{PN00} \cite{PN01} \cite{PN02}, event structures \cite{ES}, etc), and the interleaving concurrency (such as CCS \cite{CCS}, CSP \cite{CSP1} \cite{CSP2}, ACP \cite{ACP}, etc). 

In this paper, we give truly concurrent processes testing semantics following Hennessy's great work \cite{TS}, which inherits the trinity of operational semantics, axiomatic semantics and denotational semantics. 

And also, we give truly concurrent processes bisimulation semantics with the trinity of operational semantics, axiomatic semantics and denotational semantics.
\newpage\section{Preliminaries}\label{pre} 

For self-satisfactory, in this section, we introduce the preliminaries on set in \cref{set} and multiset in \cref{multiset}.

\subsection{Set}\label{set}

\begin{definition}[Set]
A set contains some objects, and let $\{-\}$ denote the contents of a set. For instance, $\mathbb{N}=\{1,2,3,\cdots\}$. Let $a\in A$ denote that $a$ is an element of the set $A$ and $a\notin A$ denote that $a$ is not an element of the set $A$. For all $a\in A$, if we can get $a\in B$, then we say that $A$ is a subset of $B$ denoted $A\subseteq B$. If $A\subseteq B$ and $B\subseteq A$, then $A=B$. We can define a new set by use of predicates on the existing sets, such that $\{n\in\mathbb{N}|\exists k\in\mathbb{N},n=2k\}$ for the set of even numbers. We can also specify a set to be the least set satisfy some inductive inference rules, for instance, we specify the set of even numbers $A$ satisfying the following rules:

$$\frac{}{0\in A}\quad\quad\frac{n\in A}{n+2\in A}$$
\end{definition}

\begin{definition}[Set composition]
The union of two sets $A$ and $B$, is denoted by $A\cup B=\{a|a\in A\mbox{ or }a\in B\}$, and the intersection of $A$ and $B$ by $A\cap B=\{a|a\in A\mbox{ and }a\in B\}$, the difference of $A$ and $B$ by $A\setminus B=\{a|a\in A\mbox{ and }a\notin B\}$. The empty set $\emptyset$ contains nothing. The set of all subsets of a set $A$ is called the powerset of $A$ denoted $2^A$.
\end{definition}

\begin{definition}[Tuple]
A tuple is a finite and ordered list of objects and denoted $\langle -\rangle$. For sets $A$ and $B$, the Cartesian product of $A$ and $B$ is denoted by $A\times B=\{\langle a,b\rangle|a\in A,b\in B\}$. $A^n$ is the $n$-fold Cartesian product of set $A$, for instance, $A^2=A\times A$. Tuples can be flattened as $A\times (B\times C)=(A\times B)\times C=A\times B\times C$ for sets $A$, $B$ and $C$.
\end{definition}

\begin{definition}[Relation]
A relation $R$ between sets $A$ and $B$ is a subset of $A\times B$, i.e., $R\subseteq A\times B$. We say that $R$ is a relation on set $A$ if $R$ is a relation between $A$ and itself, and,

\begin{itemize}
  \item $R$ is reflexive if for all $a\in A$, $aRa$ holds; it is irreflexive if for all $a\in A$, $aRa$ does not hold.
  \item $R$ is symmetric if for all $a,a'\in A$ with $aRa'$, then $a'Ra$ holds; it is antisymmetric if for all $a,a'\in A$ with $aRa'$ and $a'Ra$, then $a=a'$.
  \item $R$ is transitive if for all $a,a',a''\in A$ with $aRa'$ and $a'Ra''$, then $aRa''$ holds.
\end{itemize}
\end{definition}

\begin{definition}[Preorder, partial order, strict order]
If a relation $R$ is reflexive and transitive, we call that it is a preorder; When it is a preorder and antisymmetric, it is called a partial order, and a partially ordered set (poset) is a pair $\langle A,R\rangle$ with a set $A$ and a partial order $R$ on $A$; When it is irreflexive and transitive, it is called a strict order.
\end{definition}

\begin{definition}[Equivalence]
A relation $R$ is called an equivalence, if it is reflexive, symmetric and transitive. For an equivalent relation $R$ and a set $A$, $[a]_R=\{a'\in A|aRa'\}$ is called the equivalence class of $a\in A$.
\end{definition}

\begin{definition}[Relation composition]
For sets $A$, $B$ and $C$, and relations $R\subseteq A\times B$ and $R'\subseteq B\times C$, the relational composition denoted $R\circ R'$, is defined as the least relation $a(R\circ R')c$ satisfying $aRb$ and $bR'c$ with $a\in A$, $b\in B$ and $c\in C$. For a relation $R$ on set $A$, we denote $R^*$ for the reflexive and transitive closure of $R$, which is the least reflexive and transitive relation on $A$ that contains $R$.
\end{definition}

\begin{definition}[Function]
A function $f:A\rightarrow B$ from sets $A$ to $B$ is a relation between $A$ and $B$, i.e., for every $a\in A$, there exists one $b=f(a)\in B$, where $A$ is called the domain of $f$ and $B$ the codomain of $f$. $\sembrack{-}$ is also used as a function with $-$ a placeholder, i.e., $\sembrack{x}$ is the value of $\sembrack{-}$ for input $x$. A function $f$ is a bijection if for every $b\in B$, there exists exactly one $a\in A$ such that $b=f(a)$. For functions $f:A\rightarrow B$ and $g:B\rightarrow C$, the functional composition of $f$ and $g$ denoted $g\circ f$ such that $(g\circ f)(a)=g(f(a))$ for $a\in A$.
\end{definition}

\subsection{Multiset}\label{multiset}

\begin{definition}[Labelled poset]
A labelled poset is a tuple $\mathbf{u}=\langle S, \leq, \lambda\rangle$, where $S$ is the carrier set, $\leq$ is a partial order on $S$ and $\lambda$ is a labelling function $\lambda:S\rightarrow\Sigma$.

For a labelled poset $\mathbf{u}$, $S_{\mathbf{u}}$, $\leq_{\mathbf{u}}$ and $\lambda_{\mathbf{u}}$ denote the carrier, the partial order and the labelling of $\mathbf{u}$ respectively. The set of labelled posets is denoted $\mathsf{LP}$ and the empty labelled poset is $1$.
\end{definition}

\begin{definition}[Poset morphism]
For posets $\langle A,\leq\rangle$ and $\langle A',\leq'\rangle$ and function $\varphi:A\rightarrow A'$, $f$ is called a poset morphism if for $a_0,a_1\in A$ with $a_0\leq a_1$, then $\varphi(a_0)\leq' \varphi(a_1)$ holds.
\end{definition}

\begin{definition}[Labelled poset isomorphism]
Let $\mathbf{u}=\langle S_1,\leq_1,\lambda_1\rangle$ and $\mathbf{v}=\langle S_2,\leq_2,\lambda_2\rangle$ be labelled posets. A labelled poset morphism $\varphi$ from $\mathbf{u}=\langle S_1,\leq_1,\lambda_1\rangle$ to $\mathbf{v}=\langle S_2,\leq_2,\lambda_2\rangle$ is a poset morphism from $\langle S_1,\leq_1\rangle$ and $\langle S_2,\leq_2\rangle$ with $\lambda_2\circ \varphi=\lambda_1$. Moreover, $\varphi$ is a labelled poset isomorphism if it is a bijection with $\varphi^{-1}$ is a poset isomorphism from $\langle S_2,\leq_2,\lambda_2\rangle$ to $\langle S_1,\leq_1,\lambda_1\rangle$. We say that $\mathbf{u}=\langle S_1,\leq_1,\lambda_1\rangle$ is isomorphic to $\mathbf{v}=\langle S_2,\leq_2,\lambda_2\rangle$ denoted $\langle S_1,\leq_1,\lambda_1\rangle\sim\langle S_2,\leq_2,\lambda_2\rangle$, if there exists a poset isomorphism $\varphi$ between $\langle S_1,\leq_1,\lambda_1\rangle$ and $\langle S_2,\leq_2,\lambda_2\rangle$.
\end{definition}

\begin{definition}[Multiset]
A multiset is a kind of set of objects which may be repetitive denoted $\mset{-}$, such that $\mset{a,b,b}$ is significantly distinguishable from $\mset{a,b}$.
\end{definition}

\begin{definition}[Pomset]
A partially ordered multiset, pomset, is a $\sim$-equivalence class of posets. The set of pomsets is denoted $\mathsf{Pom}$; the empty pomset is denoted $\mathbf{1}$ and the $\sim$-equivalence class of $1$ is also denoted by $1$. If there does not exist partial orders between any two objects in a pomset, such pomset is called a step denoted $\step{-}$. The set of steps is denoted $\mathsf{Stp}$.
\end{definition}

\begin{definition}[Strict pomset]\label{strictpomset}
For a step $\step{a,b}$, there is no partial orders between $a$ and $b$. With a little abuse of concepts, a reflexive partial order $\leq$ contains two cases: one for strict partial order $<$ and the other for $=$. We let $\step{a,b}=\mset{a,b|a\leq b}\cup\mset{a,b|b\leq a}$, and each $\mset{a,b|a\leq b}$ or $\mset{a,b|b\leq a}$ is called a strict pomset. The set of strict pomsets is denoted $\mathsf{SPom}$; the empty strict pomset is denoted $\mathbf{1}$ and the $\sim$-equivalence class of $1$ is also denoted by $1$.
\end{definition}

We assume that the partial order $\leq$ can be divided into two kinds: execution order $\leq^{e}$ and communication $\leq^{c}$. In the same parallel branch, the partial orders usually execution orders and communication usually exists among different parallel branches. Of course, parallel branches can be nested. Then, we can get the following definitions naturally.

\begin{definition}[Labelled poset with communications]
A labelled poset with communications is a tuple $\mathbf{u}=\langle S, \leq^{e}, \leq^{c}, \lambda\rangle$, where $S$ is the carrier set, $\leq^{e}$ is an execution order on $S$, $\leq^{c}$ is a communication on $S$, and $\lambda$ is a labelling function $\lambda:S\rightarrow\Sigma$. We usually use $\mathbf{u},\mathbf{v}$ to denote labelled posets with communications. And the set of labelled posets with communications is denoted $\mathsf{LPC}$, and the empty labelled poset with communications is $1$.
\end{definition}

\begin{definition}[Labelled poset isomorphism]
Let $\mathbf{u}=\langle S_1,\leq^{e}_1,\leq^{c}_1,\lambda_1\rangle$ and $\mathbf{v}=\langle S_2,\leq^{e}_2, \leq^{c}_2,\lambda_2\rangle$ be labelled posets. A labelled poset morphism $h$ from $\langle S_1,\leq^{e}_1,\leq^{c}_1,\lambda_1\rangle$ to $\langle S_2,\leq^{e}_2,\leq^{c}_2,\lambda_2\rangle$ is a poset morphism from $\langle S_1,\leq^{e}_1,\leq^{c}_1\rangle$ and $\langle S_2,\leq^{e}_2,\leq^{c}_2\rangle$ with $\lambda_2\circ h=\lambda_1$. Moreover, $h$ is a labelled poset isomorphism if it is a bijection with $h^{-1}$ is a poset isomorphism from $\langle S_2,\leq^{e}_2,\leq^{c}_2,\lambda_2\rangle$ to $\langle S_1,\leq^{e}_1,\leq^{c}_1,\lambda_1\rangle$. We say that $\mathbf{u}=\langle S_1,\leq^{e}_1,\leq^{c}_1,\lambda_1\rangle$ is isomorphic to $\mathbf{v}=\langle S_2,\leq^{e}_2,\leq^{c}_2,\lambda_2\rangle$ denoted $\langle S_1,\leq^{e}_1,\leq^{c}_1,\lambda_1\rangle\sim\langle S_2,\leq^{e}_2,\leq^{c}_2,\lambda_2\rangle$, if there exists a poset isomorphism $h$ between $\langle S_1,\leq^{e}_1,\leq^{c}_1,\lambda_1\rangle$ and $\langle S_2,\leq^{e}_2,\leq^{c}_2,\lambda_2\rangle$.
\end{definition}

It is easy to see that $\sim$ is an equivalence and can be used to abstract from the carriers.

\begin{definition}[Pomset with communications]
A partially ordered multiset with communications, pomsetc, is a $\sim$-equivalence class of labelled posets with communications $\mathbf{u}$, written as $[\mathbf{u}]$, i.e., $[\mathbf{u}]=\{\mathbf{v}\in \mathsf{LPC}:\mathbf{u}\sim\mathbf{v}\}$. The set of pomsetcs is also denoted $\mathsf{Pom}$; the empty labelled poset with communications is denoted $1$ and the $\sim$-equivalence class of $1$ is denoted by $1$; the pomsetc containing exactly one action $a\in\Sigma$ is called primitive.
\end{definition}

Concurrency includes parallelism and communication, then, we can get the following definitions of Pomsetc compositions.

\begin{definition}[Pomsetc composition in parallel]
Let $U,V\in\mathsf{Pomc}$ with $U=[\mathbf{u}]$ and $V=[\mathbf{v}]$. We write $U\parallel V$ for the parallel composition of $U$ and $V$, which is the pomsetc represented by $\mathbf{u}\parallel\mathbf{v}$, where

$$S_{\mathbf{u}\parallel\mathbf{v}}=S_{\mathbf{u}}\cup S_{\mathbf{v}}
\quad\quad\leq^{e}_{\mathbf{u}\parallel\mathbf{v}}=\leq^{e}_{\mathbf{u}}\cup\leq^{e}_{\mathbf{v}}
\quad\quad\leq^{c}_{\mathbf{u}\parallel\mathbf{v}}=\leq^{c}_{\mathbf{u}}\cup\leq^{c}_{\mathbf{v}}
\quad\quad\lambda_{\mathbf{u}\parallel\mathbf{v}}(x)=\begin{cases}
\lambda_{\mathbf{u}},& x\in S_{\mathbf{u}};\\
\lambda_{\mathbf{v}},& x\in S_{\mathbf{v}}.
\end{cases}$$
\end{definition}

\begin{definition}[Pomsetc composition in communication]
Let $U,V\in\mathsf{Pomc}$ with $U=[\mathbf{u}]$ and $V=[\mathbf{v}]$. We write $U\mid V$ for the communicative composition of $U$ and $V$, which is the pomsetc represented by $\mathbf{u}\mid\mathbf{v}$, where

$$S_{\mathbf{u}\mid\mathbf{v}}=S_{\mathbf{u}}\cup S_{\mathbf{v}}
\quad\quad\leq^{e}_{\mathbf{u}\mid\mathbf{v}}=\leq^{e}_{\mathbf{u}}\cup\leq^{e}_{\mathbf{v}}
\quad\quad\leq^{c}_{\mathbf{u}\mid\mathbf{v}}=\leq^{c}_{\mathbf{u}}\cup\leq^{c}_{\mathbf{v}}\cup(S_{\mathbf{u}}\times S_{\mathbf{v}})
\quad\quad\lambda_{\mathbf{u}\mid\mathbf{v}}(x)=\begin{cases}
\lambda_{\mathbf{u}},& x\in S_{\mathbf{u}};\\
\lambda_{\mathbf{v}},& x\in S_{\mathbf{v}}.
\end{cases}$$
\end{definition}

\begin{definition}[Pomsetc composition in concurrency]
Let $U,V\in\mathsf{Pomc}$ with $U=[\mathbf{u}]$ and $V=[\mathbf{v}]$. We write $U\between V$ for the concurrent composition of $U$ and $V$, which is the pomsetc represented by $\mathbf{u}\between\mathbf{v}$, where

$$S_{\mathbf{u}\between\mathbf{v}}=S_{\mathbf{u}}\cup S_{\mathbf{v}}
\quad\leq^{e}_{\mathbf{u}\between\mathbf{v}}=\leq^{e}_{\mathbf{u}}\cup\leq^{e}_{\mathbf{v}}
\quad\leq^{c}_{\mathbf{u}\between\mathbf{v}}=\leq^{c}_{\mathbf{u}}\cup\leq^{c}_{\mathbf{v}}
\mbox{ or }\leq^{c}_{\mathbf{u}\between\mathbf{v}}=\leq^{c}_{\mathbf{u}}\cup\leq^{c}_{\mathbf{v}}\cup(S_{\mathbf{u}}\times S_{\mathbf{v}})
\quad\lambda_{\mathbf{u}\between\mathbf{v}}(x)=\begin{cases}
\lambda_{\mathbf{u}},& x\in S_{\mathbf{u}};\\
\lambda_{\mathbf{v}},& x\in S_{\mathbf{v}}.
\end{cases}$$
\end{definition}

\begin{definition}[Pomsetc composition in sequence]
Let $U,V\in\mathsf{Pomc}$ with $U=[\mathbf{u}]$ and $V=[\mathbf{v}]$. We write $U\cdot V$ for the sequential composition of $U$ and $V$, which is the pomsetc represented by $\mathbf{u}\cdot\mathbf{v}$, where

$$S_{\mathbf{u}\cdot\mathbf{v}}=S_{\mathbf{u}}\cup S_{\mathbf{v}}
\quad\quad\leq^{e}_{\mathbf{u}\cdot\mathbf{v}}=\leq^{e}_{\mathbf{u}}\cup\leq^{e}_{\mathbf{v}}\cup(S_{\mathbf{u}}\times S_{\mathbf{v}})
\quad\quad\leq^{c}_{\mathbf{u}\cdot\mathbf{v}}=\leq^{c}_{\mathbf{u}}\cup\leq^{c}_{\mathbf{v}}
\quad\quad\lambda_{\mathbf{u}\cdot\mathbf{v}}(x)=\begin{cases}
\lambda_{\mathbf{u}},& x\in S_{\mathbf{u}};\\
\lambda_{\mathbf{v}},& x\in S_{\mathbf{v}}.
\end{cases}$$
\end{definition}

The following definitions and conclusions are coming from \cite{CKA7}, we retype them.

\begin{definition}[Pomset types]
Let $U\in\mathsf{Pom}$, $U$ is sequential (resp. parallel) if there exist non-empty pomsets $U_1$ and $U_2$ such that $U=U_1\cdot U_2$ (resp. $U=U_1\parallel U_2$).
\end{definition}

\begin{definition}[Factorization]
Let $U\in\mathsf{Pom}$. (1) When $U=U_1\cdot\cdots \cdot U_i\cdot\cdots \cdot U_n$ with each $U_i$ non-sequential and non-empty, the sequence $U_1,\cdots,U_i,\cdots,U_n$ is called a sequential factorization of $U$. (2) When $U=U_1\parallel\cdots \parallel U_i\parallel\cdots \parallel U_n$ with each $U_i$ non-parallel and non-empty, the multiset $\mset{U_1,\cdots,U_i,\cdots,U_n}$ is called a parallel factorization of $U$.
\end{definition}

\begin{lemma}[Factorization]\label{LemmaFactorization}
Sequential and parallel factorizations exist uniquely.
\end{lemma}

\begin{lemma}
For $U\in\mathsf{Pomc}$, then the following two conclusions hold:

\begin{enumerate}
  \item $U$ is either sequential or parallel, and there are not other types in $U$.
  \item Sequential and parallel factorizations exist in $U$ uniquely.
\end{enumerate}
\end{lemma}

\begin{definition}[Series-parallel pomset]
The set of series-parallel pomset, or sp-pomsets denoted $\mathsf{SP}$, is the smallest set satisfying the following rules:

$$\frac{}{1\in\mathsf{SP}}
\quad\frac{a\in\Sigma}{a\in\mathsf{SP}}
\quad\frac{U,V\in\mathsf{SP}}{U\cdot V\in\mathsf{SP}}
\quad\frac{U,V\in\mathsf{SP}}{U\parallel V\in\mathsf{SP}}$$
\end{definition}

\begin{definition}[Series-communication-parallel pomsetc]
The set of series-communication-parallel pomsetcs, or scp-pomsetcs denoted $\mathsf{SCP}$, is the smallest set satisfying the following rules:

$$\frac{}{1\in\mathsf{SCP}}
\quad\frac{a\in\Sigma}{a\in\mathsf{SCP}}
\quad\frac{U,V\in\mathsf{SCP}}{U\cdot V\in\mathsf{SCP}}
\quad\frac{U,V\in\mathsf{SCP}}{U\parallel V\in\mathsf{SCP}}
\quad\frac{U,V\in\mathsf{SCP}}{U\mid V\in\mathsf{SCP}}
\quad\frac{U,V\in\mathsf{SCP}}{U\between V\in\mathsf{SCP}}$$
\end{definition}

\begin{definition}[N-shape1]\label{nshape1s}
Let $U=[\mathbf{u}]$ be a pomset. An N-shape1 in $U$ is a quadruple $u_0,u_1,u_2,u_3\in S_{\mathbf{u}}$ of distinct points such that $u_0\leq_{\mathbf{u}}u_1$, $u_2\leq_{\mathbf{u}}u_3$ and $u_0\leq_{\mathbf{u}}u_3$ and their exists no other relations among them. A pomset $U$ is N-free if it has no N-shape1s.
\end{definition}

\begin{definition}[N-shape2]\label{nshape2s}
Let $U=[\mathbf{u}]$ be a pomsetc. An N-shape2 in $U$ is a quadruple $u_0,u_1,u_2,u_3\in S_{\mathbf{u}}$ of distinct points such that $u_0\leq^{e}_{\mathbf{u}}u_1$, $u_2\leq^{e}_{\mathbf{u}}u_3$ and $u_0\leq^{c}_{\mathbf{u}}u_3$ and their exists no other relations among them.
\end{definition}

The definition of N-shape2 in Definition \ref{nshape2s} is based on the assumption that partial orders (causalities) among different parallel branches are all communications.

\begin{theorem}[N-shape1]
A pomset is series-parallel if and only if it is N-shape1-free in Definition \ref{nshape1s}.
\end{theorem}

\begin{theorem}[N-shape2]
A pomsetc is series-communication-parallel if and only if it only contains N-shape2s in Definition \ref{nshape2s}.
\end{theorem}

\begin{theorem}[Series-communication-parallelism to series-parallelism]
A series-communication-parallel pomsetc $U$ can be translated into a series-parallel pomset $U'$ if all the communications are all synchronous, i.e., for all $u_i\leq^{c} u_j$ in $U$, $u_i,u_j$ can merge into a single $\gamma(u_i,u_j)$ in $U'$, where $\gamma(u_i,u_j)$ is the communication function between $u_i$ and $u_j$.
\end{theorem}

So, in the following chapters, assume that all pomsets are series-parallel and all pomsetcs are series-communication-parallel. With a little of abuse of notions, we use $\mathsf{Pom}$, $\mathsf{Pomc}$, $\mathsf{SP}$ and $\mathsf{SCP}$ without distinctions.

\newpage\section{The Algebras}\label{algebras}

In Scott-Strachey approach \cite{DS} of denotational semantics, domains are the key concepts. In this chapter, we introduce the related concepts of $\Sigma$-domain, which are coming from algebraic semantics \cite{IACA} \cite{AlgebraicSemantics} \cite{TS}.

Firstly, we introduce $\Sigma$-algebra related concepts in \cref{sigmaalgebra}, then we introduce equational classes in \cref{ec} and inequational classes in \cref{iec}. Finally, we introduce continuous algebras in \cref{ca}.

\subsection{$\Sigma$-algebra}\label{sigmaalgebra}

\begin{definition}[Signature]
A signature $\Sigma$ consists of a finite set of function symbols (or operators) $f,g,\cdots$, where each function symbol $f$ has an arity $ar(f)$, being its number of arguments. A function symbol $a,b,c,\cdots$ of arity \emph{zero} is called a constant, a function symbol of arity one is called unary, and a function symbol of arity two is called binary.
\end{definition}

\begin{definition}[$\Sigma$-algebra]
Let $\Sigma$ be a signature. A $\Sigma$-algebra consists of $\langle A, \Sigma_A\rangle$, where:

\begin{enumerate}
  \item $A$ is the carrier set.
  \item $\Sigma_A$ is a set of functions $\{f_A:A^{ar(f)}\rightarrow A|f\in\Sigma\}$.
\end{enumerate}

Sometimes, we use $A$ to denote $\langle A, \Sigma_A\rangle$.
\end{definition}

\begin{definition}[Term algebra]
For every signature $\Sigma$, there is a particular $\Sigma$-algebra called term algebra for $\Sigma$, denoted $T_{\Sigma}$, the carriers consist of terms (strings) and the functions of the term algebras merely manipulate these terms. Let $T_{\Sigma}$ be the least set of terms satisfying:

\begin{enumerate}
  \item If $f\in\Sigma$ and $ar(f)=0$, then the term consisting of $f$ is in $T_{\Sigma}$.
  \item If $f\in\Sigma$ and $ar(f)>0$, and $t_1,\cdots,t_{ar(f)}\in T_{\Sigma}$, then $f(t_1,\cdots,t_{ar(f)})\in T_{\Sigma}$.
\end{enumerate} 
\end{definition}

\begin{definition}[$\Sigma$-homomorphism]
A mapping $\varphi:A\rightarrow B$ between two $\Sigma$-algebras $\langle A,\Sigma_A\rangle$ and $\langle B,\Sigma_B\rangle$ is a $\Sigma$-homomorphism if for every $f\in\Sigma$ and $d_1,\cdots,d_{ar(f)}\in A$, it holds that:

$$\varphi(f_A(d_1,\cdots,d_{ar(f)}))=f_B(\varphi(d_1),\cdots,\varphi(d_{ar(f)}))$$
\end{definition}

\begin{proposition}[$\Sigma$-homomorphism composition and identity $\Sigma$-homomorphism]

Let $\langle A,\Sigma_A\rangle$, $\langle B,\Sigma_B\rangle$ and $\langle C,\Sigma_C\rangle$ be $\Sigma$-algebras.

\begin{enumerate}
  \item If $\varphi:A\rightarrow B$ and $\psi:B\rightarrow C$ be $\Sigma$-homomorphisms, then their composition $\psi\circ\varphi:A\rightarrow C$ is also a $\Sigma$-homomorphism.
  \item The identity $i_A:A\rightarrow A$ is also a $\Sigma$-homomorphism.
  \item For every $\Sigma$-algebra $\langle A, \Sigma_A\rangle$, there exists a unique $\Sigma$-homomorphism $i_A:T_{\Sigma}\rightarrow A$.
\end{enumerate}
\end{proposition}

\begin{definition}[$\Sigma$-isomorphism]
A $\Sigma$-homomorphism $\varphi:A\rightarrow B$ is called $\Sigma$-isomorphic if it is a bijection.
\end{definition}

\begin{proposition}[$\Sigma$-isomorphism]
$\langle A,\Sigma_A\rangle$ and $\langle B,\Sigma_B\rangle$ are isomorphic if and only if there exist two $\Sigma$-homomorphisms, $\varphi:A\rightarrow B$ and $\psi:B\rightarrow A$, such that:

\begin{enumerate}
  \item $\varphi\circ\psi=id_B$.
  \item $\psi\circ\varphi=id_A$.
\end{enumerate}
\end{proposition}

\begin{definition}[Initiality]
Let $\mathscr{C}$ be a class of $\Sigma$-algebras. A $\Sigma$-algebra $I$ is called initial in $\mathscr{C}$ if for every $\Sigma$-algebra $J$ in $\mathscr{C}$ there exists a unique $\Sigma$-homomorphism $\varphi$ from $I$ to $J$.
\end{definition}

\begin{corollary}[Initiality]
If $I_1,I_2$ are initial in class $\mathscr{C}$ of $\Sigma$-algebras, then they are isomorphic.
\end{corollary}

\subsection{Equational Classes}\label{ec}

\begin{definition}[$\Sigma$-congruence]
Let $\langle A, \Sigma_A\rangle$ be a $\Sigma$-algebra. An equivalent relation $R$ over $A$ is a $\Sigma$-congruence, if for every $f\in\Sigma$ with arity $ar(f)$ and $\langle a_i,a_i'\rangle\in R$ for every $i$ with $0\leq i\leq ar(f)$, then $\langle f_A(a_1,\cdots,a_{ar(f)}),f_A(a_1',\cdots,a_{ar(f)}')\rangle\in R$.
\end{definition}

\begin{definition}[Equivalence classes]
Given a $\Sigma$-algebra $\langle A,\Sigma_A\rangle$, for every $a\in A$, the equivalence class of $a$ under an equivalent relation $R$, denoted $[a]_R$ with $[a]_R=\{a'|\langle a, a'\rangle\in R\}$. Let $A/R$ be the set of equivalence classes induced by $R$ over $A$ with $A/R=\{[a]_R|a\in A\}$. And for every $f\in\Sigma$, the mapping over $A/R$ is defined as:

$$f_{A/R}([a_1]_R,\cdots,[a_{ar(f)}]_R)=[f_A(a_1,\cdots,a_{ar(f)})]_R$$

For a $\Sigma$-congruence $R$ over $A$, $\langle A,\Sigma_A\rangle$ satisfies $R$ if $i_A(t)=i_A(t')$ whenever $\langle t, t'\rangle\in R$. And let $\mathscr{C}(R)$ be the class of all $\Sigma$-algebras satisfying $R$.
\end{definition}

\begin{lemma}
Let $\langle A, \Sigma_A\rangle$ be a $\Sigma$-algebra and $R$ be an equivalent relation over $A$.

\begin{enumerate}
  \item $\langle A/R,\Sigma_{A/R}\rangle$ is a $\Sigma$-algebra.
  \item The natural injection mapping $in:A\rightarrow A/R$, defined by $in(a)=[a]_R$ for $a\in A$, is a $\Sigma$-homomorphism.
\end{enumerate}
\end{lemma}

\begin{theorem}[Initiality for congruences]
For a $\Sigma$-congruence $R$, the $\Sigma$-algebra $T_{\Sigma}/R$ is initial in the class $\mathscr{C}(R)$.
\end{theorem}

By allowing the occurrences of variables $x,x_1,x_2\cdots\in X$ in a signature $\Sigma$, we can get of the extended signature denoted $\Sigma(X)$. Similarly, the term algebra $T_{\Sigma}(X)$ is extended from $T_{\Sigma}$, the elements of $T_{\Sigma}$ are called closed terms (we often refer them as processes), while the ones of $T_{\Sigma}(X)$ are called open terms. An $A$-assignment for $X$ is a mapping $\sigma_A:X\rightarrow A$, while a substitution $\sigma(X)$ is a $T_{\Sigma}(X)$-assignment by mapping each $x\in X$ to a term in $T_{\Sigma}(X)$. For each $x_i$ in a variable sequence $\underline{x}$, if $\sigma(x_i)$ is a closed term in $T_{\Sigma}$, then $\sigma$ is called a closed substitution. We write $t\sigma$ for $\sigma(t)$ and $t(\sigma\circ\sigma')=(t\sigma')\sigma$ with $t\in T_{\Sigma}(X)$, which is called an instantiation of $t$. 

\begin{theorem}[Freeness]
If $A$ is a $\Sigma$-algebra and $\sigma_A$ an $A$-assignment for $X$, then there exists a unique $\Sigma$-homomorphism $\varphi_A$ from $T_{\Sigma}(X)$ to $A$ such that $\varphi_A(x)=\sigma_A(x)$ for every $x\in X$.
\end{theorem}

\begin{lemma}[Substitution lemma]
For every $A$-assignment $\sigma_A$ and every substitution $\sigma$, the unique extension of the $A$-assignment $\sigma_A\circ\sigma$ to $T_{\Sigma}(X)$ is given by the function $\varphi(t)=\sigma_A(t\sigma)$, where $t\in T_{\Sigma}(X)$.
\end{lemma}

\begin{definition}[Equational relation]
For a $\Sigma$-algebra $\langle A,\Sigma_A\rangle$, we define an equational relation $=_A$ over $A$, which satisfying:

\begin{enumerate}
  \item It is an equivalent relation.
  \item It is closed under substitutions.
  \item It is closed under instantiations.
\end{enumerate}

A set of $\Sigma$-equations $E$ contains equations such as $t=t'$ with $t,t'\in T_{\Sigma}(X)$, $\langle A, \Sigma_A\rangle$ satisfies $E$ if $E\subseteq =_A$. Let $\mathscr{C}(E)$ be the class of $\Sigma$-algebras satisfying $E$.
\end{definition}

\begin{theorem}[Initiality for equations]
For every set of $\Sigma$-equations $E$, $\mathscr{C}(E)$ has an initial $\Sigma$-algebra.
\end{theorem}

\begin{definition}[Proof system]
A system of equational deductions by whose equations in $E$ can be used to derive equations, denoted $\mathbf{DED}(E)$. It contains the following six rules, where $t,t'\in T_{\Sigma}(X)$, $f\in\Sigma$ and $\sigma$ is a substitution:

\begin{enumerate}
  \item Reflexivity: $$\frac{}{t=t}$$
  \item Symmetry: $$\frac{t=t'}{t'=t}$$
  \item Transitivity: $$\frac{t=t',t'=t''}{t=t''}$$
  \item Substitution: $$\frac{t_1=t_1',\cdots,t_{ar(f)}=t_{ar(f)}'}{f(t_1,\cdots,t_{ar(f)})=f(t_1',\cdots,t_{ar(f)}')}$$
  \item Instantiation: $$\frac{t=t'}{t\sigma=t'\sigma}$$
  \item Equations: for every $\langle t,t'\rangle\in E$, $$\frac{}{t=t'}$$
\end{enumerate}

A proof is a sequence of deductions by use of the above six rules. If $t=t'$ is the last proof statement, denoted $\vdash_E t=t'$, then it is called a theorem of $\mathbf{DED}(E)$. And $t=_E t'$ if and only if $\vdash_E t=t'$.
\end{definition}

\begin{lemma}
For $t,t'\in T_{\Sigma}(X)$, if $\vdash_E t=t'$ and $A$ satisfies $E$ then $t=_A t'$, i.e., $E\subseteq =_A$ implies $=_E\subseteq =_A$.
\end{lemma}

\begin{corollary}[Initiality for equations]
$T_{\Sigma}/=_E$ is initial in $\mathscr{C}(E)$.
\end{corollary} 

\begin{definition}[Soundness and completeness]
Let $R$ be a relation on $T_{\Sigma}(X)$. 

\begin{enumerate}
  \item The proof system $\mathbf{DED}(E)$ is sound with respect to $R$ if $\vdash_E t=t'$ implies $\langle t,t'\rangle\in R$.
  \item The proof system $\mathbf{DED}(E)$ is complete with respect to $R$ if $\langle t,t'\rangle\in R$ implies $\vdash_E t=t'$.
\end{enumerate}
\end{definition}

\begin{theorem}[Equational logic theorem]
$\mathscr{C}(E)$ has a unique initial $\Sigma$-algebra up to $\Sigma$-isomorphism denoted by $I_E$. Then,

\begin{enumerate}
  \item $\mathbf{DED}(E)$ is sound with respect to $=_{I_E}$.
  \item $\mathbf{DED}(E)$ is complete with respect to $=_{I_E}$, restricted to $T_{\Sigma}$.
\end{enumerate}
\end{theorem}

\subsection{Inequational Classes}\label{iec}

\begin{definition}[Partial order]
Let $\langle A, \leq_A\rangle$ be a partial order. A function $f:A\rightarrow A$ is monotonic if $a\leq_A a'$ implies $f(a)\leq_A f(a')$. For $\underline{a},\underline{a}'\in A^k$, $\underline{a}\leq_A\underline{a}'$, if for each $1\leq i\leq k$, $a_i\leq a_i'$, then $f:A^k\rightarrow A$ is monotonic of $\underline{a}\leq_A\underline{a}'$ implies $f(\underline{a})\leq_A f(\underline{a}')$. A relaxed monotonicity requires that $\underline{a}\leq_A'\underline{a}'$ implies $f(\underline{a})\leq_A f(\underline{a}')$, where $\underline{a}\leq_A'\underline{a}'$ if there exists $1\leq j\leq k$, such that $a_j\leq_A a_j'$ and for every $i\neq j$, $a_i=a_i'$.
\end{definition}

\begin{definition}[$\Sigma$-po algebra]
Let $\Sigma$ be a signature. A $\Sigma$-po algebra consists of $\langle A, \leq_A, \Sigma_A\rangle$, where:

\begin{enumerate}
  \item $A$ is the carrier set.
  \item $\leq_A$ is a partial order over $A$.
  \item $\Sigma_A$ is a set of monotonic functions $\{f_A:A^{ar(f)}\rightarrow A|f\in\Sigma\}$ with respect to $\leq_A$.
\end{enumerate}

Sometimes, we use $A$ to denote $\langle A, \leq_A, \Sigma_A\rangle$.
\end{definition}

\begin{definition}[$\Sigma$-po homomorphism]
A mapping $\varphi:A\rightarrow B$ between two $\Sigma$-po algebras $\langle A,\leq_A,\Sigma_A\rangle$ and $\langle B,\leq_B,\Sigma_B\rangle$ is a $\Sigma$-po homomorphism if:

\begin{enumerate}
  \item for every $f\in\Sigma$, it holds that $\varphi(f_A(\underline{a}))=f_B(\varphi(\underline{a}))$.
  \item $a\leq_A a'$ implies $\varphi(a)\leq_A\varphi(a')$.
\end{enumerate}

A $\Sigma$-po homomorphism is $\Sigma$-po isomorphic if it is has an inverse.
\end{definition}

\begin{theorem}[Freeness]
If $A$ is a $\Sigma$-po algebra and $\sigma_A$ an $A$-assignment for $X$, then there exists a unique $\Sigma$-po homomorphism $\varphi_A$ from $T_{\Sigma}(X)$ to $A$ such that $\varphi_A(x)=\sigma_A(x)$ for every $x\in X$.
\end{theorem}

\begin{definition}[$\Sigma$-preorder]
For a $\Sigma$-po algebra $\langle A,\leq_A,\Sigma_A\rangle$, we define a $\Sigma$-preorder relation $\sqsubseteq$ over $A$, which satisfying for $a,a',a''\in A$ and $f\in\Sigma$:

\begin{enumerate}
  \item $a\sqsubseteq a$.
  \item $a\sqsubseteq a'$, $a'\sqsubseteq a''$ implies $a\sqsubseteq a''$.
  \item $a\leq_A a'$ implies $a\sqsubseteq a'$.
  \item $\underline{a}\sqsubseteq\underline{a}'$ implies $f_A(\underline{a})\sqsubseteq f_A(\underline{a}')$
\end{enumerate}

By replacing the partial order $\leq_A$ with the preorder $\sqsubseteq$, we can obtain new $\Sigma$-po algebras.
\end{definition}

\begin{definition}[Equivalence classes]
Given a $\Sigma$-po algebra $\langle A,\sqsubseteq, \Sigma_A\rangle$, let $\sim$ be the kernel of $\sqsubseteq$, for every $a\in A$, the equivalence class of $a$ under an equivalent relation $\sim$, denoted $[a]_{\sqsubseteq}$ with $[a]_{\sqsubseteq}=\{a'|\langle a, a'\rangle\in \sim\}$. Let $A/\sqsubseteq$ be the set of equivalence classes induced by $\sqsubseteq$ over $A$ with $A/\sqsubseteq=\{[a]_{\sqsubseteq}|a\in A\}$. And for every $f\in\Sigma$, the mapping over $A/\sqsubseteq$ is defined as:

$$f_{A/\sqsubseteq}([a_1]_{\sqsubseteq},\cdots,[a_{ar(f)}]_{\sqsubseteq})=[f_A(a_1,\cdots,a_{ar(f)})]_{\sqsubseteq}$$
\end{definition}

\begin{lemma}
Let $\langle A, \sqsubseteq,\Sigma_A\rangle$ be a $\Sigma$-po algebra and $\sqsubseteq$ be a preorder relation over $A$.

\begin{enumerate}
  \item $\langle A/\sqsubseteq,\sqsubseteq,\Sigma_{A/\sqsubseteq}\rangle$ is a $\Sigma$-po algebra.
  \item The natural injection mapping $in:A\rightarrow A/\sqsubseteq$, defined by $in(a)=[a]_{\sqsubseteq}$ for $a\in A$, is a $\Sigma$-po homomorphism.
\end{enumerate}
\end{lemma}

\begin{theorem}[Initiality for preoders]
Let $\mathscr{C}(\sqsubseteq)$ be the class of all $\Sigma$-po algebras satisfying $\sqsubseteq$. The $\Sigma$-po algebra $T_{\Sigma}/\sqsubseteq$ is initial in the class $\mathscr{C}(\sqsubseteq)$.
\end{theorem}

\begin{definition}[Inequations]
A set of $\Sigma$-inequations $E$ contains inequations such as $t\leq t'$ with $t,t'\in T_{\Sigma}(X)$, $\langle A, \leq_A, \Sigma_A\rangle$ satisfies $E$ if $E\subseteq \leq_A$. Let $\mathscr{C}(E)$ be the class of $\Sigma$-po algebras satisfying $E$.
\end{definition}

\begin{definition}[Proof system]
A system of inequational deductions by whose inequations in $E$ can be used to derive inequations, also denoted $\mathbf{DED}(E)$. It contains the following six rules, where $t,t'\in T_{\Sigma}(X)$, $f\in\Sigma$ and $\sigma$ is a substitution:

\begin{enumerate}
  \item Reflexivity: $$\frac{}{t\leq t}$$
  \item Transitivity: $$\frac{t\leq t',t'\leq t''}{t\leq t''}$$
  \item Substitution: $$\frac{t_1\leq t_1',\cdots,t_{ar(f)}\leq t_{ar(f)}'}{f(t_1,\cdots,t_{ar(f)})\leq f(t_1',\cdots,t_{ar(f)}')}$$
  \item Instantiation: $$\frac{t\leq t'}{t\sigma\leq t'\sigma}$$
  \item Inequations: for every $\langle t,t'\rangle\in E$, $$\frac{}{t\leq t'}$$
  \item Equations: $$\frac{t\leq t',t'\leq t}{t=t'},~~\frac{t=t'}{t\leq t',t'\leq t},~~\frac{t=t'}{t'=t}$$
\end{enumerate}

A proof is a sequence of deductions by use of the above six rules. If $t\leq t'$ is the last proof statement, denoted $\vdash_E t\leq t'$, then it is called a theorem of $\mathbf{DED}(E)$. And $t\leq_E t'$ if and only if $\vdash_E t\leq t'$.
\end{definition}

\begin{lemma}
For $t,t'\in T_{\Sigma}(X)$, if $\vdash_E t\leq t'$ and $A$ satisfies $E$ then $t\leq_A t'$, i.e., $E\subseteq \leq_A$ implies $\leq_E\subseteq \leq_A$.
\end{lemma}

\begin{corollary}[Initiality for equations]
$T_{\Sigma}/\leq_E$ is initial in $\mathscr{C}(E)$.
\end{corollary} 

\begin{definition}[Soundness and completeness]
Let $R$ be a relation on $T_{\Sigma}(X)$. 

\begin{enumerate}
  \item The proof system $\mathbf{DED}(E)$ is sound with respect to $R$ if $\vdash_E t\leq t'$ implies $\langle t,t'\rangle\in R$.
  \item The proof system $\mathbf{DED}(E)$ is complete with respect to $R$ if $\langle t,t'\rangle\in R$ implies $\vdash_E t\leq t'$.
\end{enumerate}
\end{definition}

\begin{theorem}[Inequational logic theorem]
$\mathscr{C}(E)$ has a unique initial $\Sigma$-po algebra up to $\Sigma$-po isomorphism denoted by $I_E$. Then,

\begin{enumerate}
  \item $\mathbf{DED}(E)$ is sound with respect to $\leq_{I_E}$.
  \item $\mathbf{DED}(E)$ is complete with respect to $\leq_{I_E}$, restricted to $T_{\Sigma}$.
\end{enumerate}
\end{theorem}

\begin{definition}[Substitution closed]
A $\Sigma$-preorder $R$ over $T_{\Sigma}(X)$ is substitution closed if for every substitution $\sigma$, $\langle t,t'\rangle\in R$ implies $\langle t\sigma,t'\sigma\rangle\in R$, where $t,t'\in T_{\Sigma}(X)$.
\end{definition}

\begin{proposition}
If $E$ is a set of inequations, then $\leq_E$ is the least substitution closed $\Sigma$-preorder satisfying $E$.
\end{proposition}

\begin{corollary}
The $\Sigma$-po algebra $\langle A,\leq_A,\Sigma_A\rangle$ is initial in $\mathscr{C}(E)$ if and only if it is surjective, i.e., every element in $A$ is denotable by some syntactic object in $T_{\Sigma_A}$, and $\mathbf{DED}(E)$ is sound and complete with respect to $\leq_A$ over $T_{\Sigma_A}$.
\end{corollary}

\begin{definition}[Full Abstractness]
Let $R$ be a behavioural motivated relation over terms from $T_{\Sigma_A}$, a given interpretation over A, denoted $A\sembrack{ }$, is fully abstract with respect to $R$, if for every $t,t'\in T_{\Sigma_A}$: $\langle t,t'\rangle\in R$ if and only if $A\sembrack{t}\leq A\sembrack{t'}$.
\end{definition}

\begin{proposition}
If $A$ is surjective, then it is fully abstract with respect to $R$ if and only if:

\begin{enumerate}
  \item $R$ is a $\Sigma$-preorder over $T_{\Sigma}$.
  \item $A$ is initial in $\mathscr{C}(R)$.
\end{enumerate}
\end{proposition}

\begin{corollary}
$I_E$ is fully abstract with respect to $R$ if and only if $\mathbf{DED}(E)$ is both sound and complete with respect to $R$.
\end{corollary} 

\subsection{Continuous Algebras}\label{ca}

Recursion is used to capture infinite computations. $\Sigma$-algebras whose carrier are partial orders enjoying certain continuity constraints are called $\Sigma$-domains, which are sufficient to ensure that the recursive equations always have least solutions taken as the meaning of the recursive definitions.

\subsubsection{Continuity}

\begin{definition}[Least element and upper bound]
Let $\langle A,\leq_A\rangle$ be a partial order. The least element $\bot_A$ in $A$ satisfies $\bot_A\leq_A a$ for every $a\in A$. Let $D\subseteq A$ and $a\in A$, then $a$ is an upper bound of $D$ if $d\leq_A a$ for every $d\in D$. $a$ is a least upper bound (lub) of $D$, if (1) $a$ is an upper bound of $D$; (2) if $d'$ is an upper bound of $D$ then $a\leq_A d'$. If the lub of $D$ exists, then it is unique and denoted $\lub_A D$, and sometimes $\lub D$ for short.
\end{definition}

\begin{definition}[Directed subset]
$D$ is a directed subset of $A$ if it is nonempty and for every pair of elements $d_1,d_2\in D$, the set $\{d_1,d_2\}$ has an upper bound which is also in $D$.
\end{definition}

\begin{definition}[Complete partial order (cpo)]
The partial order $\langle A,\leq_A\rangle$ is a complete partial order (cpo) if:

\begin{enumerate}
  \item It contains a least element $\bot_A$.
  \item every directed subset of $A$ has a lub.
\end{enumerate}
\end{definition}

\begin{definition}[Domination]
If $D,D'$ are directed subsets of a cpo $A$ and $D\subseteq D'$ then $\lub D\leq\lub D'$; $D'$ dominates $D$ if for every $d\in D$ there is some $d'\in D'$ such that $d\leq d'$.
\end{definition}

\begin{lemma}
If $D,D'$ are directed subsets of a cpo $A$ and $D'$ dominates $D$ then $\lub D\leq\lub D'$.
\end{lemma}

\begin{lemma}
For a doubly indexed subset of a cpo $A$ denoted $D=\{d_{ij}|i\in I,j\in J\}$, $D_i=\{d_{ij}|j\in J\}$ for each $i\in I$ and $D^j=\{d_{ij}|i\in I\}$ for each $j\in J$, if:

\begin{enumerate}
  \item $D$ is directed.
  \item Each $D_i,D^j$ are directed with lubs $d_i,d^j$ respectively.
  \item The sets $\{d_i|i\in I\}$ and $\{d^j|j\in J\}$ are also directed.
\end{enumerate}

then,

\begin{enumerate}
  \item $\lub D=\lub\{d_i|i\in I\}$.
  \item $\lub D=\lub\{d^j|j\in J\}$.
\end{enumerate}
\end{lemma}

\begin{lemma}[Cartesian product]
For any two cpos $\langle A_1,\leq_{A_1}\rangle$ and $\langle A_2,\leq_{A_2}\rangle$, $a_1,a_1'\in A_1$ and $a_2,a_2'\in A_2$, let $\langle a_1,a_2\rangle\leq\langle a_1',a_2'\rangle$ if both $a_1\leq_{A_1}a_1'$ and $a_2\leq_{A_2}a_2'$. Then, $\langle A_1\times A_2,\leq\rangle$ is a cpo.
\end{lemma}

\begin{lemma}
For any set $S$ and cpo $A$, let $(S\rightarrow A)$ be the set of all functions from $S$ to $A$. For $f,g\in (S\rightarrow A)$, let $f\leq g$ if $f(s)\leq_A g(s)$ for every $s\in S$. Then, $\langle (S\rightarrow A),\leq\rangle$ is a cpo. 
\end{lemma}

\begin{definition}[Continuous functions]
Let $\langle A,\leq_A\rangle$ and $\langle A',\leq_{A'}\rangle$ be two cpos and function $f:A\rightarrow A'$. Then $f$ is continuous if for every directed subset $D\subseteq A$, 

\begin{enumerate}
  \item $f(D)$ is directed in $A'$.
  \item $f(\lub_A D)=\lub_{A'}f(D)$
\end{enumerate}

Where $f(D)=\{f(d)|d\in D\}$
\end{definition}

\begin{lemma}
Continuous functions are monotonic.
\end{lemma}

\begin{lemma}
$f:A\rightarrow A'$ is continuous if and only if it is monotonic and for every directed subset $D\subseteq A$, it holds that $f(\lub_A D)\leq\lub_{A'}f(D)$.
\end{lemma}

\begin{proposition}[Left and right continuity]
The function $f:A_1\times A_2\rightarrow A'$ is left-continuous, if for every $a_0\in A_2$ and every directed subset $D\subseteq A_1\times A_2$ whose elements are of the form $\langle a,a_0\rangle$, it holds that $f(\lub_{A_1\times A_2} D)=\lub_{A'}f(D)$. Right-continuity can be defined correspondingly. Then, $f:A_1\times A_2\rightarrow A'$ is continuous if and only if it is both left-continuous and right-continuous.
\end{proposition}

\begin{definition}[Induced pointwise ordering]
Let $[A\rightarrow A']$ be the set of all continuous functions from the cpo $A$ to the cpo $A'$. This set can be ordered by the so-called induced pointwise ordering: $f\leq g$ if for every $a\in A$, $f(a)\leq_{A'}g(a)$.
\end{definition}

\begin{lemma}
Let $F\subseteq [A\rightarrow A']$ be a directed set of functions. Define $f:A\rightarrow A'$ by $f(a)=\lub_{A'}\{g(a)|g\in F\}$, then $f$ is well-defined and continuous.
\end{lemma}

\begin{proposition}
$[A\rightarrow A']$ is a cpo under the induced pointwise ordering.
\end{proposition}

Recursive equations will be interpreted semantically by least fixpoints.

\begin{definition}[Fixpoint]
Let $f\in[A\rightarrow A']$. The element $a\in A$ is called a fixpoint of $f$ if $a=f(a)$. It is called the least fixpoint of $f$, if $a\leq_A a'$ for every fixpoint $a'$ of $f$. Further, it is called a pre-fixpoint of $f$, if $f(a)\leq_A a$.
\end{definition}

\begin{proposition}
Every $f\in [A\rightarrow A']$ has a least fixpoint $x_f$.
\end{proposition}

\begin{proposition}
Let $Y\in[[A\rightarrow A]\rightarrow A]$ be a mapping from a function $f\in[A\rightarrow A]$ to its least fixpoint $x_f$, i.e., $Y(f)=x_f$. Then $Y$ is a continuous function.
\end{proposition}

\begin{definition}[Algebraic cpo]
Let $A$ be a cpo and $D$ a directed subset of $A$, an element $a\in A$ is compact or finite, if whenever $a\leq\lub D$ there exists some $d\in D$ such that $a\leq d$. $A$ is an algebraic cpo if for every $a\in A$, $a=\lub\{d|d\leq a,d \mbox{ compact}\}$.
\end{definition}

The compact elements are the semantic denotations of finite processes, and if an interpretation is algebraic, every recursive defined processes is semantically the limit of a directed set of finite processes. Let $\Fin(A)$ to denote the set of finite or compact elements of $A$, and for $a\in A$ $\Fin(a)=\{d\in\Fin(A)|d\leq a\}$, then the algebraicness requires $a=\lub_A \Fin(a)$. If $A,A'$ are algebraic cpos with $\langle\Fin(A),\leq_A\rangle$ and $\langle \Fin(A'),\leq_{A'}\rangle$ being isomorphic as partial orders, then $A,A'$ are isomorphic as cpos.

\subsubsection{$\Sigma$-domains} 

\begin{definition}[$\Sigma$-domains]
Let $\Sigma$ be a signature which contains a distinguished constant $\Omega$. A $\Sigma$-domain consists of $\langle A, \leq_A, \Sigma_A\rangle$, where:

\begin{enumerate}
  \item $A$ is the carrier set.
  \item $\leq_A$ is a partial order over $A$.
  \item $\langle A, \leq_A\rangle$ is an algebraic cpo.
  \item $\Sigma_A$ is a set of continuous functions $\{f_A:A^{ar(f)}\rightarrow A|f\in\Sigma\}$ with respect to $\leq_A$.
  \item $\Omega_A$ is $\bot_A$ which is the least element with respect to $\leq_A$.
\end{enumerate}

Sometimes, we use $A$ to denote $\langle A, \leq_A, \Sigma_A\rangle$.
\end{definition}

\begin{definition}[$\Sigma$-domain homomorphism]
A mapping $\varphi:A\rightarrow B$ between two $\Sigma$-domains $\langle A,\leq_A,\Sigma_A\rangle$ and $\langle B,\leq_B,\Sigma_B\rangle$ is a $\Sigma$-domain homomorphism if:

\begin{enumerate}
  \item for every $f\in\Sigma$, it holds that $\varphi(f_A(\underline{a}))=f_B(\varphi(\underline{a}))$.
  \item it is continuous with respect to $\leq$.
\end{enumerate}

A $\Sigma$-domain homomorphism is $\Sigma$-domain isomorphic if it is has an inverse.

Sometimes, we write $\Sigma$-homomorphism for $\Sigma$-domain homomorphism and $\Sigma$-isomorphism for $\Sigma$-domain isomorphism.
\end{definition}

\begin{proposition}
The $\Sigma$-domains $\langle A,\leq_A,\Sigma_A\rangle$ and $\langle B,\leq_B,\Sigma_B\rangle$ are isomorphic if and only if they are isomorphic as $\Sigma$-po algebras.
\end{proposition}

Let $\mathscr{CC}(E)$ denote the class of $\Sigma$-domains satisfying the equations $E$.

\begin{theorem}[Initiality]
$\mathscr{CC}(E)$ has an initial object, denoted $CI_E$, which is unique up to isomorphism.
\end{theorem}

When $E=\emptyset$, $CI_E$ can not give syntax for the continuous objects, as the $\Sigma$-domain $CT_{\Sigma}$ which can not be deemed as a language in the normal sense; but it can still give semantic domains. By introducing a new $\Omega$-rule to the proof system $\mathbf{DED}(E)$, where $t\in T_{\Sigma}(X)$:
$$\frac{}{\Omega\leq t}$$
we get the new proof system $\mathbf{\Omega DED}(E)$, and $t\leq_{E\Omega} t'$ if and only if $\vdash_{E\Omega} t\leq t'$.

\begin{definition}[Finitariness]
A $\Sigma$-domain $A$ is finitary, if:

\begin{enumerate}
  \item For every term $t\in T_{\Sigma}$ and $i_A:T_{\Sigma}\rightarrow A$, $i_A(t)$ is a finite element in $A$.
  \item For every finite element $a\in A$, there exists a term $t\in T_{\Sigma}$ such that $i_A(t)=a$.
\end{enumerate}
\end{definition}

\begin{theorem}
The $\Sigma$-domain $A$ is initial in $\mathscr{CC}(E)$ if and only if:

\begin{enumerate}
  \item It is finitary.
  \item $\mathbf{\Omega DED}(E)$ is sound and complete with respect to $\leq_A$, restrict to $T_{\Sigma}$.
\end{enumerate}
\end{theorem}

\subsubsection{$\Sigma$-predomains} 

\begin{definition}[$\Sigma$-predomains]
Let $\Sigma$ be a signature which contains a distinguished constant $\Omega$. A $\Sigma$-domain consists of $\langle A, \leq_A, \Sigma_A\rangle$, where:

\begin{enumerate}
  \item $A$ is the carrier set.
  \item $\leq_A$ is a partial order over $A$.
  \item $\Sigma_A$ is a set of monotonic functions $\{f_A:A^{ar(f)}\rightarrow A|f\in\Sigma\}$ with respect to $\leq_A$.
  \item $\Omega_A$ is $\bot_A$ which is the least element with respect to $\leq_A$.
\end{enumerate}

Sometimes, we use $A$ to denote $\langle A, \leq_A, \Sigma_A\rangle$.
\end{definition}

\begin{definition}[$\Sigma$-predomain homomorphism]
A mapping $\varphi:A\rightarrow B$ between two $\Sigma$-predomains $\langle A,\leq_A,\Sigma_A\rangle$ and $\langle B,\leq_B,\Sigma_B\rangle$ is a $\Sigma$-predomain homomorphism if:

\begin{enumerate}
  \item for every $f\in\Sigma$, it holds that $\varphi(f_A(\underline{a}))=f_B(\varphi(\underline{a}))$.
  \item it is monotonic with respect to $\leq$.
\end{enumerate}

A $\Sigma$-predomain homomorphism is $\Sigma$-predomain isomorphic if it is has an inverse.

Sometimes, we write $\Sigma$-homomorphism for $\Sigma$-predomain homomorphism and $\Sigma$-isomorphism for $\Sigma$-predomain isomorphism.
\end{definition}

Let $\mathscr{CP}(E)$ denote the class of $\Sigma$-predomains satisfying the equations $E$.

\begin{theorem}[Initiality]
$\mathscr{CP}(E)$ has an initial object, denoted $PI_E$, which is unique up to isomorphism.
\end{theorem}

\begin{corollary}
The $\Sigma$-predomain $\langle A,\leq_A,\Sigma_A\rangle$ is initial in $\mathscr{CP}(E)$ if and only if it is surjective, i.e., every element in $A$ is denotable by some syntactic object in $T_{\Sigma_A}$, and $\mathbf{\Omega DED}(E)$ is sound and complete with respect to $\leq_A$ over $T_{\Sigma_A}$.
\end{corollary}

$t\leq_{E\Omega} t'$ if and only if $\vdash_{E\Omega} t\leq t'$.

\begin{proposition}
$\leq_{E\Omega}$ is the least substitution closed $\Sigma$-preorder over $T_{\Sigma}(X)$ satisfying $E$ and the additional inequation $\Omega\leq x$, where $x\in T_{\Sigma}(X)$.
\end{proposition}

The following is related to generating $\Sigma$-domains from $\Sigma$-predomains by use of ideal completion.

\begin{definition}[Ideal]
Let $\langle A,\leq_A\rangle$ be a partial order. An ideal in $A$, denoted $I$, is an nonempty subset of $A$ satisfying:

\begin{enumerate}
  \item $x,y\in I$, then there exists some $z\in I$ such that $x\leq_A z$ and $y\leq_A z$, i.e., $I$ is directed.
  \item $x\in I$ and $y\leq_A x$, then $y\in I$, i.e., $I$ is downwards closed.
\end{enumerate}
\end{definition}

Let $\mathscr{I}(A)$ denote the set of ideals of $A$.

\begin{lemma}
Let $\langle A,\leq_A\rangle$ be a partial order. If $A$ has a least element then $\langle \mathscr{I}(A),\subseteq\rangle$ is an algebraic cpo.
\end{lemma}

\begin{theorem}
$\langle\mathscr{I}(A),\subseteq\rangle$ is the unique algebraic cpo (up to isomorphism) whose set of finite elements are isomorphic to $\langle A,\leq_A\rangle$ as partial order.
\end{theorem}

$in:A\rightarrow\mathscr{I}(A)$ is defined as $in(a)=\{x|x\leq_A a,a\in A\}$, and the ideal completion $\langle\mathscr{I}(A),\subseteq\rangle$ of the partial order $\langle A,\leq_A\rangle$ is denoted $\langle A,\leq_A\rangle^{\infty}$ or $A^{\infty}$.

\begin{theorem}
If $k:A\rightarrow E$ is any monotonic function from the partial order $A$ to the cpo $E$, then there exists a unique continuous function $ext(k):A^{\infty}\rightarrow E$ such that the following diagram commutes.

\begin{tikzcd}[row sep=large, column sep=large]
A \arrow{r}{in}\arrow[swap]{d}{k} & A^{\infty} \arrow[dashed]{ld}{ext(k)} \\
E 
\end{tikzcd}
\end{theorem}

\begin{lemma}
$\langle A^{\infty},\leq_{A^{\infty}},\Sigma_{A^{\infty}}\rangle$ is a $\Sigma$-domain
\end{lemma}

We write $\langle A,\leq_{A},\Sigma_{A}\rangle^{\infty}$ or $A^{\infty}$ to denote the $\Sigma$-domain $\langle A^{\infty},\leq_{A^{\infty}}, \Sigma_{A^{\infty}}\rangle$.

\begin{theorem}
If $k:A\rightarrow E$ is a $\Sigma$-predomain homomorphism from the $\Sigma$-predomain $A$ to the $\Sigma$-domain $E$, then there exists a unique $\Sigma$-domain homomorphism $ext(k):A^{\infty}\rightarrow E$ such that the following diagram commutes.

\begin{tikzcd}[row sep=large, column sep=large]
A \arrow{r}{in}\arrow[swap]{d}{k} & A^{\infty} \arrow[dashed]{ld}{ext(k)} \\
E 
\end{tikzcd}
\end{theorem}

\begin{lemma}
If $A\in\mathscr{CP}(E)$ then $A^{\infty}\in\mathscr{CP}(E)$.
\end{lemma}

\begin{proposition}
If $PI$ is initial in $\mathscr{CP}(E)$ then $PI^{\infty}$ is initial in $\mathscr{CP}(E)$.
\end{proposition} 
\newpage\section{Testing Semantics}\label{ts}

In this chapter, we generalize the testing semantics \cite{TS} to the truly concurrent processes. Firstly, we introduce the testing preorders and testing equivalences in \cref{tpte}. Then we introduce pomset labelled transition system in \cref{plts}. Finally, we introduce the operational semantics, axiomatic semantics and denotational semantics of the basic processes, recursion and abstraction in \cref{bpT}, \cref{recT} and \cref{absT}, respectively.

\subsection{Testing Preorders and Testing Equivalences}\label{tpte}

\begin{definition}[Experimental system]
A experimental system ($\mathscr{EP}$) is a collection of the form $\langle \mathsf{Proc},\mathsf{Exp},\mapsto,Success\rangle$, where

\begin{enumerate}
  \item $\mathsf{Proc}$ is an arbitrary set of processes.
  \item $\mathsf{Exp}$ is an arbitrary set of experimenters.
  \item $\mapsto\subseteq(\mathsf{Exp}\times \mathsf{Proc})\times(\mathsf{Exp}\times \mathsf{Proc})$ is the interacting relation, for the interconnection of $p\in \mathsf{Proc}$ and $e\in \mathsf{Exp}$, we denote it as $\parallel_{EP}$ to be distinguished from the parallel operator $\parallel$ in the signature in the following chapters.
  \item $Success\subseteq \mathsf{Exp}$ is the success set.
\end{enumerate} 
\end{definition}

An experiment or test, i.e., the interactions between the experimenter $e\in \mathsf{Exp}$ and $p\in \mathsf{Proc}$, is a sequence of the form:

$$e\parallel_{EP}p\mapsto e_0\parallel_{EP}p_0\mapsto e_1\parallel_{EP}p_1\mapsto\cdots\mapsto e_k\parallel_{EP}p_k\mapsto\cdots$$

The above sequence is called a computation if it is maximal, i.e., it is infinite or finite with terminal element $e_n\parallel_{EP}p_n$, and denoted $Comp(e,p)$. The possible results $Result(e,p)$ of $Comp(e,p)$ is defined by $Result(e,p)=\{\top,\bot\}$, where $\top$ denotes a successful computation and $\bot$ denotes an unsuccessful computation. There are the following natural relations between processes $p,p'\in \mathsf{Proc}$, where $e\in \mathsf{Exp}$:

\begin{enumerate}
  \item $p\pretestingeq' p'$ if for every $e\in \mathsf{Exp}$, $Result(e,p)=Result(e,p')$.
  \item $p\may e$ if $\top\in Result(e,p)$.
  \item $p\must e$ if $\bot\notin Result(e,p)$.
\end{enumerate}

This leads to the definitions of three kinds of preorders.

\begin{definition}[Testing preorders]
For an $\mathscr{EP}$ $\langle \mathsf{Proc}, \mathsf{Exp}, \rightarrow,Success\rangle$ and $p,p'\in \mathsf{Proc}$:

\begin{enumerate}
  \item $p\pretestingmay p'$ if for every $e\in \mathsf{Exp}$, $p\may e$ implies $p'\may e$.
  \item $p\pretestingmust p'$ if for every $e\in \mathsf{Exp}$, $p\must e$ implies $p'\must e$.
  \item $p\pretesting p'$ if $p\pretestingmay p'$ and $p\pretestingmust p'$.
\end{enumerate}
\end{definition}

The kernels of $\pretestingmay$, $\pretestingmust$ and $\pretesting$ are denoted $\pretestingmayeq$, $\pretestingmusteq$ and $\pretestingeq$, respectively.

\begin{lemma}
$\pretestingeq'$ and $\pretestingeq$ coincide, i.e., for $p,p'\in \mathsf{Proc}$, $p\pretestingeq' p'$ if and only if $p\pretesting p'$ and $p'\pretesting p$.
\end{lemma}

\subsection{Pomset Labelled Transition System}\label{plts}

\begin{restatable}[Pomset labelled transition system]{definition}{pomsetlts}\label{pomlts}
A pomset labelled transition system (PLTS) is a quadruple $\langle\mathsf{Proc},\mathsf{Act},\{\xrightarrow{U}|U\in\mathsf{Pom}\},\mathsf{Pred}\rangle$, where:

\begin{enumerate}
  \item $\mathsf{Proc}$ is a set of processes, ranged over by $p,q$.
  \item $\mathsf{Act}$ is a set of actions, ranged over by $a,b$.
  \item $\mathsf{Pom}$ is the set of pomsets over $\mathsf{Act}$, ranged over by $U,V$.
  \item $\xrightarrow{U}\subseteq\mathsf{Proc}\times\mathsf{Pom}\times\mathsf{Proc}$ is called a pomset transition for every $U\in\mathsf{Pom}$. We write $p\xrightarrow{U}q$ instead of $(p,q)\in\xrightarrow{U}$, and write $p\xnrightarrow{U}$ if $p\xrightarrow{U}q$ with no state $q$. Intuitively, $p\xrightarrow{U}q$ means that state $p$ can evolve into state $q$ by the execution of pomset $U$. We see that traditional single action transition $p\xrightarrow{a}q$ with $a\in\mathsf{Act}$ is a special case of pomset transition in which the pomset is primitive.
  \item For every $P\in\mathsf{Pred}$, we write $pP$ (resp. $p\neg P$) if state $p$ satisfies (resp. does not satisfy) predicate $P$. Intuitively, $pP$ means that predicate $P$ holds in state $p$.
\end{enumerate}

The binary pomset transitions $p\xrightarrow{U}q$ and unary predicates $pP$ in a PLTS are called transitions.

Note that, by replacing $U\in\mathsf{Pom}$ by $a\in\mathsf{Act}$, we can get the definition of traditional labelled transition system (LTS). When $U\in\mathsf{Stp}$, we get the special case of a PLTS, called Step Labelled Transition System (SLTS). And we use $\mathsf{Act}(p)\subseteq\mathsf{Act}$ to denote the set of actions occurring in $p$.
\end{restatable}

\begin{definition}[Finiteness conditions on a PLTS]
A PLTS is:

\begin{itemize}
  \item Finitely branching: if for every state $p$ there are only finitely many outgoing pomset transitions $p\xrightarrow{U}q$.
  \item Regular: if it is finitely branching and each state can reach only finitely many other states.
  \item Finite: if it is finitely branching and there is no infinite sequence of pomset transitions $p_0\xrightarrow{U_0}p_1\xrightarrow{U_1}\cdots$.
\end{itemize}
\end{definition}

A pomset transition system specification is a collection of inductive proof rules to derive the pomset transitions over states in $\mathsf{Proc}$.

\begin{definition}[Pomset transition system specification]
Let $\Sigma$ be a signature, and $t,t'\in\mathbb{T}(\Sigma)$. A pomset transition rule $\rho$ is of the form $\frac{H}{\alpha}$, where $H$ is the set of premises with positive premises $t\xrightarrow{U}t'$ and $tP$, and negative premises $t\xnrightarrow{U}$ and $t\neg P$; $\alpha$ is the conclusion with the form $t\xrightarrow{U}t'$ and $tP$, and $U\in\mathsf{Pom}$ and $P$ is a predicate. For the conclusion with the form of $t\xrightarrow{U}t'$, the left-hand side of the conclusion $t$ is called the source of $\rho$ and the right-hand side of the conclusion $t'$ is called the target of $\rho$. A transition rule is closed if it does not contain variables.

A pomset transition system specification (PTSS) is a set of pomset transition rules. A PTSS is positive if its transition rules do not contain negative premises.
\end{definition}

\begin{definition}[Extended pomset labelled transition system]
An extended pomset labelled transition system (EPLTS) is a quintuple $\langle\mathsf{Proc},\mathsf{Act},\{\xrightarrow{U}|U\in\mathsf{Pom}\},\mathsf{Pred},\rightarrowtail\rangle$, where:

\begin{enumerate}
  \item $\langle\mathsf{Proc},\mathsf{Act},\{\xrightarrow{U}|U\in\mathsf{Pom}\},\mathsf{Pred}, \rightarrowtail\rangle$ is a PLTS
  \item $\rightarrowtail$ is a binary relation over $\mathsf{Proc}$, the internal action relation.
\end{enumerate}

The corresponding finiteness condition and PTSS can be extended the extended ones obviously.
\end{definition}

Both the operational behaviours of processes and experimenters can be modelled by EPLTSs, and the EPLTS of experimenters can be defined as $\langle\mathsf{Exp},\mathsf{Act}\cup\{\omega,\mu,\tau\},\{\xrightarrow{U}|U\in\mathsf{Pom}\},\mathsf{Pred},\rightarrowtail\rangle$, where $\omega$ is special action and $Success=\{e\in\mathsf{Exp}|e\xrightarrow{\omega}\}$; $\tau$ is a special action, while $$\frac{e\xrightarrow{\tau}e'}{e\parallel_{EP}p\mapsto e'\parallel_{EP}p}$$ for $e,e'\in\mathsf{Exp}$ and $p\in\mathsf{Proc}$; $\mu$ is a special action, while $$\frac{e\xrightarrow{\mu}e'\quad p\surd}{e\parallel_{EP}p\mapsto e'\parallel_{EP}p}$$ for $\surd$ is the successful termination predicate, $e,e'\in\mathsf{Exp}$ and $p\in\mathsf{Proc}$. For $e,e'\in\mathsf{Exp}$, $p,p'\in\mathsf{Proc}$ and $U\in\mathsf{Pom}$, we have the following compatible interaction between experimenters and processes:

$$\frac{e\xrightarrow{U}e'\quad p\xrightarrow{U}p'}{e\parallel_{EP}p\mapsto e'\parallel_{EP}p'}$$
$$\frac{e\rightarrowtail e'}{e\parallel_{EP}p\mapsto e'\parallel_{EP}p}$$
$$\frac{p\rightarrowtail p'}{e\parallel_{EP}p\mapsto e\parallel_{EP}p'}$$

\begin{definition}
Let $L_{\mathbf{Proc}}$ and $L_{\mathsf{Exp}}$ be two compatible EPLTSs $\langle\mathsf{Proc},\mathsf{Act},\{\xrightarrow{U}|U\in\mathsf{Pom}\},\mathsf{Pred},\rightarrowtail\rangle$ and $\langle\mathsf{Exp},\mathsf{Act}\cup\{\omega,\mu,\tau\},\{\xrightarrow{U}|U\in\mathsf{Pom}\},\mathsf{Pred},\rightarrowtail\rangle$ respectively. Then $\mathscr{EP}(L_{\mathsf{Proc}},L_{\mathsf{Exp}})$ is the experimental system $\langle \mathsf{Proc},\mathsf{Exp},\mapsto,Success$, where $Success=\{e\in\mathsf{Exp}|e\xrightarrow{\omega}\}$ and $\mapsto$ is defined above as the compatible interaction between experimenters and processes.
\end{definition} 
\newpage\section{Basic Processes}\label{bpT}

In this chapter, we introduce the basic processes. Because of the elementary properties of parallelism in true concurrency, we move the parallelism and concurrency related operators into the basic processes. Firstly, we introduce the basic signature in \cref{bs5}, then the operational semantics, denotational semantics and axiomatic semantics of the basic processes are introduced in \cref{os5}, \cref{ds5} and \cref{as5}, respectively. Finally, we get the results on trinity of operational semantics, denotational semantics and axiomatic semantics in \cref{t5}.

\subsection{Basic Signature}\label{bs5}

\begin{definition}[Basic signature]
The basic signature $\Sigma^1$ consists of:

\begin{enumerate}
  \item A set of atomic actions $\mathsf{Act}$ ranged over $a,b,\cdots$.
  \item A set of pomsets $\mathsf{Pom}$ over $\mathsf{Act}$ ranged over $U,V,\cdots$.
  \item A constant $0$ denoting inaction without any behaviour.
  \item A constant $1$ denoting empty action which terminates immediately and successfully.
  \item The communication action $\gamma(a,b)$.
  \item The binary operator $\cdot$ as the sequential composition, i.e., for processes $p$ and $p'$, the process $p\cdot p'$ firstly executes $p$ followed $p'$. The process $p\cdot p'$ is abbreviated as $pp'$.
  \item The binary operator $\sharp$ as the conflict composition, i.e., the process $a\sharp b$ either executes $a$ and its successors or $b$ and its successors.
  \item The binary operator $\osharp$ as the internal conflict composition, i.e., the process $a\sharp b$ either executes $a$ and its successors or $b$ and its successors internally.
  \item The binary operator $+$ as the alternative composition, i.e., for processes $p$ and $p'$, the process $p+p'$ either executes $p$ or $p'$ alternatively.
  \item The binary operator $\oplus$ as the internal alternative composition, i.e., for processes $p$ and $p'$, the process $p+p'$ either executes $p$ or $p'$ alternatively and internally.
  \item The binary operator $\between$ as the concurrent composition, i.e., for processes $p$ and $p'$, the process $p\between p'$ means $p$ and $p'$ execute concurrently, i.e., in parallel but may be with unstructured communications.
  \item The binary operator $\parallel$ as the parallel composition, i.e., for processes $p$ and $p'$, the process $p\parallel p'$ executes $p$ and $p'$ in parallel.
  \item The binary operator $\mid$ as the communication merge, i.e., for processes $p$ and $p'$, the process $p\mid p'$ executes with synchronous communications. 
  $$a\mid b=\begin{cases}
                \gamma(a,b), & a\leq^c b;\\
                0, & \mbox{otherwise}.
            \end{cases}$$
            where $a\leq^c b$ denotes that there exists a communication between $a$ and $b$.
  \item The unary operator $\Theta$ as confliction eliminator, i.e., for process $p$, the process $\Theta(p)$ eliminates and the $\sharp$ relations between actions in $p$.
  \item The binary unless operator $\triangleleft$ as an auxiliary operator to confliction eliminator $\Theta$.
  \item The unary operator $\partial_H$ as the encapsulation, i.e., for process $p$, the process $\partial_H(p)$ renames all actions of $p$ in the set $H$ to $0$.
\end{enumerate}

Brackets are omitted whenever possible, with sequential composition $\cdot$ having a higher precedence than concurrent composition $\between$, parallel composition $\parallel$ and communication merge $\mid$. Concurrent composition $\between$, parallel composition $\parallel$ and communication merge $\mid$ have the same precedences which are higher than internal alternative composition $\oplus$ and internal conflict composition $\osharp$. While internal alternative composition $\oplus$ and internal conflict composition $\osharp$ have the same precedences which are higher than alternative composition $+$ and conflict composition $\sharp$, and alternative composition $+$ and conflict composition $\sharp$ have the same precedences.
\end{definition}

\begin{definition}[Syntax of basic process language]
The syntax of the basic process language $\mathbf{M}_1$ is given by the following BNF grammar:

$p::=~0~|~1~|~U~|~\gamma(a,b)~|~p\cdot p~|~a\sharp b~|~a\osharp b~|~p+p~|~p\oplus p~|~p\parallel p~|~p\mid p~|~p\between p~|~\Theta(p)~|~p\triangleleft p~|~\partial_H(p)~$

where $a,b\in\mathsf{Act}$, $U\in\mathbf{Pom}$, $p\in \mathsf{Proc}$.
\end{definition}

\subsection{Operational Semantics}\label{os5}

In this section, we give the operational semantics of the language $\mathbf{M}_1$. The predicate $\surd$ represents successful termination, $\xrightarrow{a}\surd$ represents successful termination after execution of the action $a\in\mathsf{Act}$, $\xrightarrow{U}\surd$ represents successful termination after execution of the action $U\in\mathsf{Pom}$ and $\xrightarrow{ }\surd$ represents successful termination without execution of the any action. The following are the PTSS of the language $\mathbf{M}_1$, where $p,q\in\mathsf{Proc}$.

The PTSS of action $1$, $a\in\mathsf{Act}$ and $U\in\mathsf{Pom}$ is as follows. Note that, there is no any transition rules for $0$.

$$\frac{}{1\xrightarrow{ }\surd}\quad\frac{}{a\xrightarrow{a}\surd}\quad\frac{}{U\xrightarrow{U}\surd}$$

The PTSS of sequential composition is as follows.

$$\frac{p\xrightarrow{U}\surd}{p\cdot q\xrightarrow{U} q}\quad\frac{p\xrightarrow{U}p'}{p\cdot q\xrightarrow{U} p'\cdot q}$$

The PTSS of alternative composition is as follows.

$$\frac{p\xrightarrow{U}\surd}{p+ q\xrightarrow{U}\surd} \quad\frac{p\xrightarrow{U}p'}{p+ q\xrightarrow{U}p'} \quad\frac{q\xrightarrow{U}\surd}{p+ q\xrightarrow{U}\surd} \quad\frac{q\xrightarrow{U}q'}{p+ q\xrightarrow{U}q'}$$

The PTSS of concurrent composition is as follows.

$$\frac{p\xrightarrow{a}\surd\quad q\xrightarrow{b}\surd}{p\between q\xrightarrow{\step{a,b}}\surd} \quad\frac{p\xrightarrow{a}p'\quad q\xrightarrow{b}\surd}{p\between q\xrightarrow{\step{a,b}}p'}$$
$$\frac{p\xrightarrow{a}\surd\quad q\xrightarrow{b}q'}{p\between q\xrightarrow{\step{a,b}}q'} \quad\frac{p\xrightarrow{a}p'\quad q\xrightarrow{b}q'}{p\between q\xrightarrow{\step{a,b}}p'\between q'}$$
$$\frac{p\xrightarrow{a}\surd\quad q\xrightarrow{b}\surd}{p\between q\xrightarrow{\gamma(a,b)}\surd} \quad\frac{p\xrightarrow{a}p'\quad q\xrightarrow{b}\surd}{p\between q\xrightarrow{\gamma(a,b)}p'}$$
$$\frac{p\xrightarrow{a}\surd\quad q\xrightarrow{b}q'}{p\between q\xrightarrow{\gamma(a,b)}q'} \quad\frac{p\xrightarrow{a}p'\quad q\xrightarrow{b}q'}{p\between q\xrightarrow{\gamma(a,b)}p'\between q'}$$

The PTSS of parallel composition is as follows.

$$\frac{p\xrightarrow{a}\surd\quad q\xrightarrow{b}\surd}{p\parallel q\xrightarrow{\step{a,b}}\surd} \quad\frac{p\xrightarrow{a}p'\quad q\xrightarrow{b}\surd}{p\parallel q\xrightarrow{\step{a,b}}p'}$$
$$\frac{p\xrightarrow{a}\surd\quad q\xrightarrow{b}q'}{p\parallel q\xrightarrow{\step{a,b}}q'} \quad\frac{p\xrightarrow{a}p'\quad q\xrightarrow{b}q'}{p\parallel q\xrightarrow{\step{a,b}}p'\between q'}$$

The PTSS of communication merge is as follows.

$$\frac{p\xrightarrow{a}\surd\quad q\xrightarrow{b}\surd}{p\mid q\xrightarrow{\gamma(a,b)}\surd} \quad\frac{p\xrightarrow{a}p'\quad q\xrightarrow{b}\surd}{p\mid q\xrightarrow{\gamma(a,b)}p'}$$
$$\frac{p\xrightarrow{a}\surd\quad q\xrightarrow{b}q'}{p\mid q\xrightarrow{\gamma(a,b)}q'} \quad\frac{p\xrightarrow{a}p'\quad q\xrightarrow{b}q'}{p\mid q\xrightarrow{\gamma(a,b)}p'\between q'}$$

The PTSS of encapsulation operator is as follows.
        
$$\frac{p\xrightarrow{a}\surd\quad a\notin H}{\partial_H(p)\xrightarrow{a}\surd}\quad\frac{p\xrightarrow{a}p'\quad a\notin H}{\partial_H(p)\xrightarrow{a}\partial_H(p')}$$

The PTSS of confliction, confliction eliminator and the auxiliary unless operator is as follows, where $\leq^e$ is the execution order.

$$\frac{p\xrightarrow{a}\surd\quad a\sharp b}{\Theta(p)\xrightarrow{a}\surd} \quad\frac{p\xrightarrow{b}\surd\quad a\sharp b}{\Theta(p)\xrightarrow{b}\surd}$$
$$\frac{p\xrightarrow{a}p'\quad a\sharp b}{\Theta(p)\xrightarrow{a}\Theta(p')} \quad\frac{p\xrightarrow{b}p'\quad a\sharp b}{\Theta(p)\xrightarrow{b}\Theta(p')}$$
$$\frac{p\xrightarrow{c}\surd \quad q\xnrightarrow{b}\quad a\sharp b\quad c\leq^e a}{p\triangleleft q\xrightarrow{c}\surd}
\quad\frac{p\xrightarrow{c}p' \quad q\xnrightarrow{b}\quad a\sharp b\quad c\leq^e a}{p\triangleleft q\xrightarrow{c}p'}$$
$$\frac{p\xrightarrow{a}\surd \quad q\xnrightarrow{b}\quad a\sharp b}{p\triangleleft q\xrightarrow{ }\surd}
\quad\frac{p\xrightarrow{a}p' \quad q\xnrightarrow{b}\quad a\sharp b}{p\triangleleft q\xrightarrow{ }p'}$$
$$\frac{p\xrightarrow{a}\surd \quad q\xnrightarrow{c}\quad a\sharp b\quad b\leq^e c}{p\triangleleft q\xrightarrow{ }\surd}
\quad\frac{p\xrightarrow{a}p' \quad q\xnrightarrow{c}\quad a\sharp b\quad b\leq^e c}{p\triangleleft q\xrightarrow{ }p'}$$
$$\frac{p\xrightarrow{c}\surd \quad q\xnrightarrow{b}\quad a\sharp b\quad a\leq^e c}{p\triangleleft q\xrightarrow{ }\surd}
\quad\frac{p\xrightarrow{c}p' \quad q\xnrightarrow{b}\quad a\sharp b\quad a\leq^e c}{p\triangleleft q\xrightarrow{ }p'}$$

The PTSS of internal confliction, confliction eliminator and the auxiliary unless operator is as follows.

$$\frac{p\xrightarrow{a}\surd\quad a\osharp b}{\Theta(p)\xrightarrow{a}\surd} \quad\frac{p\xrightarrow{b}\surd\quad a\osharp b}{\Theta(p)\xrightarrow{b}\surd}$$
$$\frac{p\xrightarrow{a}p'\quad a\osharp b}{\Theta(p)\xrightarrow{a}\Theta(p')} \quad\frac{p\xrightarrow{b}p'\quad a\osharp b}{\Theta(p)\xrightarrow{b}\Theta(p')}$$
$$\frac{p\xrightarrow{c}\surd \quad q\xnrightarrow{b}\quad a\osharp b\quad c\leq^e a}{p\triangleleft q\xrightarrow{c}\surd}
\quad\frac{p\xrightarrow{c}p' \quad q\xnrightarrow{b}\quad a\osharp b\quad c\leq^e a}{p\triangleleft q\xrightarrow{c}p'}$$
$$\frac{p\xrightarrow{a}\surd \quad q\xnrightarrow{b}\quad a\osharp b}{p\triangleleft q\xrightarrow{ }\surd}
\quad\frac{p\xrightarrow{a}p' \quad q\xnrightarrow{b}\quad a\osharp b}{p\triangleleft q\xrightarrow{ }p'}$$
$$\frac{p\xrightarrow{a}\surd \quad q\xnrightarrow{c}\quad a\osharp b\quad b\leq^e c}{p\triangleleft q\xrightarrow{ }\surd}
\quad\frac{p\xrightarrow{a}p' \quad q\xnrightarrow{c}\quad a\osharp b\quad b\leq^e c}{p\triangleleft q\xrightarrow{ }p'}$$
$$\frac{p\xrightarrow{c}\surd \quad q\xnrightarrow{b}\quad a\osharp b\quad a\leq^e c}{p\triangleleft q\xrightarrow{ }\surd}
\quad\frac{p\xrightarrow{c}p' \quad q\xnrightarrow{b}\quad a\osharp b\quad a\leq^e c}{p\triangleleft q\xrightarrow{ }p'}$$

The PTSS of internal alternative composition is as follows.

$$\frac{}{1\rightarrowtail\surd}$$
$$\frac{}{p\oplus q\rightarrowtail p} \quad\frac{}{p\oplus q\rightarrowtail q}$$
$$\frac{p\rightarrowtail p'}{p\cdot q\rightarrowtail p'\cdot q}$$
$$\frac{p\rightarrowtail p'}{p+q\rightarrowtail p'+q} \quad\frac{q\rightarrowtail q'}{p+q\rightarrowtail p+q'}$$
$$\frac{p\rightarrowtail p'}{p\between q\rightarrowtail p'\between q} \quad\frac{q\rightarrowtail q'}{p\between q\rightarrowtail p\between q'} \quad\frac{p\rightarrowtail p'\quad q\rightarrowtail q'}{p\between q\rightarrowtail p'\between q'}$$
$$\frac{p\rightarrowtail p'}{p\parallel q\rightarrowtail p'\parallel q} \quad\frac{q\rightarrowtail q'}{p\parallel q\rightarrowtail p\parallel q'} \quad\frac{p\rightarrowtail p'\quad q\rightarrowtail q'}{p\parallel q\rightarrowtail p'\parallel q'}$$
$$\frac{p\rightarrowtail p'}{p\mid q\rightarrowtail p'\mid q} \quad\frac{q\rightarrowtail q'}{p\mid q\rightarrowtail p\mid q'} \quad\frac{p\rightarrowtail p'\quad q\rightarrowtail q'}{p\mid q\rightarrowtail p'\mid q'}$$
$$\frac{p\rightarrowtail p'}{\partial_H(p)\rightarrowtail \partial_H(p')}$$
$$\frac{p\rightarrowtail p'}{\Theta(p)\rightarrowtail \Theta(p')}$$
$$\frac{p\rightarrowtail p'}{p\triangleleft q\rightarrowtail p'\triangleleft q} \quad\frac{q\rightarrowtail q'}{p\triangleleft q\rightarrowtail p\triangleleft q'}$$

\begin{definition}
For any PLTS $\langle\mathsf{Proc},\mathsf{Act},\{\xrightarrow{U}|U\in\mathsf{Pom}\},\mathsf{Pred}\rangle$, $p,p'\in\mathsf{Proc}$ and $s\in\mathsf{Pom}^*$ respectively, the following are defined:

\begin{enumerate}
  \item $D(p,s)=\{p'\in\mathsf{Proc}|p\xrightarrow{s}p'\}$, the $s$-derivatives of $p$.
  \item $L(p)=\{s|p\xrightarrow{s}\}$, the language of $p$.
  \item $D(p)=\{p'|\mbox{ for some }U\in\mathsf{Pom},p\xrightarrow{U}p'\}$, the derivatives of $p$.
  \item $S(p)=\{U|p\xrightarrow{U}\}$, the successors of $p$.
  \item $S(p,s)=\{U|p\xrightarrow{s}p'\xrightarrow{U}\}$, the successors of $p$ after $s$.
  \item $\mathscr{A}(p,s)=\{S(p')|p\xrightarrow{s}p'\}$, the Acceptance sets of $p$ after $s$.
\end{enumerate}
\end{definition}

If $\mathscr{A},\mathscr{B}$ are Acceptance sets, we write $\mathscr{A}\subset\subset\mathscr{B}$ if for every $A\in\mathscr{A}$, there exists some $B\in\mathscr{B}$ such that $B\subseteq A$.

\begin{definition}
For any PLTS $\langle\mathsf{Proc},\mathsf{Act},\{\xrightarrow{U}|U\in\mathsf{Pom}\},\mathsf{Pred}\rangle$, $p,p'\in\mathsf{Proc}$ and $s\in\mathsf{Pom}^*$ respectively, the following are defined:

\begin{enumerate}
  \item $p\ll_{\mathrm{MAY}} p'$ if $L(p)\subseteq L(p')$.
  \item $p\ll_{\mathrm{MUST}} p'$ if $\mathscr{A}(p',s)\subset\subset\mathscr{A}(p,s)$ for every $s\in\mathsf{Pom}^*$.
  \item $p\ll p'$ if both $p\ll_{\mathrm{MAY}}p'$ and $p\ll_{\mathrm{MUST}}p'$.
\end{enumerate}
\end{definition} 

\begin{lemma}
If $p\ll_{\mathrm{MUST}}p'$ then $L(p')\subseteq L(p)$.
\end{lemma}

\begin{lemma}
Both $\ll_{\mathrm{MAY}}$ and $\ll_{\mathrm{MUST}}$ are $\Sigma^1$-preorders.
\end{lemma}

For $U,U_1,\cdots,U_k,V_1,\cdots,V_n\in\mathsf{Stp}$, one essential type of test is of the form denoted $e(s,U)$ where $s$ is the sequence $V_1\cdots V_n$:

$$\tau w+V_1(\tau w+\cdots+V_n(\tau w+U)\cdots)$$

Another essential type of test is of the form denoted $e(s,A)$  where $s$ is the sequence $V_1\cdots V_n$ and $A$ is the set of steps $\{U_1,\cdots,U_k\}$:

$$\tau w+V_1(\tau w+V_2(\tau w+\cdots+V_n(U_1 w+\cdots+U_k w)\cdots))$$

Let $T$ denote the set of all experiments of the form $e(s,U)$ or $e(s,A)$. For $p,q\in\mathsf{Proc}$ and every $e\in T$, $p\pretestingmust^T q$ if $p\must e$ implies $q\must e$.

\begin{proposition}
$p\pretestingmust^{T} p'$ implies $p\ll_{\mathrm{MUST}}p'$.
\end{proposition}

\begin{theorem}[Alternative characterization of testing preorders]\label{actp}
For every $p,p'\in\mathbf{M}_1$:

\begin{enumerate}
  \item $p\pretestingmay p'$ if and only if $p\ll_{\mathrm{MAY}}p'$.
  \item $p\pretestingmust p'$ if and only if $p\ll_{\mathrm{MUST}}p'$.
  \item $p\pretesting p'$ if and only if $p\ll p'$.
\end{enumerate}
\end{theorem}

By use of the following relations:

\begin{enumerate}
  \item $p\xRightarrow{U}p'$ if $p\rightarrowtail^* q\xrightarrow{U}q'\rightarrowtail^* p'$ for some $q,q'$.
  \item $p\xRightarrow{1}p'$ if $p\rightarrowtail^* p'$.
  \item $p\xRightarrow{Us}$ if $p\xRightarrow{U}q\xRightarrow{s}p'$ for some $q$.
\end{enumerate}

and the following definitions:

\begin{enumerate}
  \item $Im(p)=\{p'|p\rightarrowtail p'\}$.
  \item $D(p,s)=\{p'\in\mathsf{Proc}|p\xRightarrow{s}p'\}$, the $s$-derivatives of $p$.
  \item $L(p)=\{s|p\xRightarrow{s}\}$, the language of $p$.
  \item $D(p)=\{p'|\mbox{ for some }U\in\mathsf{Pom},p\xRightarrow{U}p'\}$, the derivatives of $p$.
  \item $S(p)=\{U|p\xRightarrow{U}\}$, the successors of $p$.
  \item $S(p,s)=\{U|p\xRightarrow{s}p'\xRightarrow{U}\}$, the successors of $p$ after $s$.
  \item $\mathscr{A}(p,s)=\{S(p')|p\xRightarrow{s}p'\}$, the Acceptance sets of $p$ after $s$.
\end{enumerate}

we can get the modified definition of the preorders $\ll_{\mathrm{MUST}}$, $\ll_{\mathrm{MAY}}$ and $\ll$. The result theorem of alternative characterization of testing preorders in \cref{actp} still holds.

\subsection{Denotational Semantics}\label{ds5}

\begin{figure}
\begin{center}
\begin{forest}
for tree={
  circle, fill, draw,
  minimum size=6pt,
  inner sep=0pt,
  l sep=20pt,
  s sep=30pt,
  /tikz/every label/.style={font=\tiny},
}
[,name=R,label=south:$\parallel$
  [,name=C1,label=south west:$\parallel$,label=south east:$\parallel$,label=south:$+$,
    edge label={node[midway,sloped,above,font=\tiny]{$a_1$}}
    [,name=C11,
      edge label={node[midway,sloped,above,font=\tiny]{$a_2$}}
      [,name=C111,
        edge label={node[midway,left,font=\tiny]{$a_3$}}
        [,name=c1111,
          edge label={node[midway,left,font=\tiny]{$a_4$}}
        ]
      ]
    ]
    [,name=C12,
      edge label={node[midway,sloped,right, below,font=\tiny]{$a_5$}}
      [,name=C121,
        edge label={node[midway,right,font=\tiny]{$a_6$}}
        [,name=C1211,
          edge label={node[midway,right,font=\tiny]{$a_7$}}
          [,name=C12111,
            edge label={node[midway,right,font=\tiny]{$a_8$}}
          ]
        ]
      ]
    ]
    [,name=C13,
      edge label={node[midway,sloped,right,above,font=\tiny]{$a_9$}}
      [,name=C131,
        edge label={node[midway,right,font=\tiny]{$a_{10}$}}
      ]
    ]
    [,name=C14,
      edge label={node[midway,sloped,above,font=\tiny]{$a_{11}$}}
      [,name=C141,
        edge label={node[midway,right,font=\tiny]{$a_{12}$}}
      ]
    ]
  ]
  [,name=C2,
    edge label={node[midway,sloped,above,font=\tiny]{$a_{13}$}}
    [,name=C21,
      edge label={node[midway,right,font=\tiny]{$a_{14}$}}
      [,name=C211,
        edge label={node[midway,right,font=\tiny]{$a_{15}$}}
        [,name=2111,
          edge label={node[midway,right,font=\tiny]{$a_{16}$}}
        ]
      ]
    ]
  ]
]
\draw[->] (C111) to (C121);
\draw[dashed] (C14) to (C211);
\end{forest}
\end{center}
\caption{An example of $\mathbf{fPAT}$}
\label{EoffPAT}
\end{figure}
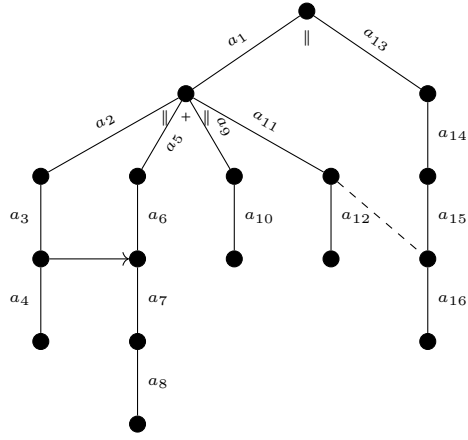

In this section, we give the denotational semantics of the language $\mathbf{M}_1$, the denotational interpretation of $\mathbf{M}_1$ is finite Pomset Acceptance Tree ($\mathbf{fPAT}$), which is a parallelism generalization of Hennessy's finite Acceptance Tree ($\mathbf{fAT}$) in \cite{TS}. In an $\mathbf{fPAT}$, every node and every edge are labelled. Intuitively, every edge can be labelled by a single action. By this way, an $\mathbf{fPAT}$ is usually unstructured, i.e., there are two kinds of branches: parallel ones and alternative ones, and there may exist communications or conflictions among actions located in different parallel branches. It is reasonable that all partial orders between actions in different parallel branches are communications. In the example of $\mathbf{fPAT}$ in \cref{EoffPAT}, there exist parallel branches denoted by $\parallel$, alternative branches denoted by $+$, communications between parallel branches denoted by lines with arrows, and conflictions between parallel branches denoted by dashed lines. Communications and conflictions between actions in different parallel branches are the causes of the unstructuredness.

The $\mathbf{fPAT}$ in \cref{EoffPAT} can be expressed by a truly concurrent process term of the language $\mathbf{M}_1$ with two unstructured constraints: $$(a_1\cdot((a_2\cdot a_3\cdot a_4)\between(a_5\cdot a_6\cdot a_7\cdot a_8)+(a_9\cdot a_{10})\between(a_{11}\cdot a_{12})))\between(a_{13}\cdot a_{14}\cdot a_{15}\cdot a_{16})\quad(a_3\leq^c a_6,a_{12}\sharp a_{16})$$

\begin{definition}[$\mathbf{fPAT}$]
$\mathbf{fPAT}$, the set of finite acceptance trees over $\mathsf{Pom}$, is the set of rooted trees whose branches are labelled by elements of $\mathsf{Pom}$, whose nodes are labelled by subsets of $\mathsf{Pom}^*$, and satisfies the following requirements:

\begin{enumerate}
  \item R1 (Determinism): For every pomset $U$, every node in the tree has most one successor alternative branch labelled by $U$. So, every node in the tree is uniquely determined by a string in $\mathsf{Pom}^*$, denoted by $L(t)$, and if $s\in L(t)$ we use $t(s)$ to denote the node identified by $s$. The set of actions labelling the successor branches of a node $n$ is called its successor set and denoted by $S(n)$. Note that, parallel branch is different to the alternative one, the successor parallel branches can be labelled by the same $U\in\mathsf{Pom}$, which means $U$ is executed several times in parallel.
  \item R2 (Finite Branching): For every $s\in L(t)$, $S(L(t),s)$ is finite, where $S(L,s)=\{U|sU\in L\}$ for $L\subseteq\mathsf{Pom}^*$ and $s\in\mathsf{Pom}^*$, so $S(t(s))=S(L(t),s)$ for any tree $t$.
  \item R3: $\mathscr{A}(n)$ is an $S(n)$-set, where $\mathscr{A}(n)$ is the acceptance set associated with the node $n$ and $S(n)$ is the $S$-set of $n$. These set labelling the nodes, called Acceptance sets. For $S\subseteq\mathsf{Pom}$ and $S$-set $\emptyset\neq\mathscr{A}\subseteq\mathsf{Pom}$, which satisfies:
      \begin{enumerate}
        \item For every $K\in\mathscr{A}$, $K\subseteq S$.
        \item For every $U\in S$, there is some $K\in\mathscr{A}$ such that $U\in K$.
        \item ($\cup$-closed) If $K_1,K_2\in\mathscr{A}$, then $K_1\cup K_2\in\mathscr{A}$.
        \item (Convex-closed) If $K_1,K_2\in\mathscr{A}$ and $K_1\subseteq K\subseteq K_2$, then $K\in\mathscr{A}$.
      \end{enumerate}
      We say $\mathscr{A}$, which is a set of subsets, is saturated if it is an $S$-set for some $S$. For $t\in\mathbf{fPAT}$, let $\mathsf{Act}(t)\subseteq\mathsf{Act}$ be the set of actions occurring in $t$, and let $\mathsf{Pom}(t)\subseteq\mathsf{Pom}$ be the set of pomsets occurring in $t$.
\end{enumerate}
\end{definition}

\begin{definition}[Partial order over $\mathbf{fPAT}$]
A partial order $\leq_{\mathbf{fPAT}}$ is defined over $\mathbf{fPAT}$ as follows, for two trees $t,t'\in\mathbf{fPAT}$, $t\leq_{\mathbf{fPAT}}t'$, if:

\begin{enumerate}
  \item $L(t)=L(t')$.
  \item For every $s\in L(t')$, $\mathscr{A}(t'(s))\subseteq\mathscr{A}(t(s))$.
\end{enumerate}
\end{definition}

\begin{lemma}
$\langle \mathbf{fPAT},\leq_{\mathbf{fPAT}}\rangle$ is a partial order.
\end{lemma}

In the following, we define functions over $\mathbf{fPAT}$ for every function symbol in $\Sigma^1$: $0_{\mathbf{fPAT}}$, $1_{\mathbf{fPAT}}$, $U_{\mathbf{fPAT}}$, $\cdot_{\mathbf{fPAT}}$, $+_{\mathbf{fPAT}}$, $\oplus_{\mathbf{fPAT}}$, $\parallel_{\mathbf{fPAT}}$, $\gamma(a,b)_{\mathbf{fPAT}}$, $\mid_{\mathbf{fPAT}}$, $\between_{\mathbf{fPAT}}$, $\partial_H(p)_{\mathbf{fPAT}}$, $\sharp_{\mathbf{fPAT}}$, $\Theta(p)_{\mathbf{fPAT}}$, $\triangleleft_{\mathbf{fPAT}}$, $\osharp_{\mathbf{fPAT}}$.

(1) $0_{\mathbf{fPAT}}$.

Let $0_{\mathbf{fPAT}}$ denote the trivial tree $\bullet\{\emptyset\}$, which consists of only one node, labelled by the set of the empty set $\emptyset$ with no successor branches, and rendered as $\bullet$.

(2) $1_{\mathbf{fPAT}}$.

Let $1_{\mathbf{fPAT}}$ denote the trivial tree $\bullet\{\{\}\}$, which consists of one node, labelled by the set of the empty set $\{\}$, and rendered as $\obullet$. Note that, $1_{\mathbf{fPAT}}$ can have successor branches.

(3) $U_{\mathbf{fPAT}}$.

Let $U_{\mathbf{fPAT}}$ denote the trivial tree:

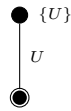
\begin{center}
\begin{forest}
  for tree={
  circle, fill, draw,
  minimum size=6pt,
  inner sep=0pt,
  l sep=20pt,
  s sep=30pt,
  /tikz/every label/.style={font=\tiny},
  }
  [, circle, fill,label=right:$\{U\}$
    [,circle,double, fill, edge label={node[midway,right,font=\tiny]{$U$}}]
  ]
\end{forest}
\end{center}

Formally, the tree $U_{\mathbf{fPAT}}$ is the tree $t$ determined by:

\begin{enumerate}
  \item $L(t)=\{1\}\cup\{U\}$.
  \item $\mathscr{A}(t(1))=\{\{U\}\}$, $\mathscr{A}(t(U))=\{\{\}\}$.
\end{enumerate}

(4) $\cdot_{\mathbf{fPAT}}$.

The function $\cdot_{\mathbf{fPAT}}$ maps the trees

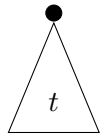
\begin{center}
\begin{forest}
  for tree={
    circle, draw,
    minimum size=6pt,
    inner sep=0pt,
    /tikz/every label/.style={font=\tiny},
    calign=center,
    parent anchor=north,
    child anchor=south,
    l sep=0pt,
    no edge
  },
  triangle/.style={
    isosceles triangle,
    draw,
    shape border rotate=90,
    minimum size=1.2cm,
    inner sep=0pt
  }
  [, circle, fill
    [ $t$, triangle ]
  ]
\end{forest}
\end{center}

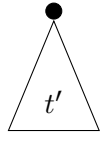
\begin{center}
\begin{forest}
  for tree={
    circle, draw,
    minimum size=6pt,
    inner sep=0pt,
    /tikz/every label/.style={font=\tiny},
    calign=center,
    parent anchor=north,
    child anchor=south,
    l sep=0pt,
    no edge
  },
  triangle/.style={
    isosceles triangle,
    draw,
    shape border rotate=90,
    minimum size=1.2cm,
    inner sep=0pt
  }
  [, circle, fill
    [$t'$, triangle]
  ]
\end{forest}
\end{center}

to the tree

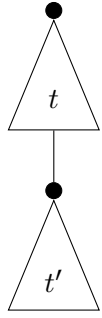
\begin{center}
\begin{forest}
  for tree={
    circle, draw,
    minimum size=6pt,
    inner sep=0pt,
    /tikz/every label/.style={font=\tiny},
    calign=center,
    parent anchor=south,
    child anchor=north,
    l sep=0pt,
  },
  triangle/.style={
    isosceles triangle,
    draw,
    shape border rotate=90,
    minimum size=1.2cm,
    inner sep=0pt
  }
  [, circle, fill
    [ $t$, triangle, no edge 
      [, circle, fill,
        [ $t'$, triangle, no edge ]
      ]
    ]
  ]
\end{forest}
\end{center}

Formally, the tree $t\cdot_{\mathbf{fPAT}}t'$ is the tree $t''$ determined by:

\begin{enumerate}
  \item $L(t'')=\{1\}\cup\{ss'|s\in L(t),s'\in L(t')\}$.
  \item $\mathscr{A}(t''(1))=\mathscr{A}(t(1))$, $\mathscr{A}(t''(\{\}_ss'))=\mathscr{A}(t(s'))$ where $\{\}_s$ is the leaf of $s$.
\end{enumerate}

(5) $+_{\mathbf{fPAT}}$.

The function $+_{\mathbf{fPAT}}$ maps the trees

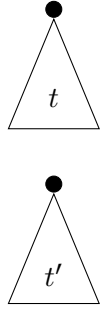
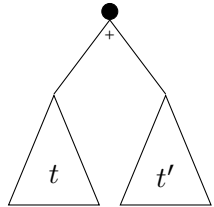
\begin{center}
\begin{forest}
  for tree={
    circle, draw,
    minimum size=6pt,
    inner sep=0pt,
    /tikz/every label/.style={font=\tiny},
    calign=center,
    parent anchor=north,
    child anchor=south,
    l sep=0pt,
    no edge
  },
  triangle/.style={
    isosceles triangle,
    draw,
    shape border rotate=90,
    minimum size=1.2cm,
    inner sep=0pt
  }
  [, circle, fill
    [ $t$, triangle ]
  ]
\end{forest}
\end{center}

\begin{center}
\begin{forest}
  for tree={
    circle, draw,
    minimum size=6pt,
    inner sep=0pt,
    /tikz/every label/.style={font=\tiny},
    calign=center,
    parent anchor=north,
    child anchor=south,
    l sep=0pt,
    no edge
  },
  triangle/.style={
    isosceles triangle,
    draw,
    shape border rotate=90,
    minimum size=1.2cm,
    inner sep=0pt
  }
  [, circle, fill
    [$t'$, triangle]
  ]
\end{forest}
\end{center}

to the tree

\begin{center}
\begin{forest}
  for tree={
    circle, draw,
    minimum size=6pt,
    inner sep=0pt,
    /tikz/every label/.style={font=\tiny},
    calign=center,
    parent anchor=south,
    child anchor=north,
    l sep=0pt,
  },
  left_triangle/.style={
    isosceles triangle,
    draw,
    shape border rotate=120,
    minimum size=1.2cm,
    inner sep=0pt,
    anchor=apex
  },
  right_triangle/.style={
    isosceles triangle,
    draw,
    shape border rotate=60,
    minimum size=1.2cm,
    inner sep=0pt,
    anchor=apex
  }
  [, circle, fill,label=south:$+$
    [$t$, left_triangle] 
    [$t'$, right_triangle]
  ]
\end{forest}
\end{center}

Let $\mathscr{B}$ be any finite collection of finite subsets of $\mathsf{Pom}$ and $A(\mathscr{B})$ be the set of actions which appear in $\mathscr{B}$. Though in general $\mathscr{B}$ is not an $A(\mathscr{B})$-set, it is always possible to extend $\mathscr{B}$ to become an $A(\mathscr{B})$-set. Let $c(\mathscr{B})$ be the least set such that:

\begin{enumerate}
  \item $\mathscr{B}\subseteq c(\mathscr{B})$.
  \item ($\cup$-closed) $X,Y\in c(\mathscr{B})$ implies $X\cup Y\in c(\mathscr{B})$.
  \item (Convex-closed) $X,Y\in c(\mathscr{B})$, $X\subseteq Z\subseteq Y$ implies $Z\in c(\mathscr{B})$.
\end{enumerate}

So, $B\in c(\mathscr{B})$ if and only if it can be derived from the following rules:

\begin{enumerate}
  \item $B\in \mathscr{B}$ implies $B\in c(\mathscr{B})$.
  \item $X,Y\in c(\mathscr{B})$ implies $X\cup Y\in c(\mathscr{B})$.
  \item $X,Y\in c(\mathscr{B})$, $X\subseteq Z\subseteq Y$ implies $Z\in c(\mathscr{B})$.
\end{enumerate}

\begin{lemma}
The following statements hold:

\begin{enumerate}
  \item $c(\mathscr{B})$ is an $A(\mathscr{B})$-set.
  \item $c(\mathscr{B})$ is the least saturated set containing $\mathscr{B}$.
\end{enumerate}
\end{lemma}

If $\mathscr{B}_1$, $\mathscr{B}_2$ are set of sets, their pointwise union is defined as: $$\mathscr{B}_1\ucup\mathscr{B}_2=\{B_1\cup B_2|B_1\in\mathscr{B}_1,B_2\in\mathscr{B}_2\}$$

Then the operators $c$ and $\ucup$ have the following properties.

c1. If $\mathscr{B}$ is an $S$-set then $c(\mathscr{B})=\mathscr{B}$.

c2. If $\mathscr{B}\subseteq\mathscr{B}'$ then $c(\mathscr{B})\subseteq c(\mathscr{B}')$.

c3. $c(\mathscr{B})=c(c(\mathscr{B}))$.

c4. $c(\mathscr{B}_1\cup c(\mathscr{B}_2\cup\mathscr{B}_3))=c(\mathscr{B}_1\cup\mathscr{B}_2\cup\mathscr{B}_3)$.

c5. $c(\mathscr{B}_1\cup\mathscr{B}_2)=c(c(\mathscr{B}_1)\cup c(\mathscr{B}_2))$.

c6. If $A\in c(\mathscr{B})$ then $B\subseteq A$ for some $B\in\mathscr{B}$.

du1. $(\mathscr{B}_1\ucup \mathscr{B}_2)\ucup \mathscr{B}_3=\mathscr{B}_1\ucup(\mathscr{B}_2\ucup \mathscr{B}_3)$.

du2. $\mathscr{B}_1\ucup \mathscr{B}_2=\mathscr{B}_2\ucup \mathscr{B}_1$.

du3. If $\mathscr{B}$ is saturated, then $\mathscr{B}\ucup \mathscr{B}=\mathscr{B}$.

du4. $\mathscr{B}_1\ucup (\mathscr{B}_2\cup \mathscr{B}_3)=(\mathscr{B}_1\ucup \mathscr{B}_2)\cup(\mathscr{B}_1\ucup \mathscr{B}_3)$.

c7. If $\mathscr{B}_1,\mathscr{B}_2$ are saturated, $\mathscr{B}_1\ucup \mathscr{B}_2$ is saturated.

c8. $c(\mathscr{B}_1\ucup \mathscr{B}_2)=c(\mathscr{B}_1)\ucup c(\mathscr{B}_2)$.

Then, formally, the tree $t+_{\mathbf{fPAT}}t'$ is the tree $t''$ determined by:

\begin{enumerate}
  \item $L(t'')=L(t)\cup L(t')$.
  \item $\mathscr{A}(t'')=\mathscr{A}(t)\ucup\mathscr{A}(t')$.
  \item $\mathscr{A}(t''(s))=c(\mathscr{A}(t(s))\cup\mathscr{A}(t'(s)))$ where $s\neq 1$, and if $s\notin L(t)$ then $\mathscr{A}(t(s))=\emptyset$.
\end{enumerate}

(6) $\oplus_{\mathbf{fPAT}}$.

The function $\oplus_{\mathbf{fPAT}}$ maps the trees

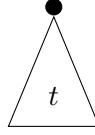
\begin{center}
\begin{forest}
  for tree={
    circle, draw,
    minimum size=6pt,
    inner sep=0pt,
    /tikz/every label/.style={font=\tiny},
    calign=center,
    parent anchor=north,
    child anchor=south,
    l sep=0pt,
    no edge
  },
  triangle/.style={
    isosceles triangle,
    draw,
    shape border rotate=90,
    minimum size=1.2cm,
    inner sep=0pt
  }
  [, circle, fill
    [ $t$, triangle ]
  ]
\end{forest}
\end{center}

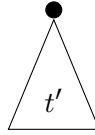
\begin{center}
\begin{forest}
  for tree={
    circle, draw,
    minimum size=6pt,
    inner sep=0pt,
    /tikz/every label/.style={font=\tiny},
    calign=center,
    parent anchor=north,
    child anchor=south,
    l sep=0pt,
    no edge
  },
  triangle/.style={
    isosceles triangle,
    draw,
    shape border rotate=90,
    minimum size=1.2cm,
    inner sep=0pt
  }
  [, circle, fill
    [$t'$, triangle]
  ]
\end{forest}
\end{center}

to the tree

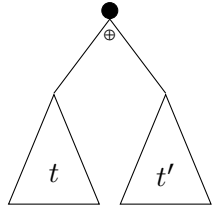
\begin{center}
\begin{forest}
  for tree={
    circle, draw,
    minimum size=6pt,
    inner sep=0pt,
    /tikz/every label/.style={font=\tiny},
    calign=center,
    parent anchor=south,
    child anchor=north,
    l sep=0pt,
  },
  left_triangle/.style={
    isosceles triangle,
    draw,
    shape border rotate=120,
    minimum size=1.2cm,
    inner sep=0pt,
    anchor=apex
  },
  right_triangle/.style={
    isosceles triangle,
    draw,
    shape border rotate=60,
    minimum size=1.2cm,
    inner sep=0pt,
    anchor=apex
  }
  [, circle, fill,label=south:$\oplus$
    [$t$, left_triangle] 
    [$t'$, right_triangle]
  ]
\end{forest}
\end{center}

Formally, the tree $t\oplus_{\mathbf{fPAT}}t'$ is the tree $t''$ determined by:

\begin{enumerate}
  \item $L(t'')=L(t)\cup L(t')$.
  \item $\mathscr{A}(t''(s))=c(\mathscr{A}(t(s))\cup\mathscr{A}(t'(s)))$ for every $s\in L(t'')$, and if $s\notin L(t)$ then $\mathscr{A}(t(s))=\emptyset$.
\end{enumerate}

(7) $\parallel_{\mathbf{fPAT}}$, $\gamma(a,b)_{\mathbf{fPAT}}$, $\mid_{\mathbf{fPAT}}$, $\between_{\mathbf{fPAT}}$ and $\partial_H(p)_{\mathbf{fPAT}}$.

The function $\parallel_{\mathbf{fPAT}}$ maps the trees

\begin{center}
\begin{forest}
  for tree={
    circle, draw,
    minimum size=6pt,
    inner sep=0pt,
    /tikz/every label/.style={font=\tiny},
    calign=center,
    parent anchor=north,
    child anchor=south,
    l sep=0pt,
    no edge
  },
  triangle/.style={
    isosceles triangle,
    draw,
    shape border rotate=90,
    minimum size=1.2cm,
    inner sep=0pt
  }
  [, circle, fill
    [ $t$, triangle ]
  ]
\end{forest}
\end{center}

\begin{center}
\begin{forest}
  for tree={
    circle, draw,
    minimum size=6pt,
    inner sep=0pt,
    /tikz/every label/.style={font=\tiny},
    calign=center,
    parent anchor=north,
    child anchor=south,
    l sep=0pt,
    no edge
  },
  triangle/.style={
    isosceles triangle,
    draw,
    shape border rotate=90,
    minimum size=1.2cm,
    inner sep=0pt
  }
  [, circle, fill
    [$t'$, triangle]
  ]
\end{forest}
\end{center}

to the tree

\begin{center}
\begin{forest}
  for tree={
    circle, draw,
    minimum size=6pt,
    inner sep=0pt,
    /tikz/every label/.style={font=\tiny},
    calign=center,
    parent anchor=south,
    child anchor=north,
    l sep=0pt,
  },
  left_triangle/.style={
    isosceles triangle,
    draw,
    shape border rotate=120,
    minimum size=1.2cm,
    inner sep=0pt,
    anchor=apex
  },
  right_triangle/.style={
    isosceles triangle,
    draw,
    shape border rotate=60,
    minimum size=1.2cm,
    inner sep=0pt,
    anchor=apex
  }
  [, circle, fill,label=south:$\parallel$
    [$t$, left_triangle] 
    [$t'$, right_triangle]
  ]
\end{forest}
\end{center}

The function $\mid_{\mathbf{fPAT}}$ maps the trees

\begin{center}
\begin{forest}
  for tree={
    circle, draw,
    minimum size=6pt,
    inner sep=0pt,
    /tikz/every label/.style={font=\tiny},
    calign=center,
    parent anchor=north,
    child anchor=south,
    l sep=0pt,
    no edge
  },
  triangle/.style={
    isosceles triangle,
    draw,
    shape border rotate=90,
    minimum size=1.2cm,
    inner sep=0pt
  }
  [, circle, fill
    [ $t$, triangle ]
  ]
\end{forest}
\end{center}

\begin{center}
\begin{forest}
  for tree={
    circle, draw,
    minimum size=6pt,
    inner sep=0pt,
    /tikz/every label/.style={font=\tiny},
    calign=center,
    parent anchor=north,
    child anchor=south,
    l sep=0pt,
    no edge
  },
  triangle/.style={
    isosceles triangle,
    draw,
    shape border rotate=90,
    minimum size=1.2cm,
    inner sep=0pt
  }
  [, circle, fill
    [$t'$, triangle]
  ]
\end{forest}
\end{center}

to the tree

\begin{center}
\begin{forest}
  for tree={
    circle, draw,
    minimum size=6pt,
    inner sep=0pt,
    /tikz/every label/.style={font=\tiny},
    calign=center,
    parent anchor=south,
    child anchor=north,
    l sep=0pt,
  },
  left_triangle/.style={
    isosceles triangle,
    draw,
    shape border rotate=120,
    minimum size=1.2cm,
    inner sep=0pt,
    anchor=apex
  },
  right_triangle/.style={
    isosceles triangle,
    draw,
    shape border rotate=60,
    minimum size=1.2cm,
    inner sep=0pt,
    anchor=apex
  }
  [, circle, fill,label=south:$\mid$
    [$t$, left_triangle] 
    [$t'$, right_triangle]
  ]
\end{forest}
\end{center}

The function $\between_{\mathbf{fPAT}}$ maps the trees

\begin{center}
\begin{forest}
  for tree={
    circle, draw,
    minimum size=6pt,
    inner sep=0pt,
    /tikz/every label/.style={font=\tiny},
    calign=center,
    parent anchor=north,
    child anchor=south,
    l sep=0pt,
    no edge
  },
  triangle/.style={
    isosceles triangle,
    draw,
    shape border rotate=90,
    minimum size=1.2cm,
    inner sep=0pt
  }
  [, circle, fill
    [ $t$, triangle ]
  ]
\end{forest}
\end{center}

\begin{center}
\begin{forest}
  for tree={
    circle, draw,
    minimum size=6pt,
    inner sep=0pt,
    /tikz/every label/.style={font=\tiny},
    calign=center,
    parent anchor=north,
    child anchor=south,
    l sep=0pt,
    no edge
  },
  triangle/.style={
    isosceles triangle,
    draw,
    shape border rotate=90,
    minimum size=1.2cm,
    inner sep=0pt
  }
  [, circle, fill
    [$t'$, triangle]
  ]
\end{forest}
\end{center}

to the tree

\begin{center}
\begin{forest}
  for tree={
    circle, draw,
    minimum size=6pt,
    inner sep=0pt,
    /tikz/every label/.style={font=\tiny},
    calign=center,
    parent anchor=south,
    child anchor=north,
    l sep=0pt,
  },
  left_triangle/.style={
    isosceles triangle,
    draw,
    shape border rotate=120,
    minimum size=1.2cm,
    inner sep=0pt,
    anchor=apex
  },
  right_triangle/.style={
    isosceles triangle,
    draw,
    shape border rotate=60,
    minimum size=1.2cm,
    inner sep=0pt,
    anchor=apex
  }
  [, circle, fill,label=south:$\between$
    [$t$, left_triangle] 
    [$t'$, right_triangle]
  ]
\end{forest}
\end{center}

Firstly, we define the parallelism of sets. In this book, every element of a set is a pomset $U\in\mathsf{Pom}$, for two such sets $B_1,B_2$, their parallelism is defined as: $$B_1\parallel B_2=\{U_1\parallel U_2|U_1\in B_1,U_2\in B_2\}$$ where $U_1\parallel U_2$ is the parallel composition of pomsets defined in \cref{multiset}. Similarly, we can define the communication of sets and the concurrency of sets: $B_1\mid B_2$ and $B_1\between B_2$, respectively. 

If $\mathscr{B}_1$, $\mathscr{B}_2$ are set of sets, their pointwise parallelism is defined as: $$\mathscr{B}_1\ucup_{\parallel}\mathscr{B}_2=\{B_1\parallel B_2|B_1\in\mathscr{B}_1,B_2\in\mathscr{B}_2\}$$

Similarly, we can define their pointwise communication and pointwise concurrency as: $\mathscr{B}_1\ucup_{\mid}\mathscr{B}_2$ and $\mathscr{B}_1\ucup_{\between}\mathscr{B}_2$.

Then the operators $\ucup_{\parallel}$, $\ucup_{\mid}$ and $\ucup_{\between}$ have the following properties.

dp1. $(\mathscr{B}_1\ucup_{\parallel} \mathscr{B}_2)\ucup_{\parallel} \mathscr{B}_3=\mathscr{B}_1\ucup_{\parallel}(\mathscr{B}_2\ucup_{\parallel} \mathscr{B}_3)$.

dp2. $\mathscr{B}_1\ucup_{\parallel} \mathscr{B}_2=\mathscr{B}_2\ucup_{\parallel} \mathscr{B}_1$.

dp3. $\mathscr{B}_1\ucup_{\parallel} (\mathscr{B}_2\cup \mathscr{B}_3)=(\mathscr{B}_1\ucup_{\parallel} \mathscr{B}_2)\cup(\mathscr{B}_1\ucup_{\parallel} \mathscr{B}_3)$.

dp4. If $\mathscr{B}_1,\mathscr{B}_2$ are saturated, $\mathscr{B}_1\ucup_{\parallel} \mathscr{B}_2$ is saturated.

dp5. $c(\mathscr{B}_1\ucup_{\parallel} \mathscr{B}_2)=c(\mathscr{B}_1)\ucup_{\parallel} c(\mathscr{B}_2)$.

dp6. $(\mathscr{B}_1\ucup_{\mid} \mathscr{B}_2)\ucup_{\mid} \mathscr{B}_3=\mathscr{B}_1\ucup_{\mid}(\mathscr{B}_2\ucup_{\mid} \mathscr{B}_3)$.

dp7. $\mathscr{B}_1\ucup_{\mid} \mathscr{B}_2=\mathscr{B}_2\ucup_{\mid} \mathscr{B}_1$.

dp8. $\mathscr{B}_1\ucup_{\mid} (\mathscr{B}_2\cup \mathscr{B}_3)=(\mathscr{B}_1\ucup_{\mid} \mathscr{B}_2)\cup(\mathscr{B}_1\ucup_{\mid} \mathscr{B}_3)$.

dp9. If $\mathscr{B}_1,\mathscr{B}_2$ are saturated, $\mathscr{B}_1\ucup_{\mid} \mathscr{B}_2$ is saturated.

dp10. $c(\mathscr{B}_1\ucup_{\mid} \mathscr{B}_2)=c(\mathscr{B}_1)\ucup_{\mid} c(\mathscr{B}_2)$.

dp11. $(\mathscr{B}_1\ucup_{\between} \mathscr{B}_2)\ucup_{\between} \mathscr{B}_3=\mathscr{B}_1\ucup_{\between}(\mathscr{B}_2\ucup_{\between} \mathscr{B}_3)$.

dp12. $\mathscr{B}_1\ucup_{\between} \mathscr{B}_2=\mathscr{B}_2\ucup_{\between} \mathscr{B}_1$.

dp13. $\mathscr{B}_1\ucup_{\between} (\mathscr{B}_2\cup \mathscr{B}_3)=(\mathscr{B}_1\ucup_{\between} \mathscr{B}_2)\cup(\mathscr{B}_1\ucup_{\between} \mathscr{B}_3)$.

dp14. If $\mathscr{B}_1,\mathscr{B}_2$ are saturated, $\mathscr{B}_1\ucup_{\between} \mathscr{B}_2$ is saturated.

dp15. $c(\mathscr{B}_1\ucup_{\between} \mathscr{B}_2)=c(\mathscr{B}_1)\ucup_{\between} c(\mathscr{B}_2)$.

Then, formally, the tree $t\parallel_{\mathbf{fPAT}}t'$ is the tree $t''$ determined by:

\begin{enumerate}
  \item $L(t'')=L(t)\parallel L(t')$.
  \item $\mathscr{A}(t'')=\mathscr{A}(t)\ucup_{\parallel}\mathscr{A}(t')$.
  \item $\mathscr{A}(t''(s))=c(\mathscr{A}(t(s))\parallel\mathscr{A}(t'(s)))$ where $s\neq 1$, and if $s\notin L(t)$ then $\mathscr{A}(t(s))=\emptyset$.
\end{enumerate}

The tree $t\mid_{\mathbf{fPAT}}t'$ is the tree $t''$ determined by:

\begin{enumerate}
  \item $L(t'')=L(t)\mid L(t')$.
  \item $\mathscr{A}(t'')=\mathscr{A}(t)\ucup_{\mid}\mathscr{A}(t')$.
  \item $\mathscr{A}(t''(s))=c(\mathscr{A}(t(s))\mid\mathscr{A}(t'(s)))$ where $s\neq 1$, and if $s\notin L(t)$ then $\mathscr{A}(t(s))=\emptyset$.
\end{enumerate}

The tree $t\between_{\mathbf{fPAT}}t'$ is the tree $t''$ determined by:

\begin{enumerate}
  \item $L(t'')=L(t)\between L(t')$.
  \item $\mathscr{A}(t'')=\mathscr{A}(t)\ucup_{\between}\mathscr{A}(t')$.
  \item $\mathscr{A}(t''(s))=c(\mathscr{A}(t(s))\between\mathscr{A}(t'(s)))$ where $s\neq 1$, and if $s\notin L(t)$ then $\mathscr{A}(t(s))=\emptyset$.
\end{enumerate}

According to the definitions of parallel composition, communication composition and concurrent composition of pomsets in \cref{multiset}, the tree 

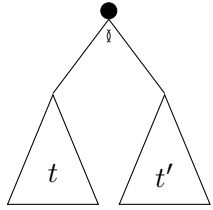
\begin{center}
\begin{forest}
  for tree={
    circle, draw,
    minimum size=6pt,
    inner sep=0pt,
    /tikz/every label/.style={font=\tiny},
    calign=center,
    parent anchor=south,
    child anchor=north,
    l sep=0pt,
  },
  left_triangle/.style={
    isosceles triangle,
    draw,
    shape border rotate=120,
    minimum size=1.2cm,
    inner sep=0pt,
    anchor=apex
  },
  right_triangle/.style={
    isosceles triangle,
    draw,
    shape border rotate=60,
    minimum size=1.2cm,
    inner sep=0pt,
    anchor=apex
  }
  [, circle, fill,label=south:$\between$
    [$t$, left_triangle] 
    [$t'$, right_triangle]
  ]
\end{forest}
\end{center}

is equivalent to the tree

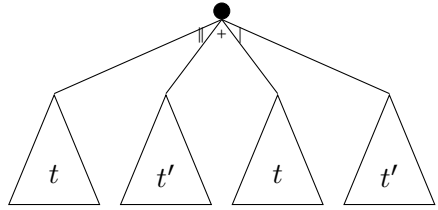
\begin{center}
\begin{forest}
  for tree={
    circle, draw,
    minimum size=6pt,
    inner sep=0pt,
    /tikz/every label/.style={font=\tiny},
    calign=center,
    parent anchor=south,
    child anchor=north,
    l sep=0pt,
  },
  left_triangle/.style={
    isosceles triangle,
    draw,
    shape border rotate=120,
    minimum size=1.2cm,
    inner sep=0pt,
    anchor=apex
  },
  right_triangle/.style={
    isosceles triangle,
    draw,
    shape border rotate=60,
    minimum size=1.2cm,
    inner sep=0pt,
    anchor=apex
  }
  [, circle, fill,label=south:$+$, label=south west:$\parallel$, label=south east:$\mid$
    [$t$, left_triangle] 
    [$t'$, right_triangle]
    [$t$, left_triangle] 
    [$t'$, right_triangle]
  ]
\end{forest}
\end{center}

Formally the tree $t\between_{\mathbf{fPAT}}t'$ is the tree $t''$ determined by:

\begin{enumerate}
  \item $L(t'')=L(t)\between L(t')=(L(t)\parallel L(t'))\cup (L(t)\mid L(t'))$.  
  \item $\mathscr{A}(t'')=\mathscr{A}(t)\ucup_{\between}\mathscr{A}(t')=(\mathscr{A}(t)\ucup_{\parallel}\mathscr{A}(t'))\ucup (\mathscr{A}(t)\ucup_{\mid}\mathscr{A}(t'))$.
  \item $\mathscr{A}(t''(s))=c(\mathscr{A}(t(s))\between\mathscr{A}(t'(s)))=(c(\mathscr{A}(t(s))\parallel\mathscr{A}(t'(s))))\cup (c(\mathscr{A}(t(s))\mid\mathscr{A}(t'(s))))$ where $s\neq 1$, and if $s\notin L(t)$ then $\mathscr{A}(t(s))=\emptyset$.
\end{enumerate}

For $a\in\mathsf{Act}(t)$ and $b\in\mathsf{Act}(t')$, $a,b$ in the tree $t\mid_{\mathbf{fPAT}}t'$ are merged into $\gamma(a,b)$ if $a\leq^c b$; into $0$, otherwise.

For $H\subseteq\mathsf{Act}(t\parallel_{\mathbf{fPAT}}t')$, $\partial_H(t\parallel_{\mathbf{fPAT}}t')$ renames every $a\in H$ into $0$; and remains themselves unchanged for all $a\notin H$ and $a\in\mathsf{Act}(t\parallel_{\mathbf{fPAT}}t')$.

Note that, in a tree $t\in\mathbf{fPAT}$, arbitrary $1$'s can be padded in the alternative branches and parallel branches, but, we leave the padding of $1$'s as the privilege of the modelling phase to explicitly present the existences of $1$'s.

(8) $\sharp_{\mathbf{fPAT}}$, $\Theta(p)_{\mathbf{fPAT}}$, $\triangleleft_{\mathbf{fPAT}}$ and $\osharp_{\mathbf{fPAT}}$.

For $a\in\mathsf{Act}(t)$ and $b\in\mathsf{Act}(t')$, there exists a confliction relation $a\sharp b$ in the tree $t\parallel_{\mathbf{fPAT}}t'$ (resp. $t\mid_{\mathbf{fPAT}}t'$), $\Theta(t\parallel_{\mathbf{fPAT}}t')_{\mathbf{fPAT}}$ (resp. $\Theta(t\mid_{\mathbf{fPAT}}t')_{\mathbf{fPAT}}$) and $\triangleleft_{\mathbf{fPAT}}$ rename $\{s'|t(1)\xrightarrow{s}\xrightarrow{a}t(s')\}$ to $1$'s and remain $t'$ unchanged, or rename $\{s'|t'(1)\xrightarrow{s}\xrightarrow{b}t'(s')\}$ to $1$'s and remain $t$ unchanged. This is reflected that the tree

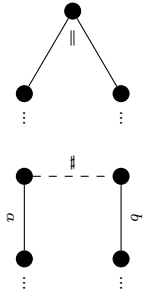
\begin{center}
\begin{forest}
for tree={
  circle, fill, draw,
  minimum size=6pt,
  inner sep=0pt,
  l sep=20pt,
  s sep=30pt,
  /tikz/every label/.style={font=\tiny},
}
[,name=R,label=south:$\parallel$
  [,name=C1,label=south:$\vdots$
    [,name=C11, no edge
      [,name=C111,label=south:$\vdots$,
      edge label={node[midway,sloped,below,font=\tiny]{$a$}}]
    ]
  ]
  [,name=C2,label=south:$\vdots$
    [,name=C21,no edge
      [,name=C211,label=south:$\vdots$,
      edge label={node[midway,sloped,above,font=\tiny]{$b$}}]
    ]
  ]
]
\draw[dashed] (C11) to node[midway,sloped,above,font=\tiny]{$\sharp$} (C21);
\end{forest}
\end{center}

is equivalent to the tree

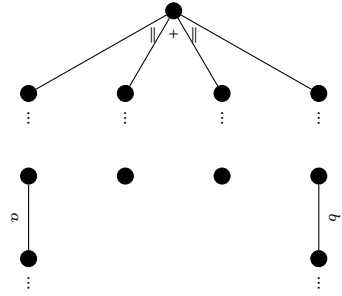
\begin{center}
\begin{forest}
for tree={
  circle, fill, draw,
  minimum size=6pt,
  inner sep=0pt,
  l sep=20pt,
  s sep=30pt,
  /tikz/every label/.style={font=\tiny},
}
[,name=R,label=south:$+$, label=south east:$\parallel$,label=south west:$\parallel$
  [,name=C1,label=south:$\vdots$
    [,name=C11, no edge
      [,name=C111,label=south:$\vdots$,
      edge label={node[midway,sloped,below,font=\tiny]{$a$}}]
    ]
  ]
  [,name=C2,label=south:$\vdots$
    [,name=C21,no edge]
  ]
  [,name=C3,label=south:$\vdots$
    [,name=C31, no edge]
  ]
  [,name=C4,label=south:$\vdots$
    [,name=C41,no edge
      [,name=C411,label=south:$\vdots$,
      edge label={node[midway,sloped,above,font=\tiny]{$b$}}]
    ]
  ]
]
\end{forest}
\end{center}

For $a\in\mathsf{Act}(t)$ and $b\in\mathsf{Act}(t')$, there exists an internal confliction relation $a\osharp b$ in the tree $t\parallel_{\mathbf{fPAT}}t'$ (resp. $t\mid_{\mathbf{fPAT}}t'$), $\Theta(t\parallel_{\mathbf{fPAT}}t')_{\mathbf{fPAT}}$ (resp. $\Theta(t\mid_{\mathbf{fPAT}}t')_{\mathbf{fPAT}}$) and $\triangleleft_{\mathbf{fPAT}}$ rename $\{s'|t(1)\xrightarrow{s}\xrightarrow{a}t(s')\}$ to $1$'s and remain $t'$ unchanged, or rename $\{s'|t'(1)\xrightarrow{s}\xrightarrow{b}t'(s')\}$ to $1$'s and remain $t$ unchanged. This is reflected that the tree

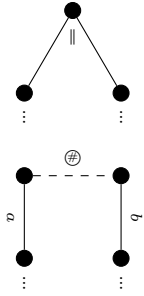
\begin{center}
\begin{forest}
for tree={
  circle, fill, draw,
  minimum size=6pt,
  inner sep=0pt,
  l sep=20pt,
  s sep=30pt,
  /tikz/every label/.style={font=\tiny},
}
[,name=R,label=south:$\parallel$
  [,name=C1,label=south:$\vdots$
    [,name=C11, no edge
      [,name=C111,label=south:$\vdots$,
      edge label={node[midway,sloped,below,font=\tiny]{$a$}}]
    ]
  ]
  [,name=C2,label=south:$\vdots$
    [,name=C21,no edge
      [,name=C211,label=south:$\vdots$,
      edge label={node[midway,sloped,above,font=\tiny]{$b$}}]
    ]
  ]
]
\draw[dashed] (C11) to node[midway,sloped,above,font=\tiny]{$\osharp$} (C21);
\end{forest}
\end{center}

is equivalent to the tree

\begin{center}
\begin{forest}
for tree={
  circle, fill, draw,
  minimum size=6pt,
  inner sep=0pt,
  l sep=20pt,
  s sep=30pt,
  /tikz/every label/.style={font=\tiny},
}
[,name=R,label=south:$\oplus$, label=south east:$\parallel$,label=south west:$\parallel$
  [,name=C1,label=south:$\vdots$
    [,name=C11, no edge
      [,name=C111,label=south:$\vdots$,
      edge label={node[midway,sloped,below,font=\tiny]{$a$}}]
    ]
  ]
  [,name=C2,label=south:$\vdots$
    [,name=C21,no edge]
  ]
  [,name=C3,label=south:$\vdots$
    [,name=C31, no edge]
  ]
  [,name=C4,label=south:$\vdots$
    [,name=C41,no edge
      [,name=C411,label=south:$\vdots$,
      edge label={node[midway,sloped,above,font=\tiny]{$b$}}]
    ]
  ]
]
\end{forest}
\end{center}

\begin{proposition}
$\langle \mathbf{fPAT},\leq_{\mathbf{fPAT}},\Sigma^1_{\mathbf{fPAT}}\rangle$ is a $\Sigma$-po algebra.
\end{proposition}

\begin{lemma}
For every $p$ in $\mathbf{M}_1$,

\begin{enumerate}
  \item $L(p)=L(\mathbf{fPAT}\sembrack{p})$.
  \item For every $s\in L(p)$, it holds that $c(\mathscr{A}(p,s))=\mathscr{A}(\mathbf{fPAT}\sembrack{p}(s))$.
\end{enumerate}
\end{lemma} 

\begin{lemma}
If $A(\mathscr{B})=A(\mathscr{A})$ then $\mathscr{B}\subset\subset\mathscr{A}$ if and only if $c(\mathscr{B})\subseteq c(\mathscr{A})$.
\end{lemma}

\begin{theorem}[Full Abstraction for $\mathbf{fPAT}$]
If $p,q\in\mathbf{M}_1$, then $p\pretesting q$ if and only if $\mathbf{fPAT}\sembrack{p}\leq_{\mathbf{fPAT}}\mathbf{fPAT}\sembrack{q}$.
\end{theorem}

\subsection{Axiomatic Semantics}\label{as5}

The proof system $\mathbf{DED}(E^1)$ of inequations of $\mathbf{M}_1$ is shown in \cref{ProofSystemE1ofM1}.

\begin{center}
    \begin{longtable}{|c|c|}
      \caption{Proof system $\mathbf{DED}(E^1)$ of $\mathbf{M}_1$}\\
      \hline No. &Axiom\\
      \hline
      \endfirsthead
      \multicolumn{2}{c}{\bfseries Continuing: Proof system $\mathbf{DED}(E^1)$ of $\mathbf{M}_1$} \\
      \hline No. &Axiom\\
      \hline
      \endhead
      \hline
      \multicolumn{2}{r}{to be continued\ldots} \\
      \endfoot
      \hline
      \multicolumn{2}{r}{end of Proof system $\mathbf{DED}(E^1)$ of $\mathbf{M}_1$} \\
      \endlastfoot
      $A1$ & $x+ y = y+ x$\\
      $A2$ & $(x+ y)+ z = x+ (y+ z)$\\
      $A3$ & $x+ x = x$\\
      $A4$ & $x+ 0 = x$\\
      $A5$ & $x\oplus y = y\oplus x$\\
      $A6$ & $(x\oplus y)\oplus z = x\oplus (y\oplus z)$\\
      $A7$ & $x\oplus x = x$\\
      $A8$ & $x\oplus y\leq x+y$\\
      $A9$ & $x+(y\oplus z)=(x+y)\oplus(x+z)$\\
      $A10$ & $x\oplus(y+z)=(x\oplus y)+(x\oplus z)$\\
      $A11$ & $(x+y)\cdot z=x\cdot z+y\cdot z$\\
      $A12$ & $(x\oplus y)\cdot z=x\cdot z\oplus y\cdot z$\\
      $A13$ & $U\cdot x+U\cdot y=U\cdot x\oplus U\cdot y$\\
      $A14$ & $U\cdot x\oplus U\cdot y=U\cdot(x\oplus y)$\\
      $A15$ & $(x\cdot y)\cdot z=x\cdot(y\cdot z)$\\
      $A16$ & $0\cdot x =0$\\
      $A17$ & $x\cdot 1=x$\\
      $A18$ & $1\cdot x=x$\\
      $P1$ & $x\between y = x\parallel y + x\mid y$\\
      $P2$ & $a\parallel (b\cdot y) = (a\parallel b)\cdot y$\\
      $P3$ & $(a\cdot x)\parallel b = (a\parallel b)\cdot x$\\
      $P4$ & $(a\cdot x)\parallel (b\cdot y) = (a\parallel b)\cdot(x\between y)$\\
      $P5$ & $(x+ y)\parallel z = (x\parallel z)+ (y\parallel z)$\\
      $P6$ & $x\parallel (y+ z) = (x\parallel y)+ (x\parallel z)$\\
      $P7$ & $(x\oplus y)\parallel z = (x\parallel z)\oplus (y\parallel z)$\\
      $P8$ & $x\parallel (y\oplus z) = (x\parallel y)\oplus (x\parallel z)$\\
      $P9$ & $0\parallel x = 0$\\
      $P10$ & $x\parallel 0 = 0$\\
      $P11$ & $x\parallel 1=x$\\
      $P12$ & $1\parallel x=x$\\
      $C1$ & $a\mid b = \gamma(a,b)$\\
      $C2$ & $a\mid (b\cdot y) = \gamma(a,b)\cdot y$\\
      $C3$ & $(a\cdot x)\mid b = \gamma(a,b)\cdot x$\\
      $C4$ & $(a\cdot x)\mid (b\cdot y) = \gamma(a,b)\cdot (x\between y)$\\
      $C5$ & $(x+ y)\mid z = (x\mid z) + (y\mid z)$\\
      $C6$ & $x\mid (y+ z) = (x\mid y)+ (x\mid z)$\\
      $C7$ & $(x\oplus y)\mid z = (x\mid z) \oplus (y\mid z)$\\
      $C8$ & $x\mid (y\oplus z) = (x\mid y)\oplus (x\mid z)$\\
      $C9$ & $0\mid x = 0$\\
      $C10$ & $x\mid 0 = 0$\\
      $C11$ & $1\mid x = 0$\\
      $C12$ & $x\mid 1 = 0$\\
      $CE1$ & $\Theta(a) = a$\\
      $CE2$ & $\Theta(U) = U$\\
      $CE3$ & $\Theta(0) = 0$\\
      $CE4$ & $\Theta(1) = 1$\\
      $CE5$ & $\Theta(x+ y) = \Theta(x) + \Theta(y)$\\
      $CE6$ & $\Theta(x\oplus y) = \Theta(x) \oplus \Theta(y)$\\
      $CE7$ & $\Theta(x\cdot y)=\Theta(x)\cdot\Theta(y)$\\
      $CE8$ & $(\exists a\in \mathsf{Act}(x),b\in\mathsf{Act(y)},a\sharp b)\quad\Theta(x\parallel y) = ((\Theta(x)\triangleleft y)\parallel y)+ ((\Theta(y)\triangleleft x)\parallel x)$\\
      $CE9$ & $(\exists a\in \mathsf{Act}(x),b\in\mathsf{Act(y)},a\sharp b)\quad\Theta(x\mid y) = ((\Theta(x)\triangleleft y)\mid y)+ ((\Theta(y)\triangleleft x)\mid x)$\\
      $CE10$ & $(\exists a\in \mathsf{Act}(x),b\in\mathsf{Act(y)},a\osharp b)\quad\Theta(x\parallel y) = ((\Theta(x)\triangleleft y)\parallel y)\oplus ((\Theta(y)\triangleleft x)\parallel x)$\\
      $CE11$ & $(\exists a\in \mathsf{Act}(x),b\in\mathsf{Act(y)},a\osharp b)\quad\Theta(x\mid y) = ((\Theta(x)\triangleleft y)\mid y)\oplus ((\Theta(y)\triangleleft x)\mid x)$\\
      $U1$ & $(a\sharp b,c\leq a)\quad c\triangleleft b = c$\\
      $U2$ & $(a\sharp b)\quad a\triangleleft b = 1$\\
      $U3$ & $(a\sharp b,b\leq c)\quad a\triangleleft c = 1$\\
      $U4$ & $(a\sharp b,b\leq c)\quad c\triangleleft a = 1$\\
      $U5$ & $(a\osharp b,c\leq a)\quad c\triangleleft b = c$\\
      $U6$ & $(a\osharp b)\quad a\triangleleft b = 1$\\
      $U7$ & $(a\osharp b,b\leq c)\quad a\triangleleft c = 1$\\
      $U8$ & $(a\osharp b,b\leq c)\quad c\triangleleft a = 1$\\
      $U9$ & $a\triangleleft 0 = a$\\
      $U10$ & $0 \triangleleft a = 0$\\
      $U11$ & $a\triangleleft 1 = a$\\
      $U12$ & $1\triangleleft a = 1$\\
      $U13$ & $(x+ y)\triangleleft z = (x\triangleleft z)+ (y\triangleleft z)$\\
      $U14$ & $(x\oplus y)\triangleleft z = (x\triangleleft z)\oplus (y\triangleleft z)$\\
      $U15$ & $(x\cdot y)\triangleleft y = (x\triangleleft y)\cdot(y\triangleleft y)$\\
      $U16$ & $(x\parallel y)\triangleleft z = (x\triangleleft z)\parallel (y\triangleleft z)$\\
      $U17$ & $(x\mid y)\triangleleft z = (x\triangleleft z)\mid (y\triangleleft z)$\\
      $U18$ & $x\triangleleft (y+ z) = (x\triangleleft y)\triangleleft z$\\
      $U19$ & $x\triangleleft (y\oplus z) = (x\triangleleft y)\triangleleft z$\\
      $U20$ & $x\triangleleft (y\cdot z)=(x\triangleleft y)\triangleleft z$\\
      $U21$ & $x\triangleleft (y\parallel z) = (x\triangleleft y)\triangleleft z$\\
      $U22$ & $x\triangleleft (y\mid z) = (x\triangleleft y)\triangleleft z$\\
      $D1$ & $(a\notin H)\quad\partial_H(a) = a$\\
      $D2$ & $(a\in H)\quad \partial_H(a) = 0$\\
      $D3$ & $\partial_H(0) = 0$\\
      $D4$ & $\partial_H(1) = 1$\\
      $D5$ & $\partial_H(x+ y) = \partial_H(x)+\partial_H(y)$\\
      $D6$ & $\partial_H(x\oplus y) = \partial_H(x)\oplus\partial_H(y)$\\
      $D7$ & $\partial_H(x\cdot y) = \partial_H(x)\cdot\partial_H(y)$\\
      $D8$ & $\partial_H(x\parallel y) = \partial_H(x)\parallel\partial_H(y)$
      \label{ProofSystemE1ofM1}
    \end{longtable}
\end{center}

\begin{proposition}
$\mathbf{fPAT}$ is in $\mathscr{C}(E^1)$.
\end{proposition}

\begin{lemma}
The mapping $i_{\mathbf{fPAT}}$ from $\mathbf{M}_1$ to $\mathbf{fPAT}$ is surjective.
\end{lemma}

If $P$ is a finite set of terms $p_1,\cdots,p_k$, let $\sum\{p|p\in P\}$ denote the term $$p_1+p_2+\cdots+p_k$$
If $P$ is a nonempty finite set of terms $p_1,\cdots,p_k$, let $\osum\{p|p\in P\}$ denote the term $$p_1\oplus p_2\oplus \cdots\oplus p_k$$

\begin{definition}[Normal form]
The normal form is defined inductively as follows:

\begin{enumerate}
  \item $0$ is a normal form.
  \item If $\mathscr{A}$ is saturated set and for every $U$ in $A(\mathscr{A})$ there is a normal form $n(U)$, then $\osum\{n(A)|A\in\mathscr{A}\}$ is a normal form, where $n(A)$ represents $\sum\{Un(U)|U\in A\}$.
\end{enumerate}
\end{definition}

\begin{lemma}
If $n,m$ are normal forms, then $\mathbf{fPAT}\sembrack{n}\leq_{\mathbf{fPAT}}\mathbf{fPAT}\sembrack{m}$ implies $n\leq_{E^1}m$.
\end{lemma}

\begin{theorem}[Normal form theorem]
For every term $p$ in $\mathbf{M}_1$, there exists a normal form $nf(p)$ such that $p=_{E^1}nf(p)$.
\end{theorem}

\begin{theorem}[Completeness]
For $p,q\in\mathbf{M}_1$, $\mathbf{fPAT}\sembrack{p}\leq_{\mathbf{fPAT}}\mathbf{fPAT}\sembrack{q}$ implies $p\leq_{E^1}q$.
\end{theorem}

\begin{theorem}[Initiality of $\mathbf{fPAT}$]
$\mathbf{fPAT}$ is initial in $\mathscr{C}(E^1)$.
\end{theorem}

\subsection{The Trinity} \label{t5} 

\begin{definition}
A partial order $\leq^{\mathrm{MUST}}_{\mathbf{fPAT}}$ is defined over $\mathbf{fPAT}$ as follows, for two trees $t,t'\in\mathbf{fPAT}$, $t\leq^{\mathrm{MUST}}_{\mathbf{fPAT}}t'$, if:

For every $s\in L(t')$, $\mathscr{A}(t'(s))\subseteq\mathscr{A}(t(s))$.
\end{definition}

\begin{lemma}
Let $\mathbf{fPAT}_{\mathbf{S}}$ denote $\langle \mathbf{fPAT}, \leq^{\mathrm{MUST}}_{\mathbf{fPAT}},\Sigma^1_{\mathbf{fPAT}}\rangle$, $\mathbf{fPAT}_{\mathbf{S}}$ is a $\Sigma^1$-po algebra.
\end{lemma}

\begin{theorem}[Full Abstraction for $\mathbf{fPAT}_{\mathbf{S}}$]
If $p,q\in\mathbf{M}_1$, then $p\pretestingmust q$ if and only if $\mathbf{fPAT}\sembrack{p}\leq^{\mathrm{MUST}}_{\mathbf{fPAT}}\mathbf{fPAT}\sembrack{q}$.
\end{theorem} 

By adding the following axiom $$A19\quad x\oplus y\leq x$$ into the proof system $\mathbf{DED}(E^1)$ in \cref{ProofSystemE1ofM1} to get the proof system $\mathbf{DED}(E^1_{\mathbf{S}})$.

\begin{theorem}[Initiality of $\mathbf{fPAT}_{\mathbf{S}}$]
$\mathbf{fPAT}_{\mathbf{S}}$ is initial in $\mathscr{C}(E^1_{\mathbf{S}})$.
\end{theorem}

\begin{definition}
Let $\mathbf{fPAT}_{\mathbf{W}}$ denote $\langle D, \leq^{\mathrm{MAY}}_{\mathbf{fPAT}_{\mathbf{W}}},\Sigma^1_{\mathbf{fPAT}_{\mathbf{W}}}\rangle$, where 

\begin{enumerate}
  \item $D$ is the set of deterministic trees in $\mathbf{fPAT}$, i.e., those trees all of whose nodes $n$ are labelled by the acceptance set $\{\{S(n)\}\}$.
  \item $t\leq^{\mathrm{MAY}}_{\mathbf{fPAT}_{\mathbf{W}}}t'$, if $L(t)\subseteq L(t')$.
  \item The various operators are defined by
  \begin{itemize}
    \item $0_{\mathbf{fPAT}_{\mathbf{W}}}$ coincides with $0_{\mathbf{fPAT}}$, $1_{\mathbf{fPAT}_{\mathbf{W}}}$ coincides with $1_{\mathbf{fPAT}}$, $U_{\mathbf{fPAT}_{\mathbf{W}}}$ coincides with $U_{\mathbf{fPAT}}$, $\cdot_{\mathbf{fPAT}_{\mathbf{W}}}$ coincides with $\cdot_{\mathbf{fPAT}}$, $\parallel_{\mathbf{fPAT}_{\mathbf{W}}}$ coincides with $\parallel_{\mathbf{fPAT}}$, $\gamma(a,b)_{\mathbf{fPAT}_{\mathbf{W}}}$ coincides with $\gamma(a,b)_{\mathbf{fPAT}}$, $\mid_{\mathbf{fPAT}_{\mathbf{W}}}$ coincides with $\mid_{\mathbf{fPAT}}$, $\between_{\mathbf{fPAT}_{\mathbf{W}}}$ coincides with $\between_{\mathbf{fPAT}}$, $\partial_H(p)_{\mathbf{fPAT}_{\mathbf{W}}}$ coincides with $\partial_H(p)_{\mathbf{fPAT}}$, $\sharp_{\mathbf{fPAT}_{\mathbf{W}}}$ coincides with $\sharp_{\mathbf{fPAT}}$, $\Theta(p)_{\mathbf{fPAT}_{\mathbf{W}}}$ coincides with $\Theta(p)_{\mathbf{fPAT}}$, $\triangleleft_{\mathbf{fPAT}_{\mathbf{W}}}$ coincides with $\triangleleft_{\mathbf{fPAT}}$, $\osharp_{\mathbf{fPAT}_{\mathbf{W}}}$ coincides with $\osharp_{\mathbf{fPAT}}$.
    \item $t+_{\mathbf{fPAT}_{\mathbf{W}}}t'$ is the tree $t''$ determined by:
    \begin{enumerate}
        \item $L(t'')=L(t)\cup L(t')$.
        \item $\mathscr{A}(t''(s))=c(\mathscr{A}(t(s))\ucup\mathscr{A}(t'(s)))$, and if $s\notin L(t)$ then $\mathscr{A}(t(s))=\emptyset$.
    \end{enumerate}
    \item $\oplus_{\mathbf{fPAT}_{\mathbf{W}}}$ coincides with $+_{\mathbf{fPAT}_{\mathbf{W}}}$.
  \end{itemize}
\end{enumerate}
\end{definition}

\begin{lemma}
$\mathbf{fPAT}_{\mathbf{W}}$ is a $\Sigma^1$-po algebra.
\end{lemma}

\begin{theorem}[Full Abstraction for $\mathbf{fPAT}_{\mathbf{W}}$]
If $p,q\in\mathbf{M}_1$, then $p\pretestingmay q$ if and only if $\mathbf{fPAT}_{\mathbf{W}}\sembrack{p}\leq^{\mathrm{MAY}}_{\mathbf{fPAT}_{\mathbf{W}}}\mathbf{fPAT}_{\mathbf{W}}\sembrack{q}$.
\end{theorem} 

By adding the following axiom $$A20\quad x\leq x\oplus y$$ into the proof system $\mathbf{DED}(E^1)$ in \cref{ProofSystemE1ofM1} to get the proof system $\mathbf{DED}(E^1_{\mathbf{W}})$.

\begin{theorem}[Initiality of $\mathbf{fPAT}_{\mathbf{W}}$]
$\mathbf{fPAT}_{\mathbf{W}}$ is initial in $\mathscr{C}(E^1_{\mathbf{W}})$.
\end{theorem} 
\newpage\section{$\Omega$}\label{omegaT}

In this chapter, we introduce the basic processes with $\Omega$. Firstly, we introduce the $\Omega$ signature in \cref{s6}, then the operational semantics, denotational semantics and axiomatic semantics of the $\Omega$ processes are introduced in \cref{os6}, \cref{ds6} and \cref{as6}, respectively. Finally, we get the results on trinity of operational semantics, denotational semantics and axiomatic semantics in \cref{t6}.

\subsection{$\Omega$ Signature}\label{s6}

\begin{definition}[$\Omega$ signature]
The $\Omega$ signature $\Sigma^2$ consists of:

\begin{enumerate}
  \item $\Sigma^2\supset \Sigma^1$.
  \item A distinguished constant $\Omega$.
\end{enumerate}
\end{definition}

\begin{definition}[Syntax of $\Omega$ process language]
The syntax of the $\Omega$ process language is given by the following BNF grammar:

$p::=\Omega~|~0~|~1~|~U~|~\gamma(a,b)~|~p\cdot p~|~a\sharp b~|~a\osharp b~|~p+p~|~p\oplus p~|~p\parallel p~|~p\mid p~|~p\between p~|~\Theta(p)~|~p\triangleleft p~|~\partial_H(p)~$

where $a,b\in\mathsf{Act}$, $U\in\mathsf{Pom}$, $p\in \mathsf{Proc}$.
\end{definition}

\subsection{Operational Semantics}\label{os6}

In this section, we give the operational semantics of the language $\mathbf{M}_2$. The predicate $\surd$ represents successful termination, $\xrightarrow{a}\surd$ represents successful termination after execution of the action $a\in\mathsf{Act}$, $\xrightarrow{U}\surd$ represents successful termination after execution of the action $U\in\mathsf{Pom}$ and $\xrightarrow{ }\surd$ represents successful termination without execution of the any action. The following are the PTSS of the language $\mathbf{M}_2$, where $p,q\in\mathsf{Proc}$.

The PTSS of action $1$, $a\in\mathsf{Act}$ and $U\in\mathsf{Pom}$ is as follows. Note that, there is no any transition rules for $0$ and $\Omega$.

$$\frac{}{1\xrightarrow{ }\surd}\quad\frac{}{a\xrightarrow{a}\surd}\quad\frac{}{U\xrightarrow{U}\surd}$$

The PTSS of sequential composition is as follows.

$$\frac{p\xrightarrow{U}\surd}{p\cdot q\xrightarrow{U} q}\quad\frac{p\xrightarrow{U}p'}{p\cdot q\xrightarrow{U} p'\cdot q}$$

The PTSS of alternative composition is as follows.

$$\frac{p\xrightarrow{U}\surd}{p+ q\xrightarrow{U}\surd} \quad\frac{p\xrightarrow{U}p'}{p+ q\xrightarrow{U}p'} \quad\frac{q\xrightarrow{U}\surd}{p+ q\xrightarrow{U}\surd} \quad\frac{q\xrightarrow{U}q'}{p+ q\xrightarrow{U}q'}$$

The PTSS of concurrent composition is as follows.

$$\frac{p\xrightarrow{a}\surd\quad q\xrightarrow{b}\surd}{p\between q\xrightarrow{\step{a,b}}\surd} \quad\frac{p\xrightarrow{a}p'\quad q\xrightarrow{b}\surd}{p\between q\xrightarrow{\step{a,b}}p'}$$
$$\frac{p\xrightarrow{a}\surd\quad q\xrightarrow{b}q'}{p\between q\xrightarrow{\step{a,b}}q'} \quad\frac{p\xrightarrow{a}p'\quad q\xrightarrow{b}q'}{p\between q\xrightarrow{\step{a,b}}p'\between q'}$$
$$\frac{p\xrightarrow{a}\surd\quad q\xrightarrow{b}\surd}{p\between q\xrightarrow{\gamma(a,b)}\surd} \quad\frac{p\xrightarrow{a}p'\quad q\xrightarrow{b}\surd}{p\between q\xrightarrow{\gamma(a,b)}p'}$$
$$\frac{p\xrightarrow{a}\surd\quad q\xrightarrow{b}q'}{p\between q\xrightarrow{\gamma(a,b)}q'} \quad\frac{p\xrightarrow{a}p'\quad q\xrightarrow{b}q'}{p\between q\xrightarrow{\gamma(a,b)}p'\between q'}$$

The PTSS of parallel composition is as follows.

$$\frac{p\xrightarrow{a}\surd\quad q\xrightarrow{b}\surd}{p\parallel q\xrightarrow{\step{a,b}}\surd} \quad\frac{p\xrightarrow{a}p'\quad q\xrightarrow{b}\surd}{p\parallel q\xrightarrow{\step{a,b}}p'}$$
$$\frac{p\xrightarrow{a}\surd\quad q\xrightarrow{b}q'}{p\parallel q\xrightarrow{\step{a,b}}q'} \quad\frac{p\xrightarrow{a}p'\quad q\xrightarrow{b}q'}{p\parallel q\xrightarrow{\step{a,b}}p'\between q'}$$

The PTSS of communication merge is as follows.

$$\frac{p\xrightarrow{a}\surd\quad q\xrightarrow{b}\surd}{p\mid q\xrightarrow{\gamma(a,b)}\surd} \quad\frac{p\xrightarrow{a}p'\quad q\xrightarrow{b}\surd}{p\mid q\xrightarrow{\gamma(a,b)}p'}$$
$$\frac{p\xrightarrow{a}\surd\quad q\xrightarrow{b}q'}{p\mid q\xrightarrow{\gamma(a,b)}q'} \quad\frac{p\xrightarrow{a}p'\quad q\xrightarrow{b}q'}{p\mid q\xrightarrow{\gamma(a,b)}p'\between q'}$$

The PTSS of encapsulation operator is as follows.

$$\frac{p\xrightarrow{a}\surd\quad a\notin H}{\partial_H(p)\xrightarrow{a}\surd}\quad\frac{p\xrightarrow{a}p'\quad a\notin H}{\partial_H(p)\xrightarrow{a}\partial_H(p')}$$

The PTSS of confliction, confliction eliminator and the auxiliary unless operator is as follows, where $\leq^e$ is the execution order.

$$\frac{p\xrightarrow{a}\surd\quad a\sharp b}{\Theta(p)\xrightarrow{a}\surd} \quad\frac{p\xrightarrow{b}\surd\quad a\sharp b}{\Theta(p)\xrightarrow{b}\surd}$$
$$\frac{p\xrightarrow{a}p'\quad a\sharp b}{\Theta(p)\xrightarrow{a}\Theta(p')} \quad\frac{p\xrightarrow{b}p'\quad a\sharp b}{\Theta(p)\xrightarrow{b}\Theta(p')}$$
$$\frac{p\xrightarrow{c}\surd \quad q\xnrightarrow{b}\quad a\sharp b\quad c\leq^e a}{p\triangleleft q\xrightarrow{c}\surd}
\quad\frac{p\xrightarrow{c}p' \quad q\xnrightarrow{b}\quad a\sharp b\quad c\leq^e a}{p\triangleleft q\xrightarrow{c}p'}$$
$$\frac{p\xrightarrow{a}\surd \quad q\xnrightarrow{b}\quad a\sharp b}{p\triangleleft q\xrightarrow{ }\surd}
\quad\frac{p\xrightarrow{a}p' \quad q\xnrightarrow{b}\quad a\sharp b}{p\triangleleft q\xrightarrow{ }p'}$$
$$\frac{p\xrightarrow{a}\surd \quad q\xnrightarrow{c}\quad a\sharp b\quad b\leq^e c}{p\triangleleft q\xrightarrow{ }\surd}
\quad\frac{p\xrightarrow{a}p' \quad q\xnrightarrow{c}\quad a\sharp b\quad b\leq^e c}{p\triangleleft q\xrightarrow{ }p'}$$
$$\frac{p\xrightarrow{c}\surd \quad q\xnrightarrow{b}\quad a\sharp b\quad a\leq^e c}{p\triangleleft q\xrightarrow{ }\surd}
\quad\frac{p\xrightarrow{c}p' \quad q\xnrightarrow{b}\quad a\sharp b\quad a\leq^e c}{p\triangleleft q\xrightarrow{ }p'}$$

The PTSS of internal confliction, confliction eliminator and the auxiliary unless operator is as follows.

$$\frac{p\xrightarrow{a}\surd\quad a\osharp b}{\Theta(p)\xrightarrow{a}\surd} \quad\frac{p\xrightarrow{b}\surd\quad a\osharp b}{\Theta(p)\xrightarrow{b}\surd}$$
$$\frac{p\xrightarrow{a}p'\quad a\osharp b}{\Theta(p)\xrightarrow{a}\Theta(p')} \quad\frac{p\xrightarrow{b}p'\quad a\osharp b}{\Theta(p)\xrightarrow{b}\Theta(p')}$$
$$\frac{p\xrightarrow{c}\surd \quad q\xnrightarrow{b}\quad a\osharp b\quad c\leq^e a}{p\triangleleft q\xrightarrow{c}\surd}
\quad\frac{p\xrightarrow{c}p' \quad q\xnrightarrow{b}\quad a\osharp b\quad c\leq^e a}{p\triangleleft q\xrightarrow{c}p'}$$
$$\frac{p\xrightarrow{a}\surd \quad q\xnrightarrow{b}\quad a\osharp b}{p\triangleleft q\xrightarrow{ }\surd}
\quad\frac{p\xrightarrow{a}p' \quad q\xnrightarrow{b}\quad a\osharp b}{p\triangleleft q\xrightarrow{ }p'}$$
$$\frac{p\xrightarrow{a}\surd \quad q\xnrightarrow{c}\quad a\osharp b\quad b\leq^e c}{p\triangleleft q\xrightarrow{ }\surd}
\quad\frac{p\xrightarrow{a}p' \quad q\xnrightarrow{c}\quad a\osharp b\quad b\leq^e c}{p\triangleleft q\xrightarrow{ }p'}$$
$$\frac{p\xrightarrow{c}\surd \quad q\xnrightarrow{b}\quad a\osharp b\quad a\leq^e c}{p\triangleleft q\xrightarrow{ }\surd}
\quad\frac{p\xrightarrow{c}p' \quad q\xnrightarrow{b}\quad a\osharp b\quad a\leq^e c}{p\triangleleft q\xrightarrow{ }p'}$$

The PTSS of internal alternative composition is as follows.

$$\frac{}{1\rightarrowtail\surd}$$
$$\frac{}{p\oplus q\rightarrowtail p} \quad\frac{}{p\oplus q\rightarrowtail q}$$
$$\frac{p\rightarrowtail p'}{p\cdot q\rightarrowtail p'\cdot q}$$
$$\frac{p\rightarrowtail p'}{p+q\rightarrowtail p'+q} \quad\frac{q\rightarrowtail q'}{p+q\rightarrowtail p+q'}$$
$$\frac{p\rightarrowtail p'}{p\between q\rightarrowtail p'\between q} \quad\frac{q\rightarrowtail q'}{p\between q\rightarrowtail p\between q'} \quad\frac{p\rightarrowtail p'\quad q\rightarrowtail q'}{p\between q\rightarrowtail p'\between q'}$$
$$\frac{p\rightarrowtail p'}{p\parallel q\rightarrowtail p'\parallel q} \quad\frac{q\rightarrowtail q'}{p\parallel q\rightarrowtail p\parallel q'} \quad\frac{p\rightarrowtail p'\quad q\rightarrowtail q'}{p\parallel q\rightarrowtail p'\parallel q'}$$
$$\frac{p\rightarrowtail p'}{p\mid q\rightarrowtail p'\mid q} \quad\frac{q\rightarrowtail q'}{p\mid q\rightarrowtail p\mid q'} \quad\frac{p\rightarrowtail p'\quad q\rightarrowtail q'}{p\mid q\rightarrowtail p'\mid q'}$$
$$\frac{p\rightarrowtail p'}{\partial_H(p)\rightarrowtail \partial_H(p')}$$
$$\frac{p\rightarrowtail p'}{\Theta(p)\rightarrowtail \Theta(p')}$$
$$\frac{p\rightarrowtail p'}{p\triangleleft q\rightarrowtail p'\triangleleft q} \quad\frac{q\rightarrowtail q'}{p\triangleleft q\rightarrowtail p\triangleleft q'}$$

\subsection{Denotational Semantics}\label{ds6}

\begin{definition}[$\mathbf{PAT}_{\mathbf{S}}$]
$\mathbf{PAT}_{\mathbf{S}}$, the set of strong acceptance trees over $\mathsf{Pom}$, is the set of rooted trees whose branches are labelled by elements of $\mathsf{Pom}$, whose nodes are either open ($\circ$) or closed ($\bullet$) labelled by subsets of $\mathscr{A}\subseteq\mathsf{Pom}^*$, and satisfies the following requirements:

\begin{enumerate}
  \item R1 (Determinism): For every pomset $U$, every node in the tree has most one successor alternative branch labelled by $U$.
  \item R2 (Finite Branching): For every closed node $n$, $S(n)$ is finite.
  \item R3: For every closed node $n$, $\mathscr{A}(n)$ is an $S(n)$-set.
  \item R4: If $n$ is open then it is a leaf.
\end{enumerate}
\end{definition}

For a tree $t\in\mathbf{PAT}_{\mathbf{S}}$, let $CL(t)\subseteq L(t)$ determine closed nodes, which is prefix-closed and may be empty. For any tree $t\in\mathbf{PAT}_{\mathbf{S}}$ and $s\in\mathsf{Pom}^*$, we write $t\downarrow s$ if $s'\in L(t)$ then $s'\in CL(t)$ for every prefix $s'$ of $s$. 

\begin{definition}[Partial order over $\mathbf{PAT}_{\mathbf{S}}$]
A partial order $\leq_{\mathbf{PAT}_{\mathbf{S}}}$ is defined over $\mathbf{PAT}_{\mathbf{S}}$ as follows, for two trees $t,t'\in\mathbf{PAT}_{\mathbf{S}}$, $t\leq_{\mathbf{PAT}_{\mathbf{S}}}t'$, if for every $s\in\mathsf{Pom}^*$, $t\downarrow s$ implies:

\begin{enumerate}
  \item $t'\downarrow s$.
  \item For every $s\in CL(t')$, $\mathscr{A}(t'(s))\subseteq\mathscr{A}(t(s))$.
\end{enumerate}
\end{definition}

\begin{lemma}
$\langle \mathbf{PAT}_{\mathbf{S}},\leq_{\mathbf{PAT}_{\mathbf{S}}}\rangle$ is an algebraic cpo.
\end{lemma}

In the following, we define functions over $\mathbf{PAT}_{\mathbf{S}}$ for every function symbol in $\Sigma^2$: $0_{\mathbf{PAT}_{\mathbf{S}}}$, $1_{\mathbf{PAT}_{\mathbf{S}}}$, $U_{\mathbf{PAT}_{\mathbf{S}}}$, $\cdot_{\mathbf{PAT}_{\mathbf{S}}}$, $+_{\mathbf{PAT}_{\mathbf{S}}}$, $\oplus_{\mathbf{PAT}_{\mathbf{S}}}$, $\parallel_{\mathbf{PAT}_{\mathbf{S}}}$, $\gamma(a,b)_{\mathbf{PAT}_{\mathbf{S}}}$, $\mid_{\mathbf{PAT}_{\mathbf{S}}}$, $\between_{\mathbf{PAT}_{\mathbf{S}}}$, $\partial_H(p)_{\mathbf{PAT}_{\mathbf{S}}}$, $\sharp_{\mathbf{PAT}_{\mathbf{S}}}$, $\Theta(p)_{\mathbf{PAT}_{\mathbf{S}}}$, $\triangleleft_{\mathbf{PAT}_{\mathbf{S}}}$, $\osharp_{\mathbf{PAT}_{\mathbf{S}}}$. They are continuous operations on the algebra cpo $\mathbf{PAT}_{\mathbf{S}}$ and mild extensions of those in $\mathbf{fPAT}$.

(1) $\Omega_{\mathbf{PAT}_{\mathbf{S}}}$.

Let $\Omega_{\mathbf{PAT}_{\mathbf{S}}}$ denote the trivial tree $\circ$, which consists of only one node, with no successor branches, and rendered as $\circ$.

(2) $0_{\mathbf{PAT}_{\mathbf{S}}}$.

Let $0_{\mathbf{PAT}_{\mathbf{S}}}$ denote the trivial tree $\bullet\{\emptyset\}$, which consists of only one node, labelled by the set of the empty set $\emptyset$ with no successor branches, and rendered as $\bullet$.

(3) $1_{\mathbf{PAT}_{\mathbf{S}}}$.

Let $1_{\mathbf{PAT}_{\mathbf{S}}}$ denote the trivial tree $\bullet\{\{\}\}$, which consists of one node, labelled by the set of the empty set $\{\}$, and rendered as $\obullet$. Note that, $1_{\mathbf{PAT}_{\mathbf{S}}}$ can have successor branches.

(4) $U_{\mathbf{PAT}_{\mathbf{S}}}$.

Formally, the tree $U_{\mathbf{PAT}_{\mathbf{S}}}$ is the tree $t$ determined by:

\begin{enumerate}
  \item $L(t)=\{1\}\cup\{U\}$.
  \item $\mathscr{A}(t(1))=\{\{U\}\}$, $\mathscr{A}(t(U))=\{\{\}\}$.
\end{enumerate}

(5) $\cdot_{\mathbf{PAT}_{\mathbf{S}}}$.

Formally, the tree $t\cdot_{\mathbf{PAT}_{\mathbf{S}}}t'$ is the tree $t''$ determined by:

\begin{enumerate}
  \item $L(t'')=\{1\}\cup\{ss'|s\in L(t),s'\in L(t')\}$.
  \item $\mathscr{A}(t''(1))=\mathscr{A}(t(1))$, $\mathscr{A}(t''(\{\}_ss'))=\mathscr{A}(t(s'))$ where $\{\}_s$ is the closed leaf of $s$.
\end{enumerate}

(6) $+_{\mathbf{PAT}_{\mathbf{S}}}$.

Formally, the tree $t+_{\mathbf{PAT}_{\mathbf{S}}}t'$ is the tree $t''$ determined by:

\begin{enumerate}
  \item $CL(t'')=\{s|s\in L(t)\cup L(t'),\mbox{for every prefix }s'\mbox{ of }s\mbox{ if }s'\in L(t)\mbox{ then }s'\in CL(t),\linebreak\mbox{ and for every prefix } s'\mbox{ of }s\mbox{ if }s'\in L(t')\mbox{ then }s'\in CL(t')\}$.
  \item $L(t'')=\{s|s\in L(t)\cup L(t'), s=s'U\mbox{ implies }s'\in CL(t'')\}$.
  \item $\mathscr{A}(t''(1))=c(\mathscr{A}(t(1))\ucup\mathscr{A}(t'(1)))$, $\mathscr{A}(t''(s))=c(\mathscr{A}(t(s))\cup\mathscr{A}(t'(s)))$ for every $s\in CL(t'')$, and if $s\notin CL(t)$ then $\mathscr{A}(t(s))=\emptyset$.
\end{enumerate}

(7) $\oplus_{\mathbf{PAT}_{\mathbf{S}}}$.

Formally, the tree $t\oplus_{\mathbf{PAT}_{\mathbf{S}}}t'$ is the tree $t''$ determined by:

\begin{enumerate}
  \item $CL(t'')=\{s|s\in L(t)\cup L(t'),\mbox{for every prefix }s'\mbox{ of }s\mbox{ if }s'\in L(t)\mbox{ then }s'\in CL(t),\linebreak\mbox{ and for every prefix } s'\mbox{ of }s\mbox{ if }s'\in L(t')\mbox{ then }s'\in CL(t')\}$.
  \item $L(t'')=\{s|s\in L(t)\cup L(t'), s=s'U\mbox{ implies }s'\in CL(t'')\}$.
  \item $\mathscr{A}(t''(s))=c(\mathscr{A}(t(s))\cup\mathscr{A}(t'(s)))$ for every $s\in CL(t'')$, and if $s\notin CL(t)$ then $\mathscr{A}(t(s))=\emptyset$.
\end{enumerate}

(8) $\parallel_{\mathbf{PAT}_{\mathbf{S}}}$, $\gamma(a,b)_{\mathbf{PAT}_{\mathbf{S}}}$, $\mid_{\mathbf{PAT}_{\mathbf{S}}}$, $\between_{\mathbf{PAT}_{\mathbf{S}}}$ and $\partial_H(p)_{\mathbf{PAT}_{\mathbf{S}}}$.

Then, formally, the tree $t\parallel_{\mathbf{PAT}_{\mathbf{S}}}t'$ is the tree $t''$ determined by:

\begin{enumerate}
  \item $CL(t'')=\{s|s\in L(t)\parallel L(t'),\mbox{for every prefix }s'\mbox{ of }s\mbox{ if }s'\in L(t)\mbox{ then }s'\in CL(t),\linebreak\mbox{ and for every prefix } s'\mbox{ of }s\mbox{ if }s'\in L(t')\mbox{ then }s'\in CL(t')\}$.
  \item $L(t'')=\{s|s\in L(t)\parallel L(t'), s=s'U\mbox{ implies }s'\in CL(t'')\}$.
  \item $\mathscr{A}(t''(1))=c(\mathscr{A}(t(1))\ucup_{\parallel}\mathscr{A}(t'(1)))$, $\mathscr{A}(t''(s))=c(\mathscr{A}(t(s))\parallel\mathscr{A}(t'(s)))$ for every $s\in CL(t'')$, and if $s\notin CL(t)$ then $\mathscr{A}(t(s))=\emptyset$.
\end{enumerate}

The tree $t\mid_{\mathbf{PAT}_{\mathbf{S}}}t'$ is the tree $t''$ determined by:

\begin{enumerate}
  \item $CL(t'')=\{s|s\in L(t)\mid L(t'),\mbox{for every prefix }s'\mbox{ of }s\mbox{ if }s'\in L(t)\mbox{ then }s'\in CL(t),\linebreak\mbox{ and for every prefix } s'\mbox{ of }s\mbox{ if }s'\in L(t')\mbox{ then }s'\in CL(t')\}$.
  \item $L(t'')=\{s|s\in L(t)\mid L(t'), s=s'U\mbox{ implies }s'\in CL(t'')\}$.
  \item $\mathscr{A}(t''(1))=c(\mathscr{A}(t(1))\ucup_{\mid}\mathscr{A}(t'(1)))$, $\mathscr{A}(t''(s))=c(\mathscr{A}(t(s))\mid\mathscr{A}(t'(s)))$ for every $s\in CL(t'')$, and if $s\notin CL(t)$ then $\mathscr{A}(t(s))=\emptyset$.
\end{enumerate}

Formally the tree $t\between_{\mathbf{PAT}_{\mathbf{S}}}t'$ is the tree $t''$ determined by:

\begin{enumerate}
  \item $CL(t'')=(CL(t)\parallel CL(t'))\cup (CL(t)\mid CL(t'))$.
  \item $L(t'')=(L(t)\parallel L(t'))\cup (L(t)\mid L(t'))$.    
  \item $\mathscr{A}(t''(1))=c(\mathscr{A}(t(1))\ucup_{\parallel}\mathscr{A}(t'(1)))\cup c(\mathscr{A}(t(1))\ucup_{\mid}\mathscr{A}(t'(1)))$, $\mathscr{A}(t''(s))=c(\mathscr{A}(t(s))\parallel\mathscr{A}(t'(s)))\cup c(\mathscr{A}(t(s))\mid\mathscr{A}(t'(s)))$  for every $s\in CL(t'')$, and if $s\notin CL(t)$ then $\mathscr{A}(t(s))=\emptyset$.
\end{enumerate}

For $a\in\mathsf{Act}(t)$ and $b\in\mathsf{Act}(t')$, $a,b$ in the tree $t\mid_{\mathbf{PAT}_{\mathbf{S}}}t'$ are merged into $\gamma(a,b)$ if $a\leq^c b$; into $0$, otherwise.

For $H\subseteq\mathsf{Act}(t\parallel_{\mathbf{PAT}_{\mathbf{S}}}t')$, $\partial_H(t\parallel_{\mathbf{PAT}_{\mathbf{S}}}t')$ renames every $a\in H$ into $0$; and remains themselves unchanged for all $a\notin H$ and $a\in\mathsf{Act}(t\parallel_{\mathbf{PAT}_{\mathbf{S}}}t')$.

(9) $\sharp_{\mathbf{PAT}_{\mathbf{S}}}$, $\Theta(p)_{\mathbf{PAT}_{\mathbf{S}}}$, $\triangleleft_{\mathbf{PAT}_{\mathbf{S}}}$ and $\osharp_{\mathbf{PAT}_{\mathbf{S}}}$.

For $a\in\mathsf{Act}(t)$ and $b\in\mathsf{Act}(t')$, there exists a confliction relation $a\sharp b$ in the tree $t\parallel_{\mathbf{PAT}_{\mathbf{S}}}t'$ (resp. $t\mid_{\mathbf{PAT}_{\mathbf{S}}}t'$), $\Theta(t\parallel_{\mathbf{PAT}_{\mathbf{S}}}t')_{\mathbf{PAT}_{\mathbf{S}}}$ (resp. $\Theta(t\mid_{\mathbf{PAT}_{\mathbf{S}}}t')_{\mathbf{PAT}_{\mathbf{S}}}$) and $\triangleleft_{\mathbf{PAT}_{\mathbf{S}}}$ rename $\{s'|t(1)\xrightarrow{s}\xrightarrow{a}t(s')\}$ to $1$'s and remain the open nodes unchanged, and remain $t'$ unchanged, or rename $\{s'|t'(1)\xrightarrow{s}\xrightarrow{b}t'(s')\}$ to $1$'s and remain the open nodes unchanged, and remain $t$ unchanged.

For $a\in\mathsf{Act}(t)$ and $b\in\mathsf{Act}(t')$, there exists an internal confliction relation $a\osharp b$ in the tree $t\parallel_{\mathbf{PAT}_{\mathbf{S}}}t'$ (resp. $t\mid_{\mathbf{PAT}_{\mathbf{S}}}t'$), $\Theta(t\parallel_{\mathbf{PAT}_{\mathbf{S}}}t')_{\mathbf{PAT}_{\mathbf{S}}}$ (resp. $\Theta(t\mid_{\mathbf{PAT}_{\mathbf{S}}}t')_{\mathbf{PAT}_{\mathbf{S}}}$) and $\triangleleft_{\mathbf{PAT}_{\mathbf{S}}}$ rename $\{s'|t(1)\xrightarrow{s}\xrightarrow{a}t(s')\}$ to $1$'s and remain the open nodes unchanged, and remain $t'$ unchanged, or rename $\{s'|t'(1)\xrightarrow{s}\xrightarrow{b}t'(s')\}$ to $1$'s and remain the open nodes unchanged, and remain $t$ unchanged.

\begin{proposition}
$\langle \mathbf{PAT}_{\mathbf{S}},\leq_{\mathbf{PAT}_{\mathbf{S}}},\Sigma^2_{\mathbf{PAT}_{\mathbf{S}}}\rangle$ is a $\Sigma^2$-domain.
\end{proposition}

\begin{theorem}[Full Abstraction for $\mathbf{PAT}_{\mathbf{S}}$]
If $p,q\in\mathbf{M}_2$, then $p\pretestingmust q$ if and only if $\mathbf{PAT}_{\mathbf{S}}\sembrack{p}\leq_{\mathbf{PAT}_{\mathbf{S}}}\mathbf{PAT}_{\mathbf{S}}\sembrack{q}$.
\end{theorem}

\subsection{Axiomatic Semantics}\label{as6}

The proof system $\mathbf{\Omega DED}(E^2_{\mathbf{S}})$ contains the inequations denoted $E^2_{\mathbf{S}}$, which include $E^1_{\mathbf{S}}$ in \cref{t5} and the inequations in \cref{ProofSystemE2ofM2}. Note that, since the inequation $\Omega\leq x$ is inherited from any proof system containing $\Omega$ $\mathbf{\Omega DED}$, the inequation $\Omega2$ is redundant.

\begin{center}
    \begin{longtable}{|c|c|}
      \caption{Proof system $\mathbf{\Omega DED}(E^2_{\mathbf{S}})$ of $\mathbf{M}_2$}\\
      \hline No. &Axiom\\
      \hline
      \endfirsthead
      \multicolumn{2}{c}{\bfseries Continuing: Proof system $\mathbf{\Omega DED}(E^2_{\mathbf{S}})$ of $\mathbf{M}_2$} \\
      \hline No. &Axiom\\
      \hline
      \endhead
      \hline
      \multicolumn{2}{r}{to be continued\ldots} \\
      \endfoot
      \hline
      \multicolumn{2}{r}{end of Proof system $\mathbf{\Omega DED}(E^2_{\mathbf{S}})$ of $\mathbf{M}_2$} \\
      \endlastfoot
      $\Omega1$ & $x+ \Omega \leq \Omega$\\
      $\Omega2$ & $x\oplus \Omega \leq \Omega$\\
      $\Omega3$ & $\Omega\cdot x \leq \Omega$\\
      $\Omega4$ & $x\between \Omega \leq \Omega$\\
      $\Omega5$ & $x\parallel \Omega \leq \Omega$\\
      $\Omega6$ & $x\mid \Omega \leq \Omega$\\
      $\Omega7$ & $\partial_H(\Omega) \leq \Omega$\\
      $\Omega8$ & $\Theta(\Omega) \leq \Omega$\\
      $\Omega9$ & $\Omega\triangleleft x \leq \Omega$\\
      $\Omega10$ & $x\triangleleft \Omega \leq \Omega$
      \label{ProofSystemE2ofM2}
    \end{longtable}
\end{center}

\begin{lemma}
$\mathbf{PAT}_{\mathbf{S}}$ is finitary.
\end{lemma}

\begin{proposition}
$\mathbf{PAT}_{\mathbf{S}}$ is in $\mathscr{CC}(E^2_{\mathbf{S}})$.
\end{proposition}

\begin{lemma}
The mapping $i_{\mathbf{PAT}_{\mathbf{S}}}$ from $\mathbf{M}_2$ to $\mathbf{PAT}_{\mathbf{S}}$ is surjective.
\end{lemma}

\begin{lemma}
$\mathbf{\Omega DED}(E^2_{\mathbf{S}})$ is sound with respect to $\leq_{\mathbf{PAT}_{\mathbf{S}}}$, restricted to $T_{\Sigma^2}$.
\end{lemma}

\begin{definition}[$\Omega$-normal form]
The $\Omega$-normal form is defined inductively as follows:

\begin{enumerate}
  \item $0$ and $\Omega$ are $\Omega$-normal forms.
  \item If $\mathscr{A}$ is saturated set and for every $U$ in $A(\mathscr{A})$ there is an $\Omega$-normal form $n(U)$, then $\osum\{n(A)|A\in\mathscr{A}\}$ is an $\Omega$-normal form, where $n(A)$ represents $\sum\{Un(U)|U\in A\}$.
\end{enumerate}
\end{definition}

\begin{theorem}[Normal form theorem]
For every term $p$ in $\mathbf{M}_2$, there exists an $\Omega$-normal form $nf(p)$ such that $p=_{E^2_{\mathbf{S}}}nf(p)$.
\end{theorem}

\begin{theorem}[Completeness]
For $p,q\in\mathbf{M}_2$, $\mathbf{PAT}_{\mathbf{S}}\sembrack{p}\leq_{\mathbf{PAT}_{\mathbf{S}}}\mathbf{PAT}_{\mathbf{S}}\sembrack{q}$ implies $p\leq_{E^2_{\mathbf{S}}}q$.
\end{theorem}

\begin{theorem}[Initiality of $\mathbf{PAT}_{\mathbf{S}}$]
$\mathbf{PAT}_{\mathbf{S}}$ is initial in $\mathscr{CC}(E^2_{\mathbf{S}})$.
\end{theorem}

\subsection{The Trinity} \label{t6} 

\begin{definition}[$\mathbf{PAT}$]
$\mathbf{PAT}$, the set of acceptance trees over $\mathsf{Pom}$, is the set of rooted trees whose branches are labelled by elements of $\mathsf{Pom}$, whose nodes are either open ($\circ$) or closed ($\bullet$) labelled by subsets of $\mathscr{A}\subseteq\mathsf{Pom}^*$, and satisfies the following requirements:

\begin{enumerate}
  \item R1 (Determinism): For every pomset $U$, every node in the tree has most one successor alternative branch labelled by $U$.
  \item R2 (Finite Branching): For every closed node $n$, $S(n)$ is finite.
  \item R3: For every closed node $n$, $\mathscr{A}(n)$ is an $S(n)$-set.
  \item R4: If $n$ is open then every descendant of $n$ is also open.
\end{enumerate}
\end{definition}

\begin{definition}
A partial order $\leq_{\mathbf{PAT}}$ is defined over $\mathbf{PAT}$ as follows, for two trees $t,t'\in\mathbf{PAT}$, $t\leq_{\mathbf{PAT}}t'$, if:

\begin{enumerate}
  \item $L(t)\subseteq L(t')$.
  \item For every $s\in \mathsf{Pom}^*$, $t\downarrow s$ implies:
  \begin{enumerate}
    \item $t'\downarrow s$.
    \item if $s\in CL(t')$ then $\mathscr{A}(t'(s))\subseteq\mathscr{A}(t(s))$.
  \end{enumerate}
\end{enumerate}
\end{definition}

\begin{lemma}
$\langle \mathbf{PAT}, \leq_{\mathbf{PAT}}\rangle$ is an algebraic cpo.
\end{lemma}

The proof system $\mathbf{\Omega DED}(E^2)$ contains the inequations denoted $E^2$, which include $E^1$ in \cref{as5} and the inequations in \cref{ProofSystemE2ofM2}.

\begin{lemma}
$\mathbf{PAT}$ is finitary.
\end{lemma}

\begin{proposition}
$\mathbf{PAT}$ is in $\mathscr{CC}(E^2)$.
\end{proposition}

\begin{lemma}
The mapping $i_{\mathbf{PAT}}$ from $\mathbf{M}_2$ to $\mathbf{PAT}$ is surjective.
\end{lemma}

\begin{lemma}
$\mathbf{\Omega DED}(E^2)$ is sound with respect to $\leq_{\mathbf{PAT}}$, restricted to $T_{\Sigma^2}$.
\end{lemma}

\begin{theorem}[Normal form theorem]
For every term $p$ in $\mathbf{M}_2$, there exists an $\Omega$-normal form $nf(p)$ such that $p=_{E^2}nf(p)$.
\end{theorem}

\begin{theorem}[Completeness]
For $p,q\in\mathbf{M}_2$, $\mathbf{PAT}\sembrack{p}\leq_{\mathbf{PAT}}\mathbf{PAT}\sembrack{q}$ implies $p\leq_{E^2}q$.
\end{theorem}

\begin{theorem}[Initiality of $\mathbf{PAT}$]
$\mathbf{PAT}$ is initial in $\mathscr{CC}(E^2)$.
\end{theorem}

\begin{definition}[$\mathbf{PAT}_{\mathbf{W}}$]
$\mathbf{PAT}_{\mathbf{W}}$, the set of weak acceptance trees over $\mathsf{Pom}$, is the set of rooted trees whose branches are labelled by elements of $\mathsf{Pom}$, whose nodes are either open ($\circ$) or closed ($\bullet$) labelled by subsets of $\mathscr{A}\subseteq\mathsf{Pom}^*$, and satisfies the following requirements:

\begin{enumerate}
  \item R1 (Determinism): For every pomset $U$, every node in the tree has most one successor alternative branch labelled by $U$.
  \item R2 (Finite Branching): For every closed node $n$, $S(n)$ is finite.
  \item R3: For every closed node $n$, $\mathscr{A}(n)$ is an $S(n)$-set.
  \item R4: Every node is open.
\end{enumerate}
\end{definition}

\begin{definition}
A partial order $\leq_{\mathbf{PAT}_{\mathbf{W}}}$ is defined over $\mathbf{PAT}_{\mathbf{W}}$ as follows, for two trees $t,t'\in\mathbf{PAT}$, $t\leq_{\mathbf{PAT}_{\mathbf{W}}}t'$, if:

$L(t)\subseteq L(t')$.
\end{definition}

\begin{lemma}
$\langle \mathbf{PAT}_{\mathbf{W}}, \leq_{\mathbf{PAT}_{\mathbf{W}}}\rangle$ is an algebraic cpo.
\end{lemma}

The proof system $\mathbf{\Omega DED}(E^2_{\mathbf{W}})$ contains the inequations denoted $E^2_{\mathbf{W}}$, which include $E^1_{\mathbf{W}}$ in \cref{t5} and the inequations in \cref{ProofSystemE2ofM2}.

\begin{lemma}
$\mathbf{PAT}_{\mathbf{W}}$ is finitary.
\end{lemma}

\begin{proposition}
$\mathbf{PAT}_{\mathbf{W}}$ is in $\mathscr{CC}(E^2_{\mathbf{W}})$.
\end{proposition}

\begin{lemma}
The mapping $i_{\mathbf{PAT}_{\mathbf{W}}}$ from $\mathbf{M}_2$ to $\mathbf{PAT}_{\mathbf{W}}$ is surjective.
\end{lemma}

\begin{lemma}
$\mathbf{\Omega DED}(E^2_{\mathbf{W}})$ is sound with respect to $\leq_{\mathbf{PAT}_{\mathbf{W}}}$, restricted to $T_{\Sigma^2}$.
\end{lemma}

\begin{theorem}[Normal form theorem]
For every term $p$ in $\mathbf{M}_2$, there exists an $\Omega$-normal form $nf(p)$ such that $p=_{E^2_{\mathbf{W}}}nf(p)$.
\end{theorem}

\begin{theorem}[Completeness]
For $p,q\in\mathbf{M}_2$, $\mathbf{PAT}_{\mathbf{W}}\sembrack{p}\leq_{\mathbf{PAT}_{\mathbf{W}}}\mathbf{PAT}_{\mathbf{W}}\sembrack{q}$ implies $p\leq_{E^2_{\mathbf{W}}}q$.
\end{theorem}

\begin{theorem}[Initiality of $\mathbf{PAT}_{\mathbf{W}}$]
$\mathbf{PAT}_{\mathbf{W}}$ is initial in $\mathscr{CC}(E^2_{\mathbf{W}})$.
\end{theorem} 
\newpage\section{Recursion}\label{recT}

In this chapter, we introduce the basic processes with recursion. Firstly, we introduce the recursive signature in \cref{rs7}, then the operational semantics and axiomatic semantics of the recursive processes are introduced in \cref{os7} and \cref{as7}, respectively. Finally, we get the results on trinity of operational semantics, denotational semantics and axiomatic semantics in \cref{t7}.

\subsection{Recursive Signature}\label{rs7}

\begin{definition}[Recursive signature]
The recursive signature $\Sigma^3$ consists of:

\begin{enumerate}
  \item $\Sigma^3\supset \Sigma^2$.
  \item A set of recursive variables $X$ ranged over $x,y,z,\cdots$.
\end{enumerate}
\end{definition}

\begin{definition}[Syntax of recursive process language]
The syntax of the recursive process language is given by the following BNF grammar:

$p::=\Omega~|~0~|~1~|~U~|~\gamma(a,b)~|~p\cdot p~|~a\sharp b~|~a\osharp b~|~x~|~p+p~|~p\oplus p~|~p\parallel p~|~p\mid p~|~p\between p~|~\Theta(p)~|~p\triangleleft p~|~\partial_H(p)~|~rec~x.p$

where $a,b\in\mathsf{Act}$, $U\in\mathsf{Pom}$, $p\in \mathsf{Proc}$, $x$ is recursive variable and $rec~x.p$ stands for the process defined by the recursive equation $x=p$. We assume that $rec~x.$ has the lowest precedence of all the operators in $\Sigma^3$.
\end{definition}

The set of recursive terms over $\Sigma^3$ is denoted $REC_{\Sigma^3}(X)$, sometimes we only write $REC_{\Sigma}(X)$. We write $FREC_{\Sigma}(X)$ to denote the subset of terms in $REC_{\Sigma}(X)$ without occurrences of $rec~x.$. We use $FV(t)$ to denote the set of variables which occur free in term $t$ and $BV(t)$ to denote the set of variables which occur bound in $t$. We use $REC_{\Sigma}$ to denote the set of closed terms and $FREC_{\Sigma}$ to denote the set of finite closed terms.

A substitution $\sigma$  will mean a mapping from $X$ to $REC_{\Sigma}(X)$, i.e., $\sigma[x\rightarrow t]$ is a substitution which maps $x$ to $t$, and $I$ denotes the identity substitution which maps every variable $x$ to itself. We write $t\sigma$ to denote the result of applying $\sigma$ to $t$. If $x$ is a variable, $t$ a term, $\sigma$ a substitution, let $new~xt\sigma$ denote the least variable $y$ in the enumeration of $X$ such that for every $z\in FV(t)$ which is different from $x,y\notin FV(\sigma(z))$.

\begin{definition}[Substitution]
For $t\in REC_{\Sigma}(X)$ and $\sigma$ a substitution, $t\sigma$ is defined inductively:

\begin{enumerate}
  \item $x\sigma=\sigma(x)$.
  \item $f(t_1,\cdots,t_{ar(f)})\sigma=f(t_1\sigma,\cdots,t_{ar(f)}\sigma)$.
  \item $rec~x.t\sigma=rec~y.(t\sigma[x\rightarrow y])$, where $y=new~xt\sigma$.
\end{enumerate}
\end{definition}

\begin{lemma}[Syntactic substitution]
For every $t\in REC_{\Sigma}(X)$, and $\sigma$ and $\sigma'$ substitutions, $(t\sigma)\sigma'=t(\sigma'\circ\sigma)$.
\end{lemma}

\begin{definition}[$\alpha$-equality]
The $\alpha$-equivalence $=_{\alpha}$ is defined as the least $\Sigma$-congruence over $REC_{\Sigma}(X)$ satisfying:

\begin{enumerate}
  \item $t[y/x]=_{\alpha}t'$ and $y\notin FV(t)$ implies $rec~x.t=_{\alpha}rec~y.t'$.
  \item $t=_{\alpha}t'$ implies $rec~x.t=_{\alpha}rec~ x.t'$.
\end{enumerate}
\end{definition}

The meaning in a $\Sigma$-domain of a term is taken to be the limit of the meaning of its finite approximations. These approximations are defined by progressively expanding out recursive subterms as follows.

\begin{definition}[Finite principal approximations]
For each $n\geq 0$ and $t\in REC_{\Sigma}(X)$, the finite principal approximations $t^n$ of $t$ is defined inductively by:

\begin{enumerate}
  \item $t^0=\Omega$.
  \item \begin{enumerate}
          \item $x^{n+1}=x$;
          \item $f(\underline{t})^{n+1}=f(\underline{t}^{n+1})$;
          \item $(rec~x.t)^{n+1}=t^{n+1}[(rec~x.t)^{n}/x]$.
        \end{enumerate}
\end{enumerate} 
\end{definition} 

Let $\App(t)=\{t^{n}|n\geq 0\}$, these approximations are related via the syntactic preorder $\preceq$.

\begin{lemma}
$n\leq m$ implies $t^{n}\preceq t^{m}$.
\end{lemma}

\begin{corollary}
For every $t\in REC_{\Sigma}(X)$, $\App(t)$ is directed with respect to $\preceq$.
\end{corollary}

Let $A$ be a $\Sigma$-domain and $\sigma_A$ an $A$-assignment, which is a mapping from each variable $x\in X$ to an element $\sigma_A(x)$ of $A$. Let $ENV_{A}$ denote the collection of $A$-assignments. Sometimes, we only write $\sigma$ for $\sigma_A$ and $ENV$ for $ENV_A$. $ENV_A$ can be ordered pointwise via: $\sigma\leq\sigma'$ if for every $x\in X$, $\sigma(x)\leq_A\sigma'(x)$. If $A$ is a domain, $ENV_A$ under this ordering is a domain. For $Y\subseteq X$, we write $\sigma_A=_{Y}\sigma'_{A}$ if $\sigma_A(y)=_{A}\sigma'_{A}(y)$ for every $y\in Y$ and $\sigma_A\leq_{Y}\sigma'_{A}$ if $\sigma_A(y)\leq_{A}\sigma'_{A}(y)$ for every $y\in Y$.

Let $A\sembrack{ }:T_{\Sigma}(X)\rightarrow(ENV_{A}\rightarrow A)$ which is a function be defined by:

\begin{enumerate}
  \item $A\sembrack{x}\sigma=\sigma(x)$.
  \item $A\sembrack{f(\underline{t})}\sigma=f_{A}(A\sembrack{\underline{t}}\sigma)$.
\end{enumerate}

\begin{proposition}
The following statements hold:

\begin{enumerate}
  \item $\sigma:T_{\Sigma}(X)\rightarrow A$ is a $\Sigma$-homomorphism.
  \item If $\varphi:T_{\Sigma}(X)\rightarrow A$ is a $\Sigma$-homomorphism satisfying $\varphi(x)=\sigma(x)$ then $\varphi$ coincides with $\sigma$.
  \item If $\sigma,\sigma'$ are $A$-assignments such that $\sigma=_{FV(t)}\sigma'$ then $\sigma(t)=\sigma'(t)$.
\end{enumerate}
\end{proposition}

Let $A\sembrack{ }:REC_{\Sigma}(X)\rightarrow(ENV_{A}\rightarrow A)$ which is well-defined be extended by:

\begin{enumerate}
  \item $A\sembrack{x}\sigma=\sigma(x)$.
  \item $A\sembrack{f(\underline{t})}\sigma=f_{A}(A\sembrack{\underline{t}}\sigma)$.
  \item $A\sembrack{rec~x.t}\sigma=Y\lambda a.A\sembrack{t}\sigma[a/x]$, where $a\in A$, $Y$ is the least fixpoint operator.
\end{enumerate}

\begin{lemma}
Let $t\in REC_{\Sigma}(X)$ and $\sigma,\sigma'$ be $A$-assignments, if $\sigma\leq_{FV(t)}\sigma'$ then $A\sembrack{t}\sigma\leq_{A}A\sembrack{t}\sigma'$.
\end{lemma}

\begin{proposition}
$A\sembrack{t\sigma}\sigma_{A}=A\sembrack{t}(\sigma_{A}\circ\sigma)$.
\end{proposition}

\begin{corollary}
If $y\notin FV(t)$, $A\sembrack{rec~x.t}=A\sembrack{rec~y.t[y/x]}$.
\end{corollary}

\begin{corollary}
$t=_{\alpha}t'$ implies $A\sembrack{t}=A\sembrack{t'}$.
\end{corollary}

\begin{corollary}
$A\sembrack{rec~x.t}=A\sembrack{t[rec~x.t/x]}$.
\end{corollary}

\begin{lemma}
For every $n\geq 0$, $A\sembrack{t^{n}}\leq A\sembrack{t}$.
\end{lemma}

\begin{theorem}[Finite Approximations]
For every $t\in REC_{\Sigma}(X)$, $A\sembrack{t}=\bigvee A\sembrack{\App(t)}$.
\end{theorem}

\begin{proposition}
Let $\varphi:REC_{\Sigma}(X)\rightarrow A$ be a $\Sigma$-homomorphism satisfying:

\begin{enumerate}
  \item It is an extension of $A$-assignment $\sigma$, i.e., $\varphi(x)=\sigma(x)$.
  \item For every $t\in REC_{\Sigma}(X)$, $\varphi(t)=\bigvee_{A}\varphi(\App(t))$.
\end{enumerate}

then $\varphi=\sigma$.
\end{proposition}

A function $\varphi:REC_{\Sigma}(X)\rightarrow[ENV_{A}\rightarrow A]$ is reasonable if:

\begin{enumerate}
  \item $\varphi(rec~x.t)=\varphi(t[rec~x.t/x])$.
  \item $\varphi(t[u/x])\sigma=\varphi(t)\sigma[\varphi(u)\sigma/x]$.
\end{enumerate}

\begin{proposition}
If $\varphi:REC_{\Sigma}(X)\rightarrow[ENV_{A}\rightarrow A]$ is a reasonable $\Sigma$-homomorphism then $A\sembrack{t}\leq\varphi(t)$ for every $t\in REC_{\Sigma}(X)$.
\end{proposition}

\subsection{Operational Semantics}\label{os7}

In this section, we give the operational semantics of the language $\mathbf{M}_3$. The predicate $\surd$ represents successful termination, $\xrightarrow{a}\surd$ represents successful termination after execution of the action $a\in\mathsf{Act}$, $\xrightarrow{U}\surd$ represents successful termination after execution of the action $U\in\mathsf{Pom}$ and $\xrightarrow{ }\surd$ represents successful termination without execution of the any action. The divergence predicate $p\uparrow$ represents that $p$ is divergent, i.e., $p$ has an infinite internal computation, $$p\rightarrowtail p_0\rightarrowtail p_1\rightarrowtail\cdots\rightarrowtail p_k\rightarrowtail\cdots$$ While the convergence predicate $p\downarrow$ represents that $p$ is not divergent $p\nuparrow$, i.e., convergent, $p$ has no infinite internal computation. The following are the PTSS of the language $\mathbf{M}_3$, where $p,q\in\mathsf{Proc}$.

The PTSS of $\Omega$ is as follows.

$$\frac{}{\Omega\uparrow}$$
$$\frac{}{\Omega\rightarrowtail\Omega}$$
$$\frac{}{\Omega\xrightarrow{1}\Omega}$$

The PTSS of action $1$, $a\in\mathsf{Act}$ and $U\in\mathsf{Pom}$ is as follows. Note that, there is no any transition rules for $0$.

$$\frac{}{0\downarrow}\quad\frac{}{1\downarrow}\quad \frac{}{a\downarrow} \quad\frac{}{U\downarrow}$$
$$\frac{}{1\xrightarrow{ }\surd}\quad\frac{}{a\xrightarrow{a}\surd}\quad\frac{}{U\xrightarrow{U}\surd}$$

The PTSS of sequential composition is as follows.

$$\frac{p\downarrow\quad q\downarrow}{p\cdot q\downarrow}$$
$$\frac{p\xrightarrow{U}\surd}{p\cdot q\xrightarrow{U} q}\quad\frac{p\xrightarrow{U}p'}{p\cdot q\xrightarrow{U} p'\cdot q}$$

The PTSS of alternative composition is as follows.

$$\frac{q\downarrow\quad q\downarrow}{p+q\downarrow}$$
$$\frac{p\xrightarrow{U}\surd}{p+ q\xrightarrow{U}\surd} \quad\frac{p\xrightarrow{U}p'}{p+ q\xrightarrow{U}p'} \quad\frac{q\xrightarrow{U}\surd}{p+ q\xrightarrow{U}\surd} \quad\frac{q\xrightarrow{U}q'}{p+ q\xrightarrow{U}q'}$$

The PTSS of concurrent composition is as follows.

$$\frac{p\downarrow\quad q\downarrow}{p\between q\downarrow}$$
$$\frac{p\xrightarrow{a}\surd\quad q\xrightarrow{b}\surd}{p\between q\xrightarrow{\step{a,b}}\surd} \quad\frac{p\xrightarrow{a}p'\quad q\xrightarrow{b}\surd}{p\between q\xrightarrow{\step{a,b}}p'}$$
$$\frac{p\xrightarrow{a}\surd\quad q\xrightarrow{b}q'}{p\between q\xrightarrow{\step{a,b}}q'} \quad\frac{p\xrightarrow{a}p'\quad q\xrightarrow{b}q'}{p\between q\xrightarrow{\step{a,b}}p'\between q'}$$
$$\frac{p\xrightarrow{a}\surd\quad q\xrightarrow{b}\surd}{p\between q\xrightarrow{\gamma(a,b)}\surd} \quad\frac{p\xrightarrow{a}p'\quad q\xrightarrow{b}\surd}{p\between q\xrightarrow{\gamma(a,b)}p'}$$
$$\frac{p\xrightarrow{a}\surd\quad q\xrightarrow{b}q'}{p\between q\xrightarrow{\gamma(a,b)}q'} \quad\frac{p\xrightarrow{a}p'\quad q\xrightarrow{b}q'}{p\between q\xrightarrow{\gamma(a,b)}p'\between q'}$$

The PTSS of parallel composition is as follows.

$$\frac{p\downarrow\quad q\downarrow}{p\parallel q\downarrow}$$
$$\frac{p\xrightarrow{a}\surd\quad q\xrightarrow{b}\surd}{p\parallel q\xrightarrow{\step{a,b}}\surd} \quad\frac{p\xrightarrow{a}p'\quad q\xrightarrow{b}\surd}{p\parallel q\xrightarrow{\step{a,b}}p'}$$
$$\frac{p\xrightarrow{a}\surd\quad q\xrightarrow{b}q'}{p\parallel q\xrightarrow{\step{a,b}}q'} \quad\frac{p\xrightarrow{a}p'\quad q\xrightarrow{b}q'}{p\parallel q\xrightarrow{\step{a,b}}p'\between q'}$$

The PTSS of communication merge is as follows.

$$\frac{p\downarrow\quad q\downarrow}{p\mid q\downarrow}$$
$$\frac{p\xrightarrow{a}\surd\quad q\xrightarrow{b}\surd}{p\mid q\xrightarrow{\gamma(a,b)}\surd} \quad\frac{p\xrightarrow{a}p'\quad q\xrightarrow{b}\surd}{p\mid q\xrightarrow{\gamma(a,b)}p'}$$
$$\frac{p\xrightarrow{a}\surd\quad q\xrightarrow{b}q'}{p\mid q\xrightarrow{\gamma(a,b)}q'} \quad\frac{p\xrightarrow{a}p'\quad q\xrightarrow{b}q'}{p\mid q\xrightarrow{\gamma(a,b)}p'\between q'}$$

The PTSS of encapsulation operator is as follows.

$$\frac{p\downarrow}{\partial_H(p)\downarrow}$$
$$\frac{p\xrightarrow{a}\surd\quad a\notin H}{\partial_H(p)\xrightarrow{a}\surd}\quad\frac{p\xrightarrow{a}p'\quad a\notin H}{\partial_H(p)\xrightarrow{a}\partial_H(p')}$$

The PTSS of confliction, confliction eliminator and the auxiliary unless operator is as follows, where $\leq$ is the execution order.

$$\frac{p\downarrow}{\Theta(p)\downarrow}\quad\frac{p\downarrow\quad q\downarrow}{p\triangleleft q\downarrow}$$
$$\frac{p\xrightarrow{a}\surd\quad a\sharp b}{\Theta(p)\xrightarrow{a}\surd} \quad\frac{p\xrightarrow{b}\surd\quad a\sharp b}{\Theta(p)\xrightarrow{b}\surd}$$
$$\frac{p\xrightarrow{a}p'\quad a\sharp b}{\Theta(p)\xrightarrow{a}\Theta(p')} \quad\frac{p\xrightarrow{b}p'\quad a\sharp b}{\Theta(p)\xrightarrow{b}\Theta(p')}$$
$$\frac{p\xrightarrow{c}\surd \quad q\xnrightarrow{b}\quad a\sharp b\quad c\leq a}{p\triangleleft q\xrightarrow{c}\surd}
\quad\frac{p\xrightarrow{c}p' \quad q\xnrightarrow{b}\quad a\sharp b\quad c\leq a}{p\triangleleft q\xrightarrow{c}p'}$$
$$\frac{p\xrightarrow{a}\surd \quad q\xnrightarrow{b}\quad a\sharp b}{p\triangleleft q\xrightarrow{ }\surd}
\quad\frac{p\xrightarrow{a}p' \quad q\xnrightarrow{b}\quad a\sharp b}{p\triangleleft q\xrightarrow{ }p'}$$
$$\frac{p\xrightarrow{a}\surd \quad q\xnrightarrow{c}\quad a\sharp b\quad b\leq c}{p\triangleleft q\xrightarrow{ }\surd}
\quad\frac{p\xrightarrow{a}p' \quad q\xnrightarrow{c}\quad a\sharp b\quad b\leq c}{p\triangleleft q\xrightarrow{ }p'}$$
$$\frac{p\xrightarrow{c}\surd \quad q\xnrightarrow{b}\quad a\sharp b\quad a\leq c}{p\triangleleft q\xrightarrow{ }\surd}
\quad\frac{p\xrightarrow{c}p' \quad q\xnrightarrow{b}\quad a\sharp b\quad a\leq c}{p\triangleleft q\xrightarrow{ }p'}$$

The PTSS of internal confliction, confliction eliminator and the auxiliary unless operator is as follows.

$$\frac{p\downarrow}{\Theta(p)\downarrow}\quad\frac{p\downarrow\quad q\downarrow}{p\triangleleft q\downarrow}$$
$$\frac{p\xrightarrow{a}\surd\quad a\osharp b}{\Theta(p)\xrightarrow{a}\surd} \quad\frac{p\xrightarrow{b}\surd\quad a\osharp b}{\Theta(p)\xrightarrow{b}\surd}$$
$$\frac{p\xrightarrow{a}p'\quad a\osharp b}{\Theta(p)\xrightarrow{a}\Theta(p')} \quad\frac{p\xrightarrow{b}p'\quad a\osharp b}{\Theta(p)\xrightarrow{b}\Theta(p')}$$
$$\frac{p\xrightarrow{c}\surd \quad q\xnrightarrow{b}\quad a\osharp b\quad c\leq a}{p\triangleleft q\xrightarrow{c}\surd}
\quad\frac{p\xrightarrow{c}p' \quad q\xnrightarrow{b}\quad a\osharp b\quad c\leq a}{p\triangleleft q\xrightarrow{c}p'}$$
$$\frac{p\xrightarrow{a}\surd \quad q\xnrightarrow{b}\quad a\osharp b}{p\triangleleft q\xrightarrow{ }\surd}
\quad\frac{p\xrightarrow{a}p' \quad q\xnrightarrow{b}\quad a\osharp b}{p\triangleleft q\xrightarrow{ }p'}$$
$$\frac{p\xrightarrow{a}\surd \quad q\xnrightarrow{c}\quad a\osharp b\quad b\leq c}{p\triangleleft q\xrightarrow{ }\surd}
\quad\frac{p\xrightarrow{a}p' \quad q\xnrightarrow{c}\quad a\osharp b\quad b\leq c}{p\triangleleft q\xrightarrow{ }p'}$$
$$\frac{p\xrightarrow{c}\surd \quad q\xnrightarrow{b}\quad a\osharp b\quad a\leq c}{p\triangleleft q\xrightarrow{ }\surd}
\quad\frac{p\xrightarrow{c}p' \quad q\xnrightarrow{b}\quad a\osharp b\quad a\leq c}{p\triangleleft q\xrightarrow{ }p'}$$

The PTSS of internal alternative composition is as follows.

$$\frac{p\downarrow\quad q\downarrow}{p\oplus q\downarrow}$$
$$\frac{}{1\rightarrowtail\surd}$$
$$\frac{}{p\oplus q\rightarrowtail p} \quad\frac{}{p\oplus q\rightarrowtail q}$$
$$\frac{p\rightarrowtail p'}{p\cdot q\rightarrowtail p'\cdot q}$$
$$\frac{p\rightarrowtail p'}{p+q\rightarrowtail p'+q} \quad\frac{q\rightarrowtail q'}{p+q\rightarrowtail p+q'}$$
$$\frac{p\rightarrowtail p'}{p\between q\rightarrowtail p'\between q} \quad\frac{q\rightarrowtail q'}{p\between q\rightarrowtail p\between q'} \quad\frac{p\rightarrowtail p'\quad q\rightarrowtail q'}{p\between q\rightarrowtail p'\between q'}$$
$$\frac{p\rightarrowtail p'}{p\parallel q\rightarrowtail p'\parallel q} \quad\frac{q\rightarrowtail q'}{p\parallel q\rightarrowtail p\parallel q'} \quad\frac{p\rightarrowtail p'\quad q\rightarrowtail q'}{p\parallel q\rightarrowtail p'\parallel q'}$$
$$\frac{p\rightarrowtail p'}{p\mid q\rightarrowtail p'\mid q} \quad\frac{q\rightarrowtail q'}{p\mid q\rightarrowtail p\mid q'} \quad\frac{p\rightarrowtail p'\quad q\rightarrowtail q'}{p\mid q\rightarrowtail p'\mid q'}$$
$$\frac{p\rightarrowtail p'}{\partial_H(p)\rightarrowtail \partial_H(p')}$$
$$\frac{p\rightarrowtail p'}{\Theta(p)\rightarrowtail \Theta(p')}$$
$$\frac{p\rightarrowtail p'}{p\triangleleft q\rightarrowtail p'\triangleleft q} \quad\frac{q\rightarrowtail q'}{p\triangleleft q\rightarrowtail p\triangleleft q'}$$

The PTSS of recursion is as follows.

$$\frac{p[rec~x.p/x]\downarrow}{rec~x.p\downarrow}$$
$$\frac{}{rec~x.p\rightarrowtail p[rec~x.p/x]}$$
$$\frac{p[rec~x.p/x]\xrightarrow{U}p'}{rec~x.p\xrightarrow{U}p'}$$

The divergence predicate and the convergence predicate can be relativized to sequences of pomsets by:

\begin{enumerate}
  \item $p\downarrow 1$ if $p\downarrow$.
  \item $p\downarrow Us$ if $p\downarrow$ and $p\xRightarrow{U}p'$ implies $p'\downarrow s$.
\end{enumerate}

While $p\uparrow s$ if $p\ndownarrow s$.

\begin{definition}
For $p,q\in\mathbf{M}_3$, let

\begin{enumerate}
  \item $p\ll_{\mathrm{MAY}}q$ if $L(p)\subseteq L(q)$.
  \item $p\ll_{\mathrm{MUST}}q$ if $p\downarrow s$ implies
  \begin{enumerate}
    \item $q\downarrow s$;
    \item $\mathscr{A}(q,s)\subset\subset\mathscr{A}(p,s)$.
  \end{enumerate}
  \item $p\ll q$ if both $p\ll_{\mathrm{MAY}}q$ and $p\ll_{\mathrm{MUST}}q$.
\end{enumerate}
\end{definition}

\begin{theorem}[Alternative characterization for $\mathbf{M}_3$]
For $p,q\in\mathbf{M}_3$,

\begin{enumerate}
  \item $p\pretestingmay q$ if and only if $p\ll_{\mathrm{MAY}}q$.
  \item $p\pretestingmust q$ if and only if $p\ll_{\mathrm{MUST}}q$.
  \item $p\pretesting q$ if and only if $p\ll q$.
\end{enumerate}
\end{theorem}

\subsection{Axiomatic Semantics}\label{as7}

\begin{definition}[Proof system $\mathbf{rDED}(E)$]
A system of inequational deductions by whose inequations in $E$ can be used to derive inequations, denoted $\mathbf{rDED}(E)$. It contains the following eight rules, where $t,t'\in REC_{\Sigma}(X)$, $f\in\Sigma$ and $\sigma$ is a substitution:

\begin{enumerate}
  \item Reflexivity: $$\frac{}{t\leq t}$$
  \item Transitivity: $$\frac{t\leq t',t'\leq t''}{t\leq t''}$$
  \item Substitution: $$a)\quad\frac{t_1\leq t_1',\cdots,t_{ar(f)}\leq t_{ar(f)}'}{f(t_1,\cdots,t_{ar(f)})\leq f(t_1',\cdots,t_{ar(f)}')}$$
      $$b)\quad\frac{t\leq t'}{rec~x.t\leq rec~x.t'}$$
  \item Instantiation: $$\frac{t\leq t'}{t\sigma\leq t'\sigma}$$
  \item Inequations: for every $\langle t,t'\rangle\in E$, $$\frac{}{t\leq t'}$$
  \item Equations: $$\frac{t\leq t',t'\leq t}{t=t'},~~\frac{t=t'}{t\leq t',t'\leq t},~~\frac{t=t'}{t'=t}$$
  \item $\Omega$-rule: $$\Omega\leq x$$
  \item $REC$: $$\frac{}{rec~x.t=t[rec~x.t/x]}$$
\end{enumerate}
\end{definition}

The following is the definition of $\omega$-Induction. The proof system $\mathbf{rDED}(E)$+$\omega$-Induction is denoted $\mathbf{\omega DED}(E)$.

$$\omega\mbox{-Induction}\quad \frac{\mbox{for every }d\in\App(t),d\leq t'}{t\leq t'}$$

We use $t\leq_{E}t'$, $t\leq_{Er}t'$ and $t\leq_{E\omega}t'$ to mean that $t\leq t'$ is derivable in the appropriate proof system. And we use $\vdash_{r}$ and $\vdash_{\omega}$ for $\vdash_{\mathbf{rDED}(\emptyset)}$ and $\vdash_{\mathbf{\omega DED}(\emptyset)}$ respectively.

\begin{lemma}
If $A\in\mathscr{CC}(E)$ then every rule in $\mathbf{\omega DED}(E)$ is sound with respect to $A$.
\end{lemma}

\begin{corollary}
For every $t,t'\in REC_{\Sigma}(X)$, if $\vdash_{\mathbf{\omega DED}(E)}t\leq t'$ then $A\sembrack{t}\leq A\sembrack{t'}$ for every interpretation in $A$ in $\mathscr{CC}(E)$.
\end{corollary}

\begin{lemma}
For every $n\geq 0$, $\vdash_{r}t^{n}\leq t$.
\end{lemma}

\begin{theorem}
For $d\in FREC_{\Sigma}$ and $q\in REC_{\Sigma}$, if $CI_{E}\sembrack{d}\leq CI_{E}\sembrack{q}$, then $\vdash_{\mathbf{rDED}(E)}d\leq q$.
\end{theorem}

\begin{theorem}
The following statements hold:

\begin{enumerate}
  \item $\mathbf{\omega DED}(E)$ is sound with respect to $\leq_{CI_{E}}$ over $REC_{\Sigma}(X)$.
  \item $\mathbf{\omega DED}(E)$ is complete with respect to $\leq_{CI_{E}}$ over $REC_{\Sigma}$.
\end{enumerate}
\end{theorem}

\begin{lemma}
The following statements hold:

\begin{enumerate}
  \item $t=_{\alpha}u$ implies $t=u$ is a theorem in $\mathbf{\omega DED}(E)$.
  \item If $\sigma(x)\leq\sigma'(x)$ is a theorem in $\mathbf{rDED}(E)$ for every $x$, then $t\sigma\leq t\sigma'$ is also a theorem.
\end{enumerate}
\end{lemma}

The following is the definition of Recursion Induction (RI).

$$RI\quad \frac{t[u/x]\leq u}{rec~x.t\leq u}$$

Let $\underline{z}=\langle z_1,\cdots,z_k\rangle$ be $k$ distinct variables and $\underline{u}=\langle u_1,\cdots,u_k\rangle$ be $k$ terms in $REC_{\Sigma}(X)$ with the property that if $z_i\in FV(u_j)$ then $i=j$, i.e., $\underline{u}$ is called noninterfering with respect to $\underline{z}$. And we use $rec~\underline{z}.\underline{u}$ to denote the sequence $\langle rec~z_1.u_1,\cdots,rec~z_k.u_k\rangle$.

The following is the definition of Scott Induction (SI).

SI: Suppose $\underline{u}$ is noninterfering with respect to $\underline{z}$ which is a sequence of distinct variables, If:

\begin{enumerate}
  \item $\vdash t[\underline{\Omega}/\underline{z}]\leq t'[\underline{\Omega}/\underline{z}]$.
  \item $t\leq t'\vdash t[\underline{u}/\underline{z}]\leq t'[\underline{u}\leq\underline{z}]$.
\end{enumerate}

then,

$\vdash t[rec~\underline{z}.\underline{u}/\underline{z}]\leq t'[rec~\underline{z}.\underline{u}/\underline{z}]$.

Let $\mathbf{PS}$ and $\mathbf{PS}'$ be two proof systems, we say that $\mathbf{PS}$ is at least as powerful as $\mathbf{PS}'$, written $\mathbf{PS}'\leq\mathbf{PS}$, if for every $t,t'\in REC_{\Sigma}(X)$,

$\vdash_{\mathbf{PS}'}t\leq t'$ implies $\vdash_{\mathbf{PS}}t\leq t'$.

\begin{proposition}
$$\mathbf{rDED}(E)\leq \mathbf{rDED}(E)+RI\leq\mathbf{rDED}(E)+SI\leq \mathbf{rDED}(E)+\omega\mbox{-Induction}$$
\end{proposition}

The following is the definition called Unique Fixpoint Induction (UFI).

$$UFI\quad \frac{t[u/x]=u}{rec~x.t=u}$$

UFI is unsound in general because it assumes every recursive equation always has a unique fixpoint. The guarded recursion has a unique fixpoint. In the guarded recursive equation $x=t$, every occurrence of $x$ in $t$ is with a prefix $U\cdot$.

\subsection{The Trinity} \label{t7} 

\begin{definition}
The head normal form (hnf) is defined inductively as follows:

\begin{enumerate}
  \item $0$ is an hnf.
  \item If $\mathscr{A}$ is saturated set, any term of the form $\osum\{\sum Up(U)|U\in A,A\in\mathscr{A}\}$ is an hnf.
\end{enumerate}
\end{definition}

\begin{theorem}[Head normal form theorem]
Every convergent term $p$ has an hnf, i.e., $p\downarrow$ implies $\vdash_{\mathbf{rDED}(E^1)}p=h$, where $h$ is an hnf.
\end{theorem}

\begin{definition}
The $\Omega$ normal form ($\Omega$-nf) is defined inductively as follows:

\begin{enumerate}
  \item $0$ and $\Omega$ are $\Omega$-nf's.
  \item If $\mathscr{A}$ is saturated set, and $d(U)$ is an $\Omega$-nf for every $U\in S(\mathscr{A})$ then $\osum\{d(A)|A\in\mathscr{A}\}$ is also an $\Omega$-nf.
\end{enumerate}
\end{definition}

\begin{corollary}
The following statements hold:

\begin{enumerate}
  \item If $d\in FREC_{\Sigma^3}$ and $d\uparrow$, then $d=_{E^2_{\mathbf{S}}}\Omega$.
  \item Every term in $FREC_{\Sigma^3}$ has an $\Omega$-nf, i.e., for every $d\in FREC_{\Sigma^3}$ there exists and $\Omega$-nf $n$ such that $d=_{E^2_{\mathbf{S}}}n$.
\end{enumerate}
\end{corollary}

\begin{proposition}
$\mathbf{\Omega DED}(E^2_{\mathbf{S}})$ is sound and complete with respect to $\pretestingmust$ over $FREC_{\Sigma^3}$.
\end{proposition}

\begin{lemma}
If $e$ is a finite experiment and $p\must e$ then $d\must e$ for some $d\in\App(p)$.
\end{lemma}

\begin{lemma}
For $d\in FREC_{\Sigma^3}$ and $p\in\mathbf{M}_3$, if $d\pretestingmust p$ then $d\pretestingmust d'$ for some $d'\in\App(p)$.
\end{lemma}

\begin{proposition}
$\pretestingmust$ is algebraic over $\mathbf{M}_3$.
\end{proposition}

\begin{proposition}
$\leq_{E^2_{\mathbf{S}}\omega}$ is algebraic over $\mathbf{M}_3$, where $t\leq_{E^2_{\mathbf{S}}\omega}t'$ denotes $\vdash_{\mathbf{\omega DED}(E^2_{\mathbf{S}})}t\leq t'$ for some $t,t'\in\mathbf{M}_3$.
\end{proposition}

\begin{theorem}[Full Abstraction]
The interpretation $\mathbf{PAT}_{\mathbf{S}}$ is fully abstract with respect to $\pretestingmust$ over $\mathbf{M}_3$.
\end{theorem}

\begin{theorem}[Soundness and Completeness]
The proof system $\mathbf{\omega DED}(E^2_{\mathbf{S}})$ is sound and complete with respect to $\pretestingmust$ over $REC_{\Sigma^3}$.
\end{theorem}

\begin{theorem}[Full Abstraction]
The interpretation $\mathbf{PAT}$ is fully abstract with respect to $\pretesting$ over $\mathbf{M}_3$.
\end{theorem}

\begin{theorem}[Soundness and Completeness]
The proof system $\mathbf{\omega DED}(E^2)$ is sound and complete with respect to $\pretesting$ over $REC_{\Sigma^3}$.
\end{theorem}

\begin{theorem}[Full Abstraction]
The interpretation $\mathbf{PAT}_{\mathbf{W}}$ is fully abstract with respect to $\pretestingmay$ over $\mathbf{M}_3$.
\end{theorem}

\begin{theorem}[Soundness and Completeness]
The proof system $\mathbf{\omega DED}(E^2_{\mathbf{W}})$ is sound and complete with respect to $\pretestingmay$ over $REC_{\Sigma^3}$.
\end{theorem} 
\newpage\section{Abstraction}\label{absT}

In this chapter, we introduce abstraction. Firstly, we introduce the whole signature of this book in \cref{ws8}, then the operational semantics, axiomatic semantics and denotational semantics of the processes are introduced in \cref{os8}, \cref{as8} and \cref{ds8}, respectively. Finally, we get the results on trinity of operational semantics, denotational semantics and axiomatic semantics in \cref{t8}.

\subsection{Whole Signature}\label{ws8}

\begin{definition}[Whole signature]
The whole signature $\Sigma^4$ consists of:

\begin{enumerate}
  \item $\Sigma^4\supset \Sigma^3$.
  \item The unary operator $\Abs_I$ as the abstraction, i.e., for process $p$, the process $\Abs_I(p)$ renames all actions of $p$ in the set $I$ to $1$.
\end{enumerate}
\end{definition}

\begin{definition}[Syntax of process language]
The syntax of the process language is given by the following BNF grammar:

$p::=\Omega~|~0~|~1~|~U~|~\gamma(a,b)~|~p\cdot p~|~a\sharp b~|~a\osharp b~|~x~|~p+p~|~p\oplus p~|~p\parallel p~|~p\mid p~|~p\between p~|~\Theta(p)~|~p\triangleleft p~|~\partial_H(p)~|~rec~x.p~|~\Abs_I(p)$

where $a,b\in\mathsf{Act}$, $U\in\mathsf{Pom}$, $p\in \mathsf{Proc}$, $x$ is recursive variable and $rec~x.p$ stands for the process defined by the recursive equation $x=p$.
\end{definition}

\subsection{Operational Semantics}\label{os8}

The PTSSs of $0$, $1$, $\Omega$, $U$, $\cdot$, $+$, $\sharp$, $\oplus$, $\osharp$, $\parallel$, $\mid$, $\between$, $\Theta$, $\triangleleft$ and $rec~x.p$ are same as the ones in \cref{os7}. The PTSS of $\Abs_I$ is as follows.

$$\frac{p\downarrow}{\Abs_I(p)\downarrow}$$
$$\frac{p\rightarrowtail\surd}{\Abs_I(p)\rightarrowtail\surd}\quad\frac{p\rightarrowtail p'}{\Abs_I(p)\rightarrowtail \Abs_I(p')}$$
$$\frac{p\xrightarrow{a}\surd\quad a\notin I}{\Abs_I(p)\xrightarrow{a}\surd}\quad\frac{p\xrightarrow{a}p'\quad a\notin I}{\Abs_I(p)\xrightarrow{a} \Abs_I(p')}$$

We continue to use the preorders $\pretestingmust$, $\pretestingmay$, $\pretesting$ and $\ll_{\mathrm{MUST}}$, $\ll_{\mathrm{MAY}}$, $\ll$, and we can get the alternative characterization theorem for $\mathbf{M}_4$.

\begin{theorem}[Alternative characterization for $\mathbf{M}_4$]
For $p,q\in\mathbf{M}_4$,

\begin{enumerate}
  \item $p\pretestingmay q$ if and only if $p\ll_{\mathrm{MAY}}q$.
  \item $p\pretestingmust q$ if and only if $p\ll_{\mathrm{MUST}}q$.
  \item $p\pretesting q$ if and only if $p\ll q$.
\end{enumerate}
\end{theorem}

\begin{corollary}
The preorders $\pretestingmust$, $\pretestingmay$, $\pretesting$ are preserved by all the operators of $\mathbf{M}_4$.
\end{corollary}

\subsection{Axiomatic Semantics}\label{as8}

Let $\Sigma$ be an arbitrary signature containing $\Sigma^1$, then $REC_{\Sigma}$ contains $\mathbf{M}_3$, which is given an operational semantics by extending that of $\mathbf{M}_3$. We continue to use the preorders $\pretestingmust$, $\pretestingmay$, $\pretesting$ and $\ll_{\mathrm{MUST}}$, $\ll_{\mathrm{MAY}}$, $\ll$, and we can get the alternative characterization theorem like $\mathbf{M}_4$ in \cref{os8}. And $E^2_{\mathbf{S}}$ is augmented to a new set of inequations $E$, in the following way, the extended proof system $\mathbf{\omega DED}(E)$ is sound and complete, i.e., by eliminating all occurrences of function symbols in $\Sigma$ which are not in $\Sigma^1$.

\begin{definition}
The $\Sigma^1$-head term is defined inductively as follows:

\begin{enumerate}
  \item $0$ is a head term.
  \item If $f\in \Sigma^1$ and $p$ are head terms then $f(p)$ is also a head term.
\end{enumerate}
\end{definition}

\begin{definition}[Reductivity]
Let $E$ be a set of inequations over $\Sigma$ containing $E^2_{\mathbf{S}}$. We say $E$ is reductive if it satisfies:

\begin{enumerate}
  \item For every $d\in FREC_{\Sigma}$ there exists a $d'\in FREC_{\Sigma^1}$ such that $d=_{E}d'$.
  \item For every $p\in REC_{\Sigma}$, $p\downarrow$ implies $p=_{Er}q$ for some head term $q$.
\end{enumerate}

Such an $E$ is called normal.
\end{definition}

\begin{theorem}[Reduction theorem]
If $E$ is reductive and $\mathbf{\Omega DED}(E)$ is sound with respect to $\pretestingmust$ over $REC_{\Sigma}$ then $\mathbf{\omega DED}(E)$ is both sound and complete with respect to $\pretestingmust$ over $REC_{\Sigma}$.
\end{theorem}

Now, let us apply the reduction theorem to $\mathbf{M}_4$.

The proof system $\mathbf{\omega DED}(E^4_{\mathbf{S}})$ contains the inequations denoted $E^4_{\mathbf{S}}$, which include $E^2_{\mathbf{S}}$ in \cref{as6} and the inequations in \cref{AxiomsofAbsI}.

\begin{center}
    \begin{longtable}{|c|c|}
      \caption{Inequations of $\Abs_I$}\\
      \hline No. &Axiom\\
      \hline
      \endfirsthead
      \multicolumn{2}{c}{\bfseries Continuing: Inequations of $\Abs_I$} \\
      \hline No. &Axiom\\
      \hline
      \endhead
      \hline
      \multicolumn{2}{r}{to be continued\ldots} \\
      \endfoot
      \hline
      \multicolumn{2}{r}{end of Inquations of $\Abs_I$} \\
      \endlastfoot
      $ABS1$ & $\Abs_I(0)=0$\\
      $ABS2$ & $\Abs_I(1)=1$\\
      $ABS3$ & $\Abs_I(a)=1\quad a\in I$\\
      $ABS4$ & $\Abs_I(a)=a\quad a\notin I$\\
      $ABS5$ & $\Abs_I(x\cdot y)=\Abs_I(x)\cdot\Abs_I(y)$\\
      $ABS6$ & $\Abs_I(x+y)=\Abs_I(x)+\Abs_I(y)$\\
      $ABS7$ & $\Abs_I(x\oplus y)=\Abs_I(x)\oplus\Abs_I(y)$\\
      $ABS8$ & $\Abs_I(x\parallel y)=\Abs_I(x)\parallel\Abs_I(y)$\\
      $ABS9$ & $\Abs_I(\Abs_I(x))=\Abs_I(x)$\\
      $ABS10$ & $\Abs_{I_1}(\Abs_{I_2}(x))=\Abs_{I_2}(\Abs_{I_1}(x))$\\
      $\Omega11$ & $\Abs_I(\Omega) \leq \Omega$
      \label{AxiomsofAbsI}
    \end{longtable}
\end{center}

\begin{lemma}
$\mathbf{\Omega DED}(E^4_{\mathbf{S}})$ is sound with respect to $\pretestingmust$ over $REC_{\Sigma^4}$.
\end{lemma}

\begin{theorem}[Soundness and Completeness]
The proof system $\mathbf{\omega DED}(E^4_{\mathbf{S}})$ is sound and complete with respect to $\pretestingmust$ over $REC_{\Sigma^4}$.
\end{theorem}

\subsection{Denotational Semantics} \label{ds8}

\begin{proposition}
If $E$ is normal then $CI_{E}$ is fully abstract with respect to $\pretestingmust$ over $REC_{\Sigma}$.
\end{proposition}

Let $\langle \mathbf{PAT}_{\mathbf{S}},\leq_{\mathbf{PAT}_{\mathbf{S}}},\Sigma^4\rangle$ be a domain, we need to show for each $f\in\Sigma^4$ there is a continuous function $f_{\Sigma^4}$ over $\mathbf{PAT}_{\mathbf{S}}$. Such a domain is called an extension of $\mathbf{PAT}_{\mathbf{S}}$ if for every $f\in \Sigma^1$, $f_{\Sigma^4}$ coincides with $f_{\mathbf{PAT}_{\mathbf{S}}}$ defined in \cref{omegaT}. 

\begin{theorem}
If $E$ is normal then there is a unique extension of $\mathbf{PAT}_{\mathbf{S}}$, up to isomorphism, which is fully abstract with respect to $\pretestingmust$ over $REC_{\Sigma}$.
\end{theorem}

\begin{corollary}
If $A$ is any extension of $\mathbf{PAT}_{\mathbf{S}}$ in $\mathscr{CC}(E)$, where $E$ is normal, then it is fully abstract with respect to $\pretestingmust$ over $REC_{\Sigma}$.
\end{corollary}

So, we only consider the new function symbols in $\Sigma^4$, i.e., $\Abs_I$.

$\Abs_I(1)=1$

$\Abs_I(\step{a,s})=\begin{cases}
                      \Abs_I(s), & a\in I;\\
                      \step{a,\Abs_I(s)}, & \mbox{otherwise}.
                    \end{cases}$
                    
$\Abs_I(as)=\begin{cases}
                      \Abs_I(s), & a\in I;\\
                      a\Abs_I(s), & \mbox{otherwise}.
                    \end{cases}$
                    
Let $\Abs_I(L)=\{\Abs_I(s)|s\in L\}$, $\Abs_I(U)=U\setminus I$ for $U\in\mathsf{Pom}$, $\Abs_I(A)=A\setminus I$ for $A\subseteq\mathsf{Act}$, and $\Abs_I(\mathscr{A})=\{\Abs_I(A)|A\in\mathscr{A}\}$ for $\mathscr{A}$ a collection of such sets $A$.

Then the tree $\Abs_I(t)$ is defined as:

\begin{enumerate}
  \item $L(\Abs_I(t))=\Abs_I(L(t))$.
  \item $CL(\Abs_I(t))=\Abs_I(CL(t))$.
  \item $\mathscr{A}(\Abs_I(t)(s))=\Abs_I(\mathscr{A}(t(s)))$.
\end{enumerate}

\begin{lemma}
The function $\Abs_I\in [\mathbf{PAT}_{\mathbf{S}}\rightarrow\mathbf{PAT}_{\mathbf{S}}]$.
\end{lemma}

\begin{theorem}[Full Abstraction]
The interpretation $\mathbf{PAT}_{\mathbf{S}}$ is fully abstract with respect to $\pretestingmust$ over $\mathbf{M}_4$.
\end{theorem}

\subsection{The Trinity}\label{t8}

The proof system $\mathbf{\omega DED}(E^4)$ contains the inequations denoted $E^4$, which include $E^2$ in \cref{t6} and the inequations in \cref{AxiomsofAbsI}.

\begin{theorem}[Full Abstraction]
The interpretation $\mathbf{PAT}$ is fully abstract with respect to $\pretesting$ over $\mathbf{M}_4$.
\end{theorem}

\begin{theorem}[Soundness and Completeness]
The proof system $\mathbf{\omega DED}(E^4)$ is sound and complete with respect to $\pretesting$ over $REC_{\Sigma^4}$.
\end{theorem}

The proof system $\mathbf{\omega DED}(E^4_{\mathbf{W}})$ contains the inequations denoted $E^4_{\mathbf{W}}$, which include $E^2_{\mathbf{W}}$ in \cref{t6} and the inequations in \cref{AxiomsofAbsI}.

\begin{theorem}[Full Abstraction]
The interpretation $\mathbf{PAT}_{\mathbf{W}}$ is fully abstract with respect to $\pretestingmay$ over $\mathbf{M}_4$.
\end{theorem}

\begin{theorem}[Soundness and Completeness]
The proof system $\mathbf{\omega DED}(E^4_{\mathbf{W}})$ is sound and complete with respect to $\pretestingmay$ over $REC_{\Sigma^4}$.
\end{theorem}

\newpage\section{Bisimulation Semantics}\label{bs} 

In this chapter, we introduce the theory of truly concurrent processes based on bisimulation semantics. Firstly, we introduce the truly concurrent bisimulations in \cref{tcb} and the signature in \cref{ts9}, then the operational semantics, denotational semantics and axiomatic semantics of the processes are introduced in \cref{os9}, \cref{ds9} and \cref{as9}, respectively. Finally, we get the results on trinity of operational semantics, denotational semantics and axiomatic semantics in \cref{t9}.

\subsection{Truly Concurrent Bisimulations}\label{tcb}

Let us recall the definition of PLTS in \cref{pomlts}.

\pomsetlts*

For $n\geq 0$ and $s=U_1\cdots U_n\in\mathsf{Pom}^*$, then $p\xrightarrow{s}q$ if and only if there exist processes $p_0,\cdots,p_n$ such that $p=p_0\xrightarrow{U_1}p_1\xrightarrow{U_2}\cdots\xrightarrow{U_{n-1}}p_{n-1}\xrightarrow{U_n}p_n=q$. For $p\in\mathsf{Proc}$ and $U\in\mathsf{Pom}$, we define:

$$\mathsf{initials}(p)\triangleq\{U\in\mathsf{Pom}|\exists q:p\xrightarrow{U}q\}$$
$$\mathsf{sort}(p)\triangleq\{U\in\mathsf{Pom}|\exists s\in\mathsf{Pom}^*,p',p''\in\mathsf{Proc}:p\xrightarrow{s}p'\xrightarrow{U}p''\}$$
$$\mathsf{derivatives}(p,U)\triangleq\{q|p\xrightarrow{U}q\}$$

A PLTS is sort-finite if and only if $\mathsf{sort}(p)$ is finite for every $p\in\mathsf{Proc}$.

\begin{definition}[Pomset synchronization tree]
The set of countably branching pomset synchronization trees over $\mathsf{Pom}$, denoted $\mathbf{PST}_{\infty}(\mathsf{Pom})$, is the set of infinitary terms generated by the following inductive definition:

$$\frac{\{U_i\in\mathsf{Pom},t_i\in\mathbf{PST}_{\infty}(\mathsf{Pom})\}_{i\in I}}{\sum_{i\in I}U_i:t_i[+\Omega]\in\mathbf{PST}_{\infty}(\mathsf{Pom})}$$

where $I$ is a countable index set, $0=\sum_{i\in\emptyset}U_i:t_i$ which stands for the one-node synchronization tree, a representation of an inactive process, $\Omega=\sum_{i\in\emptyset}U_i:t_i+\Omega$ which stands that the behaviours of the synchronization tree are completely unspecified, $[+\Omega]$ means optional inclusion of $\Omega$ as a summand.
\end{definition} 

Let $\mathbf{PST}(\mathsf{Pom})$ denote the finite synchronization tree, i.e., the above set of terms is built using only finite summations. The set of synchronization trees $\mathbf{PST}_{\infty}(\mathsf{Pom})$ can be turned into a PLTS with divergence by stipulating that, for $t\in\mathbf{PST}_{\infty}(\mathsf{Pom})$:

\begin{itemize}
  \item $t\uparrow$ if and only if $\Omega$ is a summand of $t$.
  \item $t\xrightarrow{U_i}t_i$ if and only if $U_i:t_i$ is a summand of $t$.
\end{itemize} 

In the following, we give the definitions of truly concurrent bisimulations.

\begin{definition}[Configuration]
Let $\mathcal{P}=\langle\mathsf{Proc},\mathsf{Act},\{\xrightarrow{U}|U\in\mathsf{Pom}\},\mathsf{Pred}\rangle$ be a PLTS. A (finite) configuration in $\mathcal{P}$ is a (finite) consistent subset of actions $C\subseteq \mathcal{P}$, closed with respect to $\xrightarrow{U}$ (i.e. $\lceil C\rceil=C$). The set of finite configurations of $\mathcal{P}$ is denoted by $\mathcal{C}(\mathcal{P})$. For $p\in\mathsf{Proc}$, the corresponding configuration is denoted $C(p)$.
\end{definition}

\begin{definition}[Pomset, step bisimulation]\label{PSB}
Let $\langle\mathsf{Proc},\mathsf{Act},\{\xrightarrow{U}|U\in\mathsf{Pom}\},\mathsf{Pred}\rangle$ be a PLTS and $p,q\in\mathsf{Proc}$. A pomset bisimulation is a relation $R\subseteq\mathsf{Proc}\times\mathsf{Proc}$, such that:

\begin{enumerate}
  \item If $(p,q)\in R$, and $C(p)\xrightarrow{U}C(p')$ then $C(q)\xrightarrow{U}C(q')$, with $U\in \mathsf{Pom}$, and $(p',q')\in R$.
  \item If $(p,q)\in R$, and $C(q)\xrightarrow{U}C(q')$ then $C(p)\xrightarrow{U}C(p')$, with $U\in \mathsf{Pom}$, and $(p',q')\in R$.
\end{enumerate}

We say that $p$, $q$ are pomset bisimilar, written $p\sim_P q$, if there exists a pomset bisimulation $R$, such that $(p,q)\in R$. By replacing pomset transitions with steps, we can get the definition of step bisimulation. When $p$ and $q$ are step bisimilar, we write $p\sim_S q$.
\end{definition}

\begin{definition}[Posetal product]
Let $\langle\mathsf{Proc},\mathsf{Act},\{\xrightarrow{U}|U\in\mathsf{Pom}\},\mathsf{Pred}\rangle$ be a PLTS. Given two processes $p,q\in\mathsf{Proc}$, the posetal product of their configurations, denoted $\mathcal{C}(\mathcal{P})\overline{\times}\mathcal{C}(\mathcal{P})$, is defined as

$$\{(C(p),f,C(q))|C(p)\in\mathcal{C}(\mathcal{P}),C(q)\in\mathcal{C}(\mathcal{P}),f:C(p)\rightarrow C(q) \mbox{ isomorphism}\}$$

A subset $R\subseteq\mathcal{C}(\mathcal{P})\overline{\times}\mathcal{C}(\mathcal{P})$ is called a posetal relation. We say that $R$ is downward closed when for any $(C(p),f,C(q)),(C(p'),f',C(q'))\in \mathcal{C}(\mathcal{P})\overline{\times}\mathcal{C}(\mathcal{P})$, if $(C(p),f,C(q))\subseteq (C(p'),f',C(q'))$ pointwise and $(C(p'),f',C(q'))\in R$, then $(C(p),f,C(q))\in R$.

For $f:U\rightarrow U$, we define $f[a\mapsto a]:U\cup\{a\}\rightarrow U\cup\{a\}$, $z\in U\cup\{a\}$,(1)$f[a\mapsto a](z)=a$,if $z=a$;(2)$f[a\mapsto a](z)=f(z)$, otherwise. Where $U\in \mathsf{Pom}$, $a\in \mathsf{Act}$.
\end{definition}

\begin{definition}[(Hereditary) history-preserving bisimulation]\label{HHPB}
Let $\mathcal{P}=\langle\mathsf{Proc},\mathsf{Act},\{\xrightarrow{U}|U\in\mathsf{Pom}\},\mathsf{Pred}\rangle$ be a PLTS and $p,q\in\mathsf{Proc}$. A history-preserving (hp-) bisimulation is a posetal relation $R\subseteq\mathcal{C}(\mathcal{P})\overline{\times}\mathcal{C}(\mathcal{P})$ such that:

\begin{enumerate}
  \item If $(C(p),f,C(q))\in R$, and $C(p)\xrightarrow{a} C(p')$, then $C(q)\xrightarrow{a} C(q')$, with $(C(p'),f[a\mapsto a],C(q'))\in R$.
  \item If $(C(p),f,C(q))\in R$, and $C(q)\xrightarrow{a} C(q')$, then $C(p)\xrightarrow{a} C(p')$, with $(C(p'),f[a\mapsto a],C(q'))\in R$.
\end{enumerate}

$p,q$ are history-preserving (hp-)bisimilar and are written $p\sim_{HP}q$ if there exists an hp-bisimulation $R$ such that $(\emptyset,\emptyset,\emptyset)\in R$.

A hereditary history-preserving (hhp-)bisimulation is a downward closed hp-bisimulation. $p,q$ are hereditary history-preserving (hhp-)bisimilar and are written $p\sim_{HHP}q$.
\end{definition}

In the following, we give the definitions of truly concurrent prebisimulations.

\begin{definition}[Pomset, step prebisimulation]\label{PSPB}
Let $\langle\mathsf{Proc},\mathsf{Act},\{\xrightarrow{U}|U\in\mathsf{Pom}\},\mathsf{Pred}\rangle$ be a PLTS with divergence $\uparrow$ and $p,q\in\mathsf{Proc}$. Let $\mathsf{Rel}(\mathsf{Proc})$ denote the set of binary relations over $\mathsf{Proc}$. Define the functional $F:\mathsf{Rel}(\mathsf{Proc})\rightarrow\mathsf{Rel}(\mathsf{Proc})$, such that:

\begin{align*}
\begin{aligned}
F(R)=&\{(p,q)|\forall U\in\mathsf{Pom}\\
  &\bullet C(p)\xrightarrow{U}C(p')\mbox{ implies } C(q)\xrightarrow{U}C(q'), \mbox{ and }(p',q')\in R\\
  &\bullet p\downarrow\mbox{ implies }\bigg(q\downarrow\mbox{ and }\Big(C(q)\xrightarrow{U}C(q')\mbox{ implies } C(p)\xrightarrow{U}C(p'),\mbox{ and }(p',q')\in R\Big)\bigg)\}\\
\end{aligned}
\end{align*}

A relation $R$ is a pomset prebisimulation if and only if $R\subseteq F(R)$. We say that $p$, $q$ are pomset prebisimilar, written $p\prepomset q$, if there exists a pomset prebisimulation $R$, such that $(p,q)\in R$. By replacing pomset transitions with steps, we can get the definition of step prebisimulation. When $p$ and $q$ are step prebisimilar, we write $p\prestep q$.

The kernels of $\prepomset$ and $\prestep$ are denoted $\prepomseteq$ and $\prestepeq$.
\end{definition}

\begin{definition}[(Hereditary) history-preserving prebisimulation]\label{HHPPB}
Let $\langle\mathsf{Proc},\mathsf{Act},\{\xrightarrow{U}|U\in\mathsf{Pom}\},\mathsf{Pred}\rangle$ be a PLTS with divergence $\uparrow$ and $p,q\in\mathsf{Proc}$. Let $\mathsf{Rel}(\mathcal{C}(\mathcal{P})\overline{\times}\mathcal{C}(\mathcal{P}))$ denote the set of posetal relations over $\mathsf{Proc}$. Define the functional $F:\mathsf{Rel}(\mathcal{C}(\mathcal{P})\overline{\times}\mathcal{C}(\mathcal{P}))\rightarrow\mathsf{Rel}(\mathcal{C}(\mathcal{P})\overline{\times}\mathcal{C}(\mathcal{P}))$, such that:

\begin{align*}
\begin{aligned}
F(R)=&\{(C(p),f,C(q))|\forall a\in\mathsf{Act}\\
  &\bullet C(p)\xrightarrow{a}C(p')\mbox{ implies } C(q)\xrightarrow{a}C(q'), \mbox{ and }(C(p'),f[a\mapsto a],C(q'))\in R\\
  &\bullet p\downarrow\mbox{ implies }\\
  &\bigg(q\downarrow\mbox{ and }\Big(C(q)\xrightarrow{a}C(q')\mbox{ implies } C(p)\xrightarrow{a}C(p'),\mbox{ and }(C(p'),f[a\mapsto a],C(q'))\in R\Big)\bigg)\}\\
\end{aligned}
\end{align*}

A relation $R$ is an hp-prebisimulation if and only if $R\subseteq F(R)$. We say that $p$, $q$ are hp-prebisimilar, written $p\prehp q$, if there exists an hp-prebisimulation $R$, such that $(\emptyset,\emptyset,\emptyset)\in R$. A hereditary history-preserving (hhp-)prebisimulation is a downward closed hp-prebisimulation. $p,q$ are hereditary history-preserving (hhp-)prebisimilar and are written $p\prehhp q$.

The kernels of $\prehp$ and $\prehhp$ are denoted $\prehpeq$ and $\prehhpeq$.
\end{definition}

By use of the functional $F$ in \cref{PSPB} and \cref{HHPPB}, we can obtain a pomset behavioural preorder which applies itself inductively as follows:

$$\prepomsets{0}\triangleq\mathsf{Proc}\times\mathsf{Proc}$$
$$\prepomsets{n+1}\triangleq F(\prepomsets{n})$$
$$\prepomsets{\omega}\triangleq\bigcap_{n\geq 0}\prepomsets{n}$$

Similarly, we can get the definitions of $\presteps{\omega}$, $\prehps{\omega}$ and $\prehhps{\omega}$.

\begin{definition}[Finitary pomset prebisimulation]
Let $\langle\mathsf{Proc},\mathsf{Act},\{\xrightarrow{U}|U\in\mathsf{Pom}\},\mathsf{Pred}\rangle$ be a PLTS and $p,q\in\mathsf{Proc}$. The finitary pomset prebisimulation $\prepomset^F$ over $\mathsf{Proc}$ is defined as follows: $p\prepomset^F q$ if and only if, for every $t\in\mathbf{PST}(\mathsf{Pom})$, $t\prepomset p$ implies $t\prepomset q$.
\end{definition} 

\begin{definition}[Finitary step prebisimulation]
Let $\langle\mathsf{Proc},\mathsf{Act},\{\xrightarrow{U}|U\in\mathsf{Pom}\},\mathsf{Pred}\rangle$ be a PLTS and $p,q\in\mathsf{Proc}$. The finitary step prebisimulation $\prestep^F$ over $\mathsf{Proc}$ is defined as follows: $p\prestep^F q$ if and only if, for every $t\in\mathbf{PST}(\mathsf{Pom})$, $t\prestep p$ implies $t\prestep q$.
\end{definition}

\begin{definition}[Finitary hp-prebisimulation]
Let $\langle\mathsf{Proc},\mathsf{Act},\{\xrightarrow{U}|U\in\mathsf{Pom}\},\mathsf{Pred}\rangle$ be a PLTS and $p,q\in\mathsf{Proc}$. The finitary hp-prebisimulation $\prehp^F$ over $\mathsf{Proc}$ is defined as follows: $p\prehp^F q$ if and only if, for every $t\in\mathbf{PST}(\mathsf{Pom})$, $t\prehp p$ implies $t\prehp q$.
\end{definition}

\begin{definition}[Finitary hhp-prebisimulation]
Let $\langle\mathsf{Proc},\mathsf{Act},\{\xrightarrow{U}|U\in\mathsf{Pom}\},\mathsf{Pred}\rangle$ be a PLTS and $p,q\in\mathsf{Proc}$. The finitary hhp-prebisimulation $\prehhp^F$ over $\mathsf{Proc}$ is defined as follows: $p\prehhp^F q$ if and only if, for every $t\in\mathbf{PST}(\mathsf{Pom})$, $t\prehhp p$ implies $t\prehhp q$.
\end{definition} 

They are obvious that $\prepomset~\subseteq~\prepomsets{\omega}~\subseteq~\prepomset^F$, $\prestep~\subseteq~\presteps{\omega}~\subseteq~\prestep^F$, $\prehp~\subseteq~\prehps{\omega}~\subseteq~\prehp^F$ and $\prehhp~\subseteq~\prehhps{\omega}~\subseteq~\prehhp^F$. Moreover, the inclusions are strict for infinitely branching PLTSs, and collapse to equalities for finitely branching PLTSs. 

Let $\langle\mathsf{Proc},\mathsf{Act},\{\xrightarrow{U}|U\in\mathsf{Pom}\},\mathsf{Pred}\rangle$ be a PLTS and $p,q\in\mathsf{Proc}$. For every $P\subseteq\mathsf{Pom}$, we define the following relations.

$\{\pretesting^{P}_{(P,n)}|n\geq 0\}$

$p\pretesting^{P}_{(P,0)}q\mbox{ if and only if } \mathrm{true}$
\begin{align*}
\begin{aligned}
&p\pretesting^{P}_{(P,n+1)}q\mbox{ if and only if }\\
  &\bullet \forall U\in P,C(p)\xrightarrow{U}C(p')\mbox{ implies } C(q)\xrightarrow{U}C(q'), \mbox{ and }p'\pretesting^{P}_{(P,n)}q'\\
  &\bullet \mathsf{initials(p)}\subseteq P\mbox{ and }p\downarrow\mbox{ imply }\\
  &\bigg(\mathsf{initials(q)}\subseteq P\mbox,q\downarrow,\Big(\forall U\in P, C(q)\xrightarrow{U}C(q')\mbox{ implies } C(p)\xrightarrow{U}C(p'),\mbox{ and }p'\pretesting^{P}_{(P,n)}q'\Big)\bigg)\\
\end{aligned}
\end{align*}

Similarly, we can get the definition of $\{\pretesting^{P}_{(S,n)}|n\geq 0\}$, $\{\pretesting^{P}_{(HP,n)}|n\geq 0\}$ and $\{\pretesting^{P}_{(HHP,n)}|n\geq 0\}$.

\begin{proposition}
For every $n\geq 0$ and $P\subseteq\mathsf{Pom}$, the following statements hold:

\begin{enumerate}
  \item $\pretesting^{P}_{(P,n)}$, $\pretesting^{P}_{(S,n)}$, $\pretesting^{P}_{(HP,n)}$ and $\pretesting^{P}_{(HHP,n)}$ are all preorders.
  \item For $p,q\in\mathsf{Proc}$, 
  \begin{enumerate}
    \item $p\pretesting^{P}_{(P,n+1)}q\mbox{ implies } p\pretesting^{P}_{(P,n)}q$.
    \item $p\pretesting^{P}_{(S,n+1)}q\mbox{ implies } p\pretesting^{P}_{(S,n)}q$.
    \item $p\pretesting^{P}_{(HP,n+1)}q\mbox{ implies } p\pretesting^{P}_{(HP,n)}q$.
    \item $p\pretesting^{P}_{(HHP,n+1)}q\mbox{ implies } p\pretesting^{P}_{(HHP,n)}q$.
  \end{enumerate}
  \item Assume that $P\subseteq P'\subseteq\mathsf{Pom}$, then for $p,q\in\mathsf{Proc}$,
  \begin{enumerate}
    \item $p\pretesting^{P'}_{(P,n)}q\mbox{ implies } p\pretesting^{P}_{(P,n)}q$.
    \item $p\pretesting^{P'}_{(S,n)}q\mbox{ implies } p\pretesting^{P}_{(S,n)}q$.
    \item $p\pretesting^{P'}_{(HP,n)}q\mbox{ implies } p\pretesting^{P}_{(HP,n)}q$.
    \item $p\pretesting^{P'}_{(HHP,n)}q\mbox{ implies } p\pretesting^{P}_{(HHP,n)}q$.
  \end{enumerate}
\end{enumerate}
\end{proposition}

We define:

$$\pretesting^{P}_{(P,\omega)}\triangleq\bigcap_{n\geq 0}\pretesting^{P}_{(P,n)}$$
$$\pretesting^{P}_{(S,\omega)}\triangleq\bigcap_{n\geq 0}\pretesting^{P}_{(S,n)}$$
$$\pretesting^{P}_{(HP,\omega)}\triangleq\bigcap_{n\geq 0}\pretesting^{P}_{(HP,n)}$$
$$\pretesting^{P}_{(HHP,\omega)}\triangleq\bigcap_{n\geq 0}\pretesting^{P}_{(HHP,n)}$$
$$p\pretesting^{fin}_{(P,\omega)}q\triangleq\forall P\subseteq_{fin}\mathsf{Pom}.p\pretesting^{P}_{(P,\omega)}q$$
$$p\pretesting^{fin}_{(S,\omega)}q\triangleq\forall P\subseteq_{fin}\mathsf{Pom}.p\pretesting^{P}_{(S,\omega)}q$$
$$p\pretesting^{fin}_{(HP,\omega)}q\triangleq\forall P\subseteq_{fin}\mathsf{Pom}.p\pretesting^{P}_{(HP,\omega)}q$$
$$p\pretesting^{fin}_{(HHP,\omega)}q\triangleq\forall P\subseteq_{fin}\mathsf{Pom}.p\pretesting^{P}_{(HHP,\omega)}q$$

where $P\subseteq_{fin}\mathsf{Pom}$ means that $P$ is a finite subset of $\mathsf{Pom}$.

\begin{lemma}
For every $t\in\mathbf{PST}(\mathsf{Pom})$ and $p\in\mathsf{Proc}$, the following statements hold:

\begin{enumerate}
  \item $t\prepomset p\mbox{ if and only if } t\pretesting^{fin}_{(P,\omega)}p$.
  \item $t\prestep p\mbox{ if and only if } t\pretesting^{fin}_{(S,\omega)}p$.
  \item $t\prehp p\mbox{ if and only if } t\pretesting^{fin}_{(HP,\omega)}p$.
  \item $t\prehhp p\mbox{ if and only if } t\pretesting^{fin}_{(HHP,\omega)}p$.
\end{enumerate}
\end{lemma}

\begin{lemma}
The following statements hold:

\begin{enumerate}
  \item The preorder $\pretesting^{fin}_{(P,\omega)}$ is finitary.
  \item The preorder $\pretesting^{fin}_{(S,\omega)}$ is finitary.
  \item The preorder $\pretesting^{fin}_{(HP,\omega)}$ is finitary.
  \item The preorder $\pretesting^{fin}_{(HHP,\omega)}$ is finitary.
\end{enumerate}
\end{lemma}

\begin{theorem}
For $p,q\in\mathsf{Proc}$ in any transition system, the following statements hold:

\begin{enumerate}
  \item $p\prepomset^{F} q\mbox{ if and only if } p\pretesting^{fin}_{(P,\omega)}q$.
  \item $p\prestep^{F} q\mbox{ if and only if } p\pretesting^{fin}_{(S,\omega)}q$.
  \item $p\prehp^{F} q\mbox{ if and only if } p\pretesting^{fin}_{(HP,\omega)}q$.
  \item $p\prehhp^{F} q\mbox{ if and only if } p\pretesting^{fin}_{(HHP,\omega)}q$.
\end{enumerate}
\end{theorem}

\begin{corollary}
For $p,q\in\mathsf{Proc}$ in any sort-finite transition system, the following statements hold:

\begin{enumerate}
  \item $p\prepomset^{F} q\mbox{ if and only if } p\prepomsets{\omega}q$.
  \item $p\prestep^{F} q\mbox{ if and only if } p\presteps{\omega}q$.
  \item $p\prehp^{F} q\mbox{ if and only if } p\prehps{\omega}q$.
  \item $p\prehhp^{F} q\mbox{ if and only if } p\prehhps{\omega}q$.
\end{enumerate}
\end{corollary} 

\subsection{The Signature}\label{ts9}

\begin{definition}[The signature]
The signature $\Sigma$ consists of:

\begin{enumerate}
  \item A set of atomic actions $\mathsf{Act}$ ranged over $a,b,\cdots$.
  \item A set of pomsets $\mathsf{Pom}$ over $\mathsf{Act}$ ranged over $U,V,\cdots$.
  \item A constant $0$ denoting inaction without any behaviour.
  \item A constant $1$ denoting empty action which terminates immediately and successfully.
  \item A distinguished constant $\Omega$.
  \item The communication action $\gamma(a,b)$.
  \item The binary operator $\cdot$ as the sequential composition, i.e., for processes $p$ and $p'$, the process $p\cdot p'$ firstly executes $p$ followed $p'$. The process $p\cdot p'$ is abbreviated as $pp'$.
  \item The binary operator $\sharp$ as the conflict composition, i.e., the process $a\sharp b$ either executes $a$ and its successors or $b$ and its successors.
  \item The binary operator $+$ as the alternative composition, i.e., for processes $p$ and $p'$, the process $p+p'$ either executes $p$ or $p'$ alternatively.
  \item The binary operator $\between$ as the concurrent composition, i.e., for processes $p$ and $p'$, the process $p\between p'$ means $p$ and $p'$ execute concurrently, i.e., in parallel but may be with unstructured communications.
  \item The binary operator $\parallel$ as the parallel composition, i.e., for processes $p$ and $p'$, the process $p\parallel p'$ executes $p$ and $p'$ in parallel.
  \item The binary operator $\leftmerge$ as the left parallel composition, i.e., for processes $p$ and $p'$, the process $p\parallel p'$ executes $p$ and $p'$ in left parallel.
  \item The binary operator $\mid$ as the communication merge, i.e., for processes $p$ and $p'$, the process $p\mid p'$ executes with synchronous communications. 
  $$a\mid b=\begin{cases}
                \gamma(a,b), & a\leq^c b;\\
                0, & \mbox{otherwise}.
            \end{cases}$$
            where $a\leq^c b$ denotes that there exists a communication between $a$ and $b$.
  \item The unary operator $\Theta$ as confliction eliminator, i.e., for process $p$, the process $\Theta(p)$ eliminates and the $\sharp$ relations between actions in $p$.
  \item The binary unless operator $\triangleleft$ as an auxiliary operator to confliction eliminator $\Theta$.
  \item The unary operator $\partial_H$ as the encapsulation, i.e., for process $p$, the process $\partial_H(p)$ renames all actions of $p$ in the set $H$ to $0$.
  \item A set of recursive variables $X$ ranged over $x,y,z,\cdots$.
\end{enumerate}

Brackets are omitted whenever possible, with sequential composition $\cdot$ having a higher precedence than concurrent composition $\between$, parallel composition $\parallel$, left parallel composition $\leftmerge$ and communication merge $\mid$. Concurrent composition $\between$, parallel composition $\parallel$, left parallel composition $\leftmerge$ and communication merge $\mid$ have the same precedences which are higher than alternative composition $+$ and conflict composition $\sharp$, and alternative composition $+$ and conflict composition $\sharp$ have the same precedences. And we assume that $rec~x.$ has the lowest precedence of all the operators in $\Sigma$.
\end{definition}

\begin{definition}[Syntax of basic process language]
The syntax of the basic process language $\mathbf{M}$ is given by the following BNF grammar:

$p::=\Omega~|~0~|~1~|~U~|~\gamma(a,b)~|~p\cdot p~|~a\sharp b~|~x~|~p+p~|~p\parallel p~|~p\leftmerge p~|~p\mid p~|~p\between p~|~\Theta(p)~|~p\triangleleft p~|~\partial_H(p)~|~rec~x.p~$

where $a,b\in\mathsf{Act}$, $U\in\mathbf{Pom}$, $p\in \mathsf{Proc}$, $x$ is recursive variable and $rec~x.p$ stands for the process defined by the recursive equation $x=p$.
\end{definition}

\subsection{Operational Semantics}\label{os9}

In this section, we give the operational semantics of the language $\mathbf{M}$. The predicate $\surd$ represents successful termination, $\xrightarrow{a}\surd$ represents successful termination after execution of the action $a\in\mathsf{Act}$, $\xrightarrow{U}\surd$ represents successful termination after execution of the action $U\in\mathsf{Pom}$ and $\xrightarrow{ }\surd$ represents successful termination without execution of the any action. The divergence predicate $p\uparrow$ represents that $p$ is divergent, i.e., $p$ has an infinite internal computation. While the convergence predicate $p\downarrow$ represents that $p$ is not divergent $p\nuparrow$, i.e., convergent, $p$ has no infinite internal computation. The following are the PTSS of the language $\mathbf{M}$, where $p,q\in\mathsf{Proc}$.

The PTSS of $\Omega$ is as follows.

$$\frac{}{\Omega\uparrow}$$
$$\frac{}{\Omega\xrightarrow{1}\Omega}$$

The PTSS of action $1$, $a\in\mathsf{Act}$ and $U\in\mathsf{Pom}$ is as follows. Note that, there is no any transition rules for $0$.

$$\frac{}{0\downarrow}\quad\frac{}{1\downarrow}\quad \frac{}{a\downarrow} \quad\frac{}{U\downarrow}$$
$$\frac{}{1\xrightarrow{ }\surd}\quad\frac{}{a\xrightarrow{a}\surd}\quad\frac{}{U\xrightarrow{U}\surd}$$

The PTSS of sequential composition is as follows.

$$\frac{p\downarrow\quad q\downarrow}{p\cdot q\downarrow}$$
$$\frac{p\xrightarrow{U}\surd}{p\cdot q\xrightarrow{U} q}\quad\frac{p\xrightarrow{U}p'}{p\cdot q\xrightarrow{U} p'\cdot q}$$

The PTSS of alternative composition is as follows.

$$\frac{q\downarrow\quad q\downarrow}{p+q\downarrow}$$
$$\frac{p\xrightarrow{U}\surd}{p+ q\xrightarrow{U}\surd} \quad\frac{p\xrightarrow{U}p'}{p+ q\xrightarrow{U}p'} \quad\frac{q\xrightarrow{U}\surd}{p+ q\xrightarrow{U}\surd} \quad\frac{q\xrightarrow{U}q'}{p+ q\xrightarrow{U}q'}$$

The PTSS of concurrent composition is as follows.

$$\frac{p\downarrow\quad q\downarrow}{p\between q\downarrow}$$
$$\frac{p\xrightarrow{a}\surd\quad q\xrightarrow{b}\surd}{p\between q\xrightarrow{\step{a,b}}\surd} \quad\frac{p\xrightarrow{a}p'\quad q\xrightarrow{b}\surd}{p\between q\xrightarrow{\step{a,b}}p'}$$
$$\frac{p\xrightarrow{a}\surd\quad q\xrightarrow{b}q'}{p\between q\xrightarrow{\step{a,b}}q'} \quad\frac{p\xrightarrow{a}p'\quad q\xrightarrow{b}q'}{p\between q\xrightarrow{\step{a,b}}p'\between q'}$$
$$\frac{p\xrightarrow{a}\surd\quad q\xrightarrow{b}\surd}{p\between q\xrightarrow{\gamma(a,b)}\surd} \quad\frac{p\xrightarrow{a}p'\quad q\xrightarrow{b}\surd}{p\between q\xrightarrow{\gamma(a,b)}p'}$$
$$\frac{p\xrightarrow{a}\surd\quad q\xrightarrow{b}q'}{p\between q\xrightarrow{\gamma(a,b)}q'} \quad\frac{p\xrightarrow{a}p'\quad q\xrightarrow{b}q'}{p\between q\xrightarrow{\gamma(a,b)}p'\between q'}$$

The PTSS of parallel composition is as follows.

$$\frac{p\downarrow\quad q\downarrow}{p\parallel q\downarrow}$$
$$\frac{p\xrightarrow{a}\surd\quad q\xrightarrow{b}\surd}{p\parallel q\xrightarrow{\step{a,b}}\surd} \quad\frac{p\xrightarrow{a}p'\quad q\xrightarrow{b}\surd}{p\parallel q\xrightarrow{\step{a,b}}p'}$$
$$\frac{p\xrightarrow{a}\surd\quad q\xrightarrow{b}q'}{p\parallel q\xrightarrow{\step{a,b}}q'} \quad\frac{p\xrightarrow{a}p'\quad q\xrightarrow{b}q'}{p\parallel q\xrightarrow{\step{a,b}}p'\between q'}$$

The PTSS of left parallel composition is as follows, where $\leq^e$ is the execution order.

$$\frac{p\downarrow\quad q\downarrow}{p\leftmerge q\downarrow}$$
$$\frac{p\xrightarrow{a}\surd\quad q\xrightarrow{b}\surd\quad a\leq^e b}{p\leftmerge q\xrightarrow{\mset{a,b}}\surd} \quad\frac{p\xrightarrow{a}p'\quad q\xrightarrow{b}\surd\quad a\leq^e b}{p\leftmerge q\xrightarrow{\mset{a,b}}p'}$$
$$\frac{p\xrightarrow{a}\surd\quad q\xrightarrow{b}q'\quad a\leq^e b}{p\leftmerge q\xrightarrow{\mset{a,b}}q'} \quad\frac{p\xrightarrow{a}p'\quad q\xrightarrow{b}q'\quad a\leq^e b}{p\leftmerge q\xrightarrow{\mset{a,b}}p'\between q'}$$

The PTSS of communication merge is as follows.

$$\frac{p\downarrow\quad q\downarrow}{p\mid q\downarrow}$$
$$\frac{p\xrightarrow{a}\surd\quad q\xrightarrow{b}\surd}{p\mid q\xrightarrow{\gamma(a,b)}\surd} \quad\frac{p\xrightarrow{a}p'\quad q\xrightarrow{b}\surd}{p\mid q\xrightarrow{\gamma(a,b)}p'}$$
$$\frac{p\xrightarrow{a}\surd\quad q\xrightarrow{b}q'}{p\mid q\xrightarrow{\gamma(a,b)}q'} \quad\frac{p\xrightarrow{a}p'\quad q\xrightarrow{b}q'}{p\mid q\xrightarrow{\gamma(a,b)}p'\between q'}$$

The PTSS of encapsulation operator is as follows.

$$\frac{p\downarrow}{\partial_H(p)\downarrow}$$
$$\frac{p\xrightarrow{a}\surd\quad a\notin H}{\partial_H(p)\xrightarrow{a}\surd}\quad\frac{p\xrightarrow{a}p'\quad a\notin H}{\partial_H(p)\xrightarrow{a}\partial_H(p')}$$

The PTSS of confliction, confliction eliminator and the auxiliary unless operator is as follows, where $\leq^e$ is the execution order.

$$\frac{p\downarrow}{\Theta(p)\downarrow}\quad\frac{p\downarrow\quad q\downarrow}{p\triangleleft q\downarrow}$$
$$\frac{p\xrightarrow{a}\surd\quad a\sharp b}{\Theta(p)\xrightarrow{a}\surd} \quad\frac{p\xrightarrow{b}\surd\quad a\sharp b}{\Theta(p)\xrightarrow{b}\surd}$$
$$\frac{p\xrightarrow{a}p'\quad a\sharp b}{\Theta(p)\xrightarrow{a}\Theta(p')} \quad\frac{p\xrightarrow{b}p'\quad a\sharp b}{\Theta(p)\xrightarrow{b}\Theta(p')}$$
$$\frac{p\xrightarrow{c}\surd \quad q\xnrightarrow{b}\quad a\sharp b\quad c\leq^e a}{p\triangleleft q\xrightarrow{c}\surd}
\quad\frac{p\xrightarrow{c}p' \quad q\xnrightarrow{b}\quad a\sharp b\quad c\leq^e a}{p\triangleleft q\xrightarrow{c}p'}$$
$$\frac{p\xrightarrow{a}\surd \quad q\xnrightarrow{b}\quad a\sharp b}{p\triangleleft q\xrightarrow{ }\surd}
\quad\frac{p\xrightarrow{a}p' \quad q\xnrightarrow{b}\quad a\sharp b}{p\triangleleft q\xrightarrow{ }p'}$$
$$\frac{p\xrightarrow{a}\surd \quad q\xnrightarrow{c}\quad a\sharp b\quad b\leq^e c}{p\triangleleft q\xrightarrow{ }\surd}
\quad\frac{p\xrightarrow{a}p' \quad q\xnrightarrow{c}\quad a\sharp b\quad b\leq^e c}{p\triangleleft q\xrightarrow{ }p'}$$
$$\frac{p\xrightarrow{c}\surd \quad q\xnrightarrow{b}\quad a\sharp b\quad a\leq^e c}{p\triangleleft q\xrightarrow{ }\surd}
\quad\frac{p\xrightarrow{c}p' \quad q\xnrightarrow{b}\quad a\sharp b\quad a\leq^e c}{p\triangleleft q\xrightarrow{ }p'}$$

The PTSS of recursion is as follows.

$$\frac{p[rec~x.p/x]\downarrow}{rec~x.p\downarrow}$$
$$\frac{p[rec~x.p/x]\xrightarrow{U}p'}{rec~x.p\xrightarrow{U}p'}$$

\subsection{Denotational Semantics}\label{ds9}

\begin{definition}[Partial order over $\mathbf{PST}(\mathsf{Pom})$]
A partial order $\leq_{\mathbf{PST}(\mathsf{Pom})}$ is defined over $\mathbf{PST}(\mathsf{Pom})$ as follows, for two trees $t_1,t_2\in\mathbf{PST}(\mathsf{Pom})$, $t_1\leq_{\mathbf{PST}(\mathsf{Pom})}t_2$, if:

\begin{enumerate}
  \item If $\langle U,t_1'\rangle\in t_1$, then $t_1'\leq_{\mathbf{PST}(\mathsf{Pom})}t_2'$ for some $\langle U, t_2'\rangle\in t_2$.
  \item If $\bot\in t_2$, then $\bot\in t_1$.
  \item If $\langle U,t_2'\rangle\in t_2$, then either $\bot\in t_1$ or $t_1'\leq_{\mathbf{PST}(\mathsf{Pom})}t_2'$ for some $\langle U, t_1'\rangle\in t_1$.
\end{enumerate}
\end{definition}

\begin{lemma}
$\langle \mathbf{PST}(\mathsf{Pom}),\leq_{\mathbf{PST}(\mathsf{Pom})}\rangle$ is an algebraic cpo.
\end{lemma}

Similarly to the functions over $\mathbf{PAT}$ in the previous chapters, we can define functions over $\mathbf{PST}(\mathsf{Pom})$ for every function symbol in $\Sigma$: $\Omega_{\mathbf{PST}(\mathsf{Pom})}$, $0_{\mathbf{PST}(\mathsf{Pom})}$, $1_{\mathbf{PST}(\mathsf{Pom})}$, $U_{\mathbf{PST}(\mathsf{Pom})}$, $\cdot_{\mathbf{PST}(\mathsf{Pom})}$, $+_{\mathbf{PST}(\mathsf{Pom})}$, $\parallel_{\mathbf{PST}(\mathsf{Pom})}$, $\gamma(a,b)_{\mathbf{PST}(\mathsf{Pom})}$, $\mid_{\mathbf{PST}(\mathsf{Pom})}$, $\between_{\mathbf{PST}(\mathsf{Pom})}$, $\partial_H(p)_{\mathbf{PST}(\mathsf{Pom})}$, $\sharp_{\mathbf{PST}(\mathsf{Pom})}$, $\Theta(p)_{\mathbf{PST}(\mathsf{Pom})}$, $\triangleleft_{\mathbf{PST}(\mathsf{Pom})}$. They are all continuous operations on the algebra cpo $\mathbf{PST}(\mathsf{Pom})$.

\begin{proposition}
$\langle \mathbf{PST}(\mathsf{Pom}),\leq_{\mathbf{PST}(\mathsf{Pom})},\Sigma_{\mathbf{PST}(\mathsf{Pom})}\rangle$ is a $\Sigma$-domain.
\end{proposition}

\begin{theorem}[Full Abstraction for $\mathbf{PST}(\mathsf{Pom})$]
If $p,q\in\mathbf{M}$, then $p\prepomset^F q$ if and only if $\mathbf{PST}(\mathsf{Pom})\sembrack{p}\leq_{\mathbf{PST}(\mathsf{Pom})}\mathbf{PST}(\mathsf{Pom})\sembrack{q}$.
\end{theorem}

\subsection{Axiomatic Semantics}\label{as9}

The proof system $\mathbf{DED}(E)$ of inequations of $\mathbf{M}$ is shown in \cref{ProofSystemEofM}.

\begin{center}
    \begin{longtable}{|c|c|}
      \caption{Proof system $\mathbf{DED}(E)$ of $\mathbf{M}$}\\
      \hline No. &Axiom\\
      \hline
      \endfirsthead
      \multicolumn{2}{c}{\bfseries Continuing: Proof system $\mathbf{DED}(E)$ of $\mathbf{M}$} \\
      \hline No. &Axiom\\
      \hline
      \endhead
      \hline
      \multicolumn{2}{r}{to be continued\ldots} \\
      \endfoot
      \hline
      \multicolumn{2}{r}{end of Proof system $\mathbf{DED}(E)$ of $\mathbf{M}$} \\
      \endlastfoot
      $A1$ & $x+ y = y+ x$\\
      $A2$ & $(x+ y)+ z = x+ (y+ z)$\\
      $A3$ & $x+ x = x$\\
      $A4$ & $x+ 0 = x$\\
      $A5$ & $(x+y)\cdot z=x\cdot z+y\cdot z$\\
      $A6$ & $(x\cdot y)\cdot z=x\cdot(y\cdot z)$\\
      $A7$ & $0\cdot x =0$\\
      $A8$ & $x\cdot 1=x$\\
      $A9$ & $1\cdot x=x$\\
      $P1$ & $x\between y = x\parallel y + x\mid y$\\
      $P2$ & $a\parallel (b\cdot y) = (a\parallel b)\cdot y$\\
      $P3$ & $(a\cdot x)\parallel b = (a\parallel b)\cdot x$\\
      $P4$ & $(a\cdot x)\parallel (b\cdot y) = (a\parallel b)\cdot(x\between y)$\\
      $P5$ & $(x+ y)\parallel z = (x\parallel z)+ (y\parallel z)$\\
      $P6$ & $x\parallel (y+ z) = (x\parallel y)+ (x\parallel z)$\\
      $P7$ & $0\parallel x = 0$\\
      $P8$ & $x\parallel 0 = 0$\\
      $P9$ & $x\parallel 1=x$\\
      $P10$ & $1\parallel x=x$\\
      $C1$ & $a\mid b = \gamma(a,b)$\\
      $C2$ & $a\mid (b\cdot y) = \gamma(a,b)\cdot y$\\
      $C3$ & $(a\cdot x)\mid b = \gamma(a,b)\cdot x$\\
      $C4$ & $(a\cdot x)\mid (b\cdot y) = \gamma(a,b)\cdot (x\between y)$\\
      $C5$ & $(x+ y)\mid z = (x\mid z) + (y\mid z)$\\
      $C6$ & $x\mid (y+ z) = (x\mid y)+ (x\mid z)$\\
      $C7$ & $0\mid x = 0$\\
      $C8$ & $x\mid 0 = 0$\\
      $C9$ & $1\mid x = 0$\\
      $C10$ & $x\mid 1 = 0$\\
      $CE1$ & $\Theta(a) = a$\\
      $CE2$ & $\Theta(U) = U$\\
      $CE3$ & $\Theta(0) = 0$\\
      $CE4$ & $\Theta(1) = 1$\\
      $CE5$ & $\Theta(x+ y) = \Theta(x) + \Theta(y)$\\
      $CE6$ & $\Theta(x\cdot y)=\Theta(x)\cdot\Theta(y)$\\
      $CE7$ & $(\exists a\in \mathsf{Act}(x),b\in\mathsf{Act(y)},a\sharp b)\quad\Theta(x\parallel y) = ((\Theta(x)\triangleleft y)\parallel y)+ ((\Theta(y)\triangleleft x)\parallel x)$\\
      $CE8$ & $(\exists a\in \mathsf{Act}(x),b\in\mathsf{Act(y)},a\sharp b)\quad\Theta(x\mid y) = ((\Theta(x)\triangleleft y)\mid y)+ ((\Theta(y)\triangleleft x)\mid x)$\\
      $U1$ & $(a\sharp b,c\leq a)\quad c\triangleleft b = c$\\
      $U2$ & $(a\sharp b)\quad a\triangleleft b = 1$\\
      $U3$ & $(a\sharp b,b\leq c)\quad a\triangleleft c = 1$\\
      $U4$ & $(a\sharp b,b\leq c)\quad c\triangleleft a = 1$\\
      $U5$ & $a\triangleleft 0 = a$\\
      $U6$ & $0 \triangleleft a = 0$\\
      $U7$ & $a\triangleleft 1 = a$\\
      $U8$ & $1\triangleleft a = 1$\\
      $U9$ & $(x+ y)\triangleleft z = (x\triangleleft z)+ (y\triangleleft z)$\\
      $U10$ & $(x\cdot y)\triangleleft y = (x\triangleleft y)\cdot(y\triangleleft y)$\\
      $U11$ & $(x\parallel y)\triangleleft z = (x\triangleleft z)\parallel (y\triangleleft z)$\\
      $U12$ & $(x\mid y)\triangleleft z = (x\triangleleft z)\mid (y\triangleleft z)$\\
      $U13$ & $x\triangleleft (y+ z) = (x\triangleleft y)\triangleleft z$\\
      $U14$ & $x\triangleleft (y\cdot z)=(x\triangleleft y)\triangleleft z$\\
      $U15$ & $x\triangleleft (y\parallel z) = (x\triangleleft y)\triangleleft z$\\
      $U16$ & $x\triangleleft (y\mid z) = (x\triangleleft y)\triangleleft z$\\
      $D1$ & $(a\notin H)\quad\partial_H(a) = a$\\
      $D2$ & $(a\in H)\quad \partial_H(a) = 0$\\
      $D3$ & $\partial_H(0) = 0$\\
      $D4$ & $\partial_H(1) = 1$\\
      $D5$ & $\partial_H(x+ y) = \partial_H(x)+\partial_H(y)$\\
      $D6$ & $\partial_H(x\cdot y) = \partial_H(x)\cdot\partial_H(y)$\\
      $D7$ & $\partial_H(x\parallel y) = \partial_H(x)\parallel\partial_H(y)$\\
      $\Omega1$ & $x+ \Omega \leq \Omega$\\
      $\Omega2$ & $\Omega\cdot x \leq \Omega$\\
      $\Omega3$ & $x\between \Omega \leq \Omega$\\
      $\Omega4$ & $x\parallel \Omega \leq \Omega$\\
      $\Omega5$ & $x\mid \Omega \leq \Omega$\\
      $\Omega6$ & $\partial_H(\Omega) \leq \Omega$\\
      $\Omega7$ & $\Theta(\Omega) \leq \Omega$\\
      $\Omega8$ & $\Omega\triangleleft x \leq \Omega$\\
      $\Omega9$ & $x\triangleleft \Omega \leq \Omega$
      \label{ProofSystemEofM}
    \end{longtable}
\end{center}

The proof system $\mathbf{DED}(E')$ of inequations of $\mathbf{M}$ which include those shown in \cref{ProofSystemEofM} and those shown in \cref{ProofSystemEofM2}.

\begin{center}
    \begin{longtable}{|c|c|}
      \caption{Inequations of $\leftmerge$}\\
      \hline No. &Axiom\\
      \hline
      \endfirsthead
      \multicolumn{2}{c}{\bfseries Continuing: Inequations of $\leftmerge$} \\
      \hline No. &Axiom\\
      \hline
      \endhead
      \hline
      \multicolumn{2}{r}{to be continued\ldots} \\
      \endfoot
      \hline
      \multicolumn{2}{r}{end of Inequations of $\leftmerge$} \\
      \endlastfoot
      $P11$ & $x\parallel y = x\leftmerge y + y\leftmerge x$\\
      $P12$ & $a\leftmerge (b\cdot y) = (a\leftmerge b)\cdot y$\\
      $P13$ & $(a\cdot x)\leftmerge b = (a\leftmerge b)\cdot x$\\
      $P14$ & $(a\cdot x)\leftmerge (b\cdot y) = (a\leftmerge b)\cdot(x\between y)$\\
      $P15$ & $(x+ y)\leftmerge z = (x\leftmerge z)+ (y\leftmerge z)$\\
      $P16$ & $0\leftmerge x = 0$\\
      $P17$ & $x\leftmerge 1=x$\\
      $P18$ & $1\leftmerge x=x$\\
      $CE9$ & $(\exists a\in \mathsf{Act}(x),b\in\mathsf{Act(y)},a\sharp b)\quad\Theta(x\leftmerge y) = ((\Theta(x)\triangleleft y)\leftmerge y)+ ((\Theta(y)\triangleleft x)\leftmerge x)$\\
      $U17$ & $(x\leftmerge y)\triangleleft z = (x\triangleleft z)\parallel (y\triangleleft z)$\\
      $U18$ & $x\triangleleft (y\leftmerge z) = (x\triangleleft y)\triangleleft z$\\
      $D8$ & $\partial_H(x\leftmerge y) = \partial_H(x)\leftmerge\partial_H(y)$\\
      $\Omega10$ & $\Omega\leftmerge x \leq \Omega$
      \label{ProofSystemEofM2}
    \end{longtable}
\end{center}

\begin{definition}
The head normal form (hnf) is defined inductively as follows:

\begin{enumerate}
  \item $0$ and $\Omega$ are hnf's.
  \item If $p$ is an hnf, any term of the form $\sum \{Up|U\in \mathsf{Pom}\}$ is an hnf.
\end{enumerate}
\end{definition}

\begin{theorem}[Head normal form theorem]
Every convergent term $p$ has an hnf, i.e., $p\downarrow$ implies $\vdash_{\mathbf{rDED}(E)}p=h$, where $h$ is an hnf.
\end{theorem}

\begin{proposition}
$\mathbf{\Omega DED}(E)$ is sound and complete with respect to $\prepomset^F$ over $FREC_{\Sigma}$.
\end{proposition}

\begin{theorem}[Full Abstraction]
The interpretation $\mathbf{PST}(\mathsf{Pom})$ is fully abstract with respect to $\prepomset^F$ over $\mathbf{M}$.
\end{theorem}

\begin{theorem}[Soundness and Completeness]
The proof system $\mathbf{\omega DED}(E)$ is sound and complete with respect to $\prepomset^F$ over $REC_{\Sigma}$.
\end{theorem}

\subsection{The Trinity}\label{t9}

\begin{proposition}
$\mathbf{\Omega DED}(E)$ is sound and complete with respect to $\prestep^F$ over $FREC_{\Sigma}$.
\end{proposition}

\begin{theorem}[Full Abstraction]
The interpretation $\mathbf{PST}(\mathsf{Pom})$ is fully abstract with respect to $\prestep^F$ over $\mathbf{M}$.
\end{theorem}

\begin{theorem}[Soundness and Completeness]
The proof system $\mathbf{\omega DED}(E)$ is sound and complete with respect to $\prestep^F$ over $REC_{\Sigma}$.
\end{theorem}

\begin{proposition}
$\mathbf{\Omega DED}(E)$ is sound and complete with respect to $\prehp^F$ over $FREC_{\Sigma}$.
\end{proposition}

\begin{theorem}[Full Abstraction]
The interpretation $\mathbf{PST}(\mathsf{Pom})$ is fully abstract with respect to $\prehp^F$ over $\mathbf{M}$.
\end{theorem}

\begin{theorem}[Soundness and Completeness]
The proof system $\mathbf{\omega DED}(E)$ is sound and complete with respect to $\prehp^F$ over $REC_{\Sigma}$.
\end{theorem}

\begin{proposition}
$\mathbf{\Omega DED}(E')$ is sound and complete with respect to $\prehhp^F$ over $FREC_{\Sigma}$.
\end{proposition}

If all pomsets $U\in\mathsf{Pom}$ in $\mathbf{PST}(\mathsf{Pom})$ are strict (see \cref{strictpomset}), the $\mathbf{PST}(\mathsf{Pom})$ is called strict, denoted $\mathbf{PST}(\mathsf{SPom})$. Every $\mathbf{PST}(\mathsf{Pom})$ has a unique $\mathbf{PST}(\mathsf{SPom})$.

\begin{theorem}[Full Abstraction]
The interpretation $\mathbf{PST}(\mathsf{SPom})$ is fully abstract with respect to $\prehhp^F$ over $\mathbf{M}$.
\end{theorem}

\begin{theorem}[Soundness and Completeness]
The proof system $\mathbf{\omega DED}(E')$ is sound and complete with respect to $\prehhp^F$ over $REC_{\Sigma}$.
\end{theorem} 
\bibliographystyle{elsarticle-num}
\newpage\bibliography{Refs-CFMAI}

\begin{thebibliography}{10}
\expandafter\ifx\csname url\endcsname\relax
  \def\url#1{\texttt{#1}}\fi
\expandafter\ifx\csname urlprefix\endcsname\relax\def\urlprefix{URL }\fi
\expandafter\ifx\csname href\endcsname\relax
  \def\href#1#2{#2} \def\path#1{#1}\fi

\bibitem{APTC}
Y.~Wang, Algebraic theory for true concurrency, Elsevier AP, 2023.

\bibitem{APTC2}
Y.~Wang, Handbook of truly concurrent process algebra, Elsevier MK, 2023.

\bibitem{PN00}
C.~A. Petri, Non-sequential processes, GMD-ISF Report 77~(5) (1977).

\bibitem{PN01}
C.~A. Petri, General net theory. communication disciplines, in: proc. Joint IBM
  University of Newcastle Seminar, B. Shaw ed., Newcastle GB, 1976.

\bibitem{PN02}
C.~A. Petri, Concurrency as a basis of systems thinking, in: Proc. from 5th
  Scandinavian Logic Symposium, 1979, pp. 143--162.

\bibitem{ES}
G.~Winskel, Event structures, in: advanced course on Petri nets, Springer,
  1986, pp. 325--392.

\bibitem{CCS}
R.~Milner, A calculus of communicating systems, Springer, 1980.

\bibitem{CSP1}
C.~A.~R. Hoare, Communicating sequential processes, Communications of the ACM
  21~(8) (1978) 666--677.
\newblock \href {https://doi.org/10.1145/359576.359585}
  {\path{doi:10.1145/359576.359585}}.

\bibitem{CSP2}
C.~A.~R. Hoare, Communicating sequential processes, Prentice Hall International
  Series in Computer Science, Prentice-Hall, 1985.

\bibitem{ACP}
W.~Fokkink, Introduction to process algebra, 2nd Edition, Springer-Verlag,
  2007.

\bibitem{TS}
M.~Hennessy, Algebraic theory of processes, Foundations of Computing, MIT
  Press, Cambridge, MA, 1988.

\bibitem{CKA7}
T.~Kapp\'{e}, Concurrent kleene algebra: completeness and decidability
  (doctoral dissertation, Ph.D. thesis, UCL (University College London) (2020).

\bibitem{DS}
D.~Scott, C.~Strachey, Towards a mathematical semantics for computer languages,
  in: J.~Fox (Ed.), Proceedings of the Symposium on Computers and Automata,
  Vol.~21 of MRI Symposium Proceedings, Polytechnic Press, Brooklyn, New York,
  1971, pp. 19--46.

\bibitem{IACA}
J.~A. Goguen, J.~W. Thatcher, E.~G. Wagner, J.~B. Wright, Initial algebra
  semantics and continuous algebras, J. ACM 24~(1) (1977) 68--95.
\newblock \href {https://doi.org/10.1145/321992.321997}
  {\path{doi:10.1145/321992.321997}}.

\bibitem{AlgebraicSemantics}
I.~Guessarian, Algebraic Semantics, Vol.~99 of Lecture Notes in Computer
  Science, Springer, 1981.
\newblock \href {https://doi.org/10.1007/3-540-10284-7}
  {\path{doi:10.1007/3-540-10284-7}}.

\end{thebibliography}

\end{document}